\documentstyle[12pt,epsf]{article}
\renewcommand{\theequation}{\thesection.\arabic{equation}}
\newlength{\extraspace}
\newlength{\extraspaces}
\newcommand{\be}{\begin{equation}
\addtolength{\abovedisplayskip}{\extraspaces}
\addtolength{\belowdisplayskip}{\extraspaces}
\addtolength{\abovedisplayshortskip}{\extraspace}
\addtolength{\belowdisplayshortskip}{\extraspace}}
\newcommand{\ee}{\end{equation}}
 
\newcommand{\ba}{\begin{eqnarray}
\addtolength{\abovedisplayskip}{\extraspaces}
\addtolength{\belowdisplayskip}{\extraspaces}
\addtolength{\abovedisplayshortskip}{\extraspace}
\addtolength{\belowdisplayshortskip}{\extraspace}}
\newcommand{\ea}{\end{eqnarray}}

\newcommand{\bas}{\begin{eqnarray*}
\addtolength{\abovedisplayskip}{\extraspaces}
\addtolength{\belowdisplayskip}{\extraspaces}
\addtolength{\abovedisplayshortskip}{\extraspace}
\addtolength{\belowdisplayshortskip}{\extraspace}}
\newcommand{\eas}{\end{eqnarray*}}
 
\newcounter{subequation}[equation]
\makeatletter

\expandafter\let\expandafter
\reset@font\csname reset@font\endcsname

\def\subeqnarray{\arraycolsep1pt
    \def\@eqnnum\stepcounter##1{\stepcounter{subequation}%
        {\reset@font\rm(\theequation\alph{subequation})}}
\jot5mm     \eqnarray}

\def\subarray{\arraycolsep1pt
    \def\@eqnnum\stepcounter##1{\stepcounter{subequation}%
        {\reset@font\rm(\alph{subequation})}}
\jot5mm     \eqnarray}

\makeatother

\newcommand{\newappendix}[1]{
\vspace{15mm}
\pagebreak[3]
\addtocounter{section}{1}
\setcounter{equation}{0}
\setcounter{subsection}{0}
\renewcommand{\theequation}{\Alph{section}.\arabic{equation}}
\begin{flushleft}
{\large\bf Appendix \Alph{section}: #1}
\end{flushleft}
\nopagebreak
\medskip
\nopagebreak}

\def\eqll#1{\\ (#1)\vadjust{\penalty10000\vskip-4.4ex}}
 
\newcommand{\newsection}[1]{
\vspace{15mm}
\pagebreak[3]
\addtocounter{section}{1}
\setcounter{equation}{0}
\setcounter{subsection}{0}
 
\begin{flushleft}
{\large\bf \thesection. #1}
\end{flushleft}
\nopagebreak
\medskip
\nopagebreak}
 
\newcommand{\newsubsection}[1]{
\vspace{1cm}
\pagebreak[3]
 
\addtocounter{subsection}{1}
\noindent{ \bf \thesection.\arabic{subsection} #1}
\nopagebreak
\vspace{2mm}
\nopagebreak}

\newcommand{\newappsection}[1]{
\vspace{1cm}
\pagebreak[3]

\addtocounter{subsection}{1}
\noindent{ \bf \Alph{section}.\arabic{subsection} #1}
\nopagebreak
\vspace{2mm}
\nopagebreak}

\newcommand{\C}{\mbox{$\,${\sf I}\hspace{-1.2ex}{\bf C}}}

\newcommand{\R}{\mbox{\rm I\hspace{-.4ex}R}}

\newcommand{\1}{\mbox{1\hspace{-.8ex}1}}
\newcommand{\bra}{\langle}
\newcommand{\ket}{\rangle}
\newcommand{\ra}{\rightarrow}

\newcommand{\rra}{\ \longrightarrow \ }

\newcommand{\is}{ &\! =\! & }
\newcommand{\nonum}{\nonumber \\[1.5mm]}
\newcommand{\sspace}{\makebox[1cm]{ }}
\newcommand{\bspace}{\makebox[2cm]{ }}
\newcommand{\nspace}{\!\!\!\!\!\!\!\!\!\!}

\newcommand{\Tr}{{\rm Tr}}

\newcommand{\inv}{^{-1}}

\newcommand{\th}{{\theta}}
\newcommand{\lb}{\lambda}
\newcommand{\sh}{{\rm sh}}
\newcommand{\ch}{{\rm ch}}

\newcommand{\fbar}{{\overline{f}}}

\newcommand{\qbar}{{\overline{q}}}

\newcommand{\Tbar}{{\overline{T}}}
\newcommand{\Lbar}{{\overline{L}}}

\newcommand{\cD}{{\cal D}}
\newcommand{\cE}{{\cal E}}

\newcommand{\cM}{{\cal M}}

\newcommand{\cP}{{\cal P}}

\newcommand{\cT}{{\cal T}}

\newcommand{\cV}{{\cal V}}

\newcommand{\eps}{\epsilon}

\newcommand{\thh}{\widehat{\theta}}

\newcommand{\cVh}{\widehat{\cV}}

\newcommand{\Obra}{{\langle \Theta|}}
\newcommand{\Oket}{{|\Omega\rangle}}

\renewcommand{\P}{\,\mbox{I\hspace{-0.5ex}P}}

\def\ft#1#2{{\textstyle {\frac{#1}{#2}} }}
\newcommand{\n}{{\sc n}}
\newcommand{\DI}{{\cal D}_{\rm I}}
\begin{document}
%
\begin{titlepage}
%
\renewcommand{\thefootnote}{\fnsymbol{footnote}}
\mbox{ }
\vspace{-10mm}
\begin{flushright}
{\small LPTENS-99/56}\\
{\small December 1999}
\end{flushright} 
\bigskip

\begin{center}
{\LARGE An Algebraic Bootstrap for Dimensionally Reduced }\\[4mm]
{\LARGE Quantum Gravity}\\[1.5cm]
{\bf M. Niedermaier\footnote{E-mail: nie@prospero.phyast.pitt.edu}}\\[2mm]
{\small\sl Department of Physics }\\
{\small\sl 100 Allen Hall, University of Pittsburgh} \\
{\small\sl Pittsburgh, PA 15260, USA}\\[8mm]
\setcounter{footnote}{3}
{\bf H. Samtleben\footnote{E-mail: henning@lpt.ens.fr}}\\[2mm]
{\small\sl Laboratoire de Physique Th{\'e}orique}\\
{\small\sl de l' Ecole Normale Sup{\'e}rieure}
\footnote{UMR 8548: Unit{\'e} Mixte de Recherche du Centre
National de la Recherche Scientifique et de l'Ecole Normale
Sup{\'e}rieure.}  \\
{\small\sl 24 Rue Lhomond, 75231 Paris C{\'e}dex 05,
France}
\vspace{0.7cm}
\end{center}

\begin{abstract}
Cylindrical gravitational waves of Einstein gravity are described by
an integrable system (Ernst system) whose quantization is a long
standing problem. We propose to bootstrap the quantum theory along
the following lines: The quantum theory is described in terms of
matrix elements e.g.~of the metric operator between
spectral-transformed multi-vielbein configurations. These matrix
elements are computed exactly as solutions of a recursive system of
functional equations, which in turn is derived from an underlying
quadratic algebra.  The Poisson algebra emerging in its classical
limit links the spectral-transformed vielbein and the non-local
conserved charges and can be derived from first principles within the
Ernst system.

Among the noteworthy features of the quantum theory are: (i) The issue
of (non-)renormalizability is sidestepped and (ii) there is an
apparently unavoidable ``spontaneous'' breakdown of the $SL(2,\R)$
symmetry that is a remnant of the 4D diffeomorphism invariance in the
compactified dimensions.
\end{abstract}
\vfill

\renewcommand{\thefootnote}{\arabic{footnote}}
\setcounter{footnote}{0}

\end{titlepage}

\newsection{Introduction}

Attempts to construct a quantum theory of gravity based on a
functional integral formulation have so far been unsuccessful.
Initially this was thought to be a breakdown only of its perturbative
expansion. Meanwhile various reasonably looking discretized versions
of the functional integral have (in all likelihood) failed to produce
a continuum limit of the desired form. This may indicate the necessity
to incorporate specific forms of matter, it may indicate a failure of
the functional integral approach, or the analogy to the quantization
of conventional field theories may just be physically misleading
altogether.  In any case it seems desirable to locate the source of
the problem more clearly by studying model situations sufficiently
complex to make (non-)renormalizability an issue, but tractable enough
to be mathematically controllable.

An intriguing such system is the Ernst system, an infinite dimensional
subsector (the two Killing vector field reduction) of the full phase
space of general relativity; see \cite{Erns68}--\cite{Nico91} and
references therein. Together with its abelian truncation it has become
a prominent testing ground to explore certain quantum issues of
gravity; see e.g.~\cite{Kuch71}--\cite{KorSam97b}.  The reduced phase
space is equivalent to that of a two-dimensional diffeomorphism
invariant field theory. The latter couples 2D gravity via a dilaton
field $\rho$ to a 2D matter system equivalent to the noncompact
$O(1,2)$ nonlinear sigma-model. This means, the matter degrees of
freedom are mappings $n:\Sigma \ra H_2$ from the 2D spacetime manifold
$\Sigma$ to the hyperboloid $H_2 = \{n = (n^0,n^1,n^2) \in
\R^{1,2}\,|\, n\cdot n = (n^0)^2 - (n^1)^2 - (n^2)^2 =1, \,n^0 >0\}$.
If $h_{\mu\nu}$ denotes the (Lorentzian) metric on $\Sigma$ and
$R^{(2)}(h)$ its scalar curvature, the action of the two-dimensional
system is given by
\be
S = \int_{\Sigma} d^2 x 
\sqrt{-h}\left\{ - \rho R^{(2)}(h) + 
\frac{1}{2}\, \rho \, h^{\mu\nu} \partial_{\mu} n \cdot 
\partial_{\nu}n\right\}\;,
\label{i1}
\ee  
where `$\;\cdot\;$' is the bilinear form on $\R^{1,2}$.  The relation
to the coset action principles usually employed in the literature and
that to the original Ernst variables is described in appendix~A. 

Depending on the signature of $\Sigma$, this sector physically
describes either stationary axisymmetric solutions or gravitational
waves with additional symmetries. The latter case comprises --
depending on the norm of $\partial_{\mu}\rho$ -- the Gowdy universes,
colliding plane wave solutions, and (the case considered here)
cylindrical gravitational waves. The original Einstein-Rosen waves
form a collinearly polarized subsector, they have $n^1 \equiv 0$,
and are thus described by the abelian $O(1,1)$ subtheory of (\ref{i1}).

At first sight, the action (\ref{i1}) seems perfectly amenable to
conventional quantum field theoretical techniques.  Upon closer
inspection however, one is quickly led to address the following
questions:
\begin{itemize}
\item[(i)] Can one expect 2D conformal invariance to be unbroken?
\item[(ii)] What is the physics of the flat space
sigma-model with target space $H_2$? 
\item[(iii)] Is the theory (\ref{i1}) renormalizable in perturbation
theory?  
\end{itemize}
Let us briefly comment on these issues: 

\begin{itemize}
\item[(i)] 
Flat space non-linear sigma-models are known to exhibit dynamical mass
generation, destroying the 2D conformal invariance of the classical
theory.  In the gauge $h_{\mu\nu} = e^{2 \sigma} \eta_{\mu\nu}$ (where
$\eta_{\mu\nu}$ is the flat 2D Minkowski metric) the classical system
(\ref{i1}) likewise exhibits a conformal symmetry, and off-hand there
is no reason why it should not again be broken in the quantum
theory. Rather, there are strong indications for a dynamical breaking
of the conformal symmetry from other 2D quantum gravity models
\cite{KlKoPo93,AmbGho94}. In any case, it seems advisable not to
employ 2D conformal invariance as a guiding principle to construct the
quantum theory.
\item[(ii)] Although sigma-models with a non-compact target space have
been studied for some time \cite{GomHa84}--\cite{vHol87}, even basic
qualitative features are unknown. Perturbative renormalizability in
the presence of an infrared cutoff should largely parallel that of the
compact case \cite{Zinn89}. However the free field Fock space carrying
a nonlinear realization of the $O(1,N)$ group action has indefinite
metric \cite{vHol87} and the projection onto a physical state space is
not fully understood. Likewise the extent to which the results
\cite{Davi81,Elit83} on the infrared finiteness carry over has to be
examined. One might hope a non-perturbative construction to be
feasible via the lattice approach. Specifically, the results of a
lowest order large $N$ analysis of the $O(1,N)$ models
\cite{GomHa84,Ha85,MorNoj86} suggest that a non-trivial continuum
limit theory might emerge when using spacelike hyperbolic variables
$n\cdot n = - 1/g_0,\,g_0 > 0$, and sending the bare coupling constant
$g_0$ to zero. (Classically this is the region where the Hamiltonian
density fails to be positive semi-definite.)  Relying on these
indications one would further guess that the resulting QFT is massive,
has unbroken $O(1,N)$ symmetry, a positive definite physical Hilbert
space, and a unitary S-matrix. However this scenario has not been
corroborated so far.

\item[(iii)] Here, renormalizability should mean in particular that
the target space geometry $H_2$ is left intact by the renormalization
process.  In conformal gauge one may take $\rho$ as a loop counting
parameter, in which case the model is 1-loop renormalizable in the
background field expansion \cite{dWGNR92}. For higher loops however
one cannot expect off-hand that the results for the generalized
Riemannian sigma-models \cite{Frie85,CFMP85} used in the context of
string theory, employing a much weaker notion of renormalizability,
will carry over.  Let us emphasize that the answer is not automatic
even if one takes the (ultraviolet and infrared) renormalizability of
the non-compact sigma-models in flat space for granted. One way to see
this is to fix the 2D diffeomorphism invariance in (\ref{i1})
completely and to identify $\rho$ with one of the coordinates on
$\Sigma$ (this corresponds to the 4D Weyl coordinates). Then
(\ref{i1}) becomes a flat space action, though with an explicit
coordinate dependence which in particular destroys Poincar\'{e}
invariance.  Clearly the presumed renormalizability of the
Poincar\'{e} invariant flat space theory is not very indicative for
the behavior of the other system.
\end{itemize}

In summary none of the conventional quantum field theoretical
techniques presents itself to construct a quantum theory for
(\ref{i1}).  We propose therefore to bootstrap the quantum theory
from structures linked to its classical integrability. In upshot the
quantum theory is described in terms of matrix elements e.g.~of the
metric operator between certain spectral-transformed multi-vielbein
configurations. These matrix elements are described exactly as
solutions of a recursive system of functional equations, which in turn
are derived from an underlying dynamical algebra.  The Poisson algebra
emerging in its classical limit links the spectral-transformed
vielbein and the non-local conserved charges and can be derived from
first principles within the Ernst system. Schematically one can
summarize the approach as follows:
$$
\begin{tabular}{c} 
Dynamical algebra  
\end{tabular}
\;\;\longrightarrow \;\;
\begin{tabular}{c} 
Functional equations \\
for sequences of \\
meromorphic functions
\end{tabular}
\;\;\longrightarrow\;\; 
\begin{tabular}{c} 
Exact matrix elements \\
in (renormalized) \\
quantum theory
\end{tabular}
$$
We first briefly describe the ingredients of the above scheme and then
comment on why we expect it to yield a viable quantum theory for the
Ernst system.

The data for the dynamical algebra ${\cal D}$ are: A solution $R$ of
the Yang-Baxter equation, a real parameter $\beta$, and a choice of
$*$-operation. There are two sets of generators $T^{\pm}(\theta)_a^b$
and $W_a(\theta)$, where $\theta \in \C$, and the indices $a,b$ refer
to a basis in a finite dimensional vector space. The algebra ${\cal D}
= {\cal D}(R,\beta,*)$ associated with the data is basically the most
general simple associative algebra, where ``simple'' means that all
ideals have been divided out. In the case at hand the data are as
follows: $R$ is the rational $sl_2$ $R$-matrix multiplied with a
scalar function that ensures a suitable unitarity and crossing
condition.  The parameter $\beta$ vanishes, which corresponds to the
case where the algebra has an enlarged center. The latter roughly
speaking ensures that the quantum theory has the same number of
dynamical degrees of freedom as the classical theory. For any other
value of $\beta$ this correspondence would be violated and degrees of
freedom transversal to the reduced phase space would become dynamical
(without describing a consistent bigger portion of the full phase
space). The possible $*$-structures turn out to fall into several
equivalence classes; the proper one is selected by matching it against
that of the classical Poisson algebra.

Once the correct dynamical algebra ${\cal D}$ has been identified one
considers linear functionals ${\cal D} \ni X \rightarrow \langle
\Theta |X| \Omega \rangle \in \C$ over ${\cal D}$, where the vectors
$\langle \Theta|$ and $|\Omega\rangle$ are characterized by the
conditions: $T^+(\theta)_a^b|\Omega\rangle = \gamma^+{}_a^b
|\Omega\rangle$ and $\langle\Theta |T^-(\theta)_a^b =
\,\gamma^-{}_a^b \langle\Theta|$.  The $\gamma^{\pm}$ are
numerical matrices subject to certain consistency conditions, and in
general $\langle\Theta |\Omega \rangle =0$.  Given such a functional
one can introduce the sequence of functions
\begin{equation}
f_{a_\n\ldots a_1}(\theta_\n,\ldots,\theta_1)
=\langle \Theta| W_{a_\n}(\theta_\n)\ldots W_{a_1}(\theta_1)  
|\Omega\rangle \;,\quad \n \geq 1\,.
\label{i2}
\end{equation}
The relations of the dynamical algebra then imply that this sequence
satisfies a recursive system of functional equations, where the
consistency of the underlying algebra ensures the consistency of the
functional equations.  Conversely the original functional over ${\cal
D}$ provides an abstract solution to the functional equations. The
functional equations can be grouped into two sets (I) and (II). The
set (I) characterizes the functions (\ref{i2}) for fixed $\n$, the
second set (II) prescribes how the solutions of (I) are arranged into
sequences. Essentially (II) stipulates that the functions (\ref{i2})
have simple poles whenever two $\th$-variables differ by a fixed
purely imaginary number (which in physical units equals $i l_{\sc
pl}^2/l_{\sc z}$, $l_{\sc pl}$ being the Planck length and $l_{\sc z}$
the unit length along the symmetry axis) and that the residues at
these poles are linked to a function (\ref{i2}) in $\n-2$ variables.

Having arrived at the functional equations one can in principle forget
about their derivation and take them as the starting point. The aim
then is to construct explicit solutions in the form of sequences of
meromorphic functions $f^{(\n)} = f_{a_\n\ldots
a_1}(\theta_\n,\ldots,\theta_1)$. To understand their physical
interpretation the analogy to the form factor approach \cite{Smir92}
to relativistic integrable QFTs is useful. For these systems a similar
interplay between a dynamical algebra and a system of functional
equations (the so-called form factor equations) exists
\cite{Nied95,Nied98}. The solutions $f^{(\n)}$ in that case describe
the form factors of the QFT, i.e.~matrix elements of some local
operator between the physical vacuum and an asymptotic multi-particle
state. In the case at hand of course a conventional quantum field
theoretical framework is not available; in particular scattering
states in the usual sense are unlikely to exist.  Nevertheless the
functions $f^{(\n)}$ can still be interpreted as the matrix elements
of some operator between a ground state and a ``multi-$W_a(\th)$''
configuration. From the analysis of the semi-classical limit one finds
that the $W_a(\th)$ are the quantum counterparts of a
spectral-transformed vielbein variable. Moreover the analogy to the
$\C\!\P_1$ system (cf.~appendix A) suggests to view them as
confined degrees of freedom rather than generators of scattering
states. Similarly as in the form factor approach the identification of
the operator whose matrix elements are obtained in that way requires
external input. Of course ``identification'' here to some extent just
means making contact to a conventional quantum field theoretical
formulation where the operator in question is constructed as a
composite operator from a set of fundamental field operators. As
explained before such a more conventional formulation is presently not
available for the Ernst system, so that the identification of the
operator underlying a sequence $f^{(\n)}$ in this sense has to be left
for future work.

Note that at no stage any renormalization procedure entered.  This is
a known, yet striking, feature of the form factor approach.
Mathematically it can be understood in terms of the extreme rigidity
of the underlying dynamical algebra, which simply does not allow for
interesting continuous automorphisms that could account for a
renormalization process. The same is true in the present setting and
suggests that the functions $f^{(\n)}$ indeed are exact matrix
elements that do not require renormalization.

Physically our most important finding is an apparently unavoidable
``spontaneous'' breakdown of the global $SL(2,\R)$ invariance that is
a remnant of the original 4D diffeomorphism invariance in the Killing
coordinates. Technically this emerges because the consistency
conditions on the matrices $\gamma^{\pm}$ mentioned before
eq.~(\ref{i2}) do not admit a $SL(2,\R)$ invariant solution. As a
consequence the functions $f^{(\n)}$ are invariant only under a
maximal compact $SO(2)$ subgroup.  Despite the trivial technical
origin the symmetry breaking is a genuine dynamical feature intimately
linked to the structure of the dynamical algebra; in particular it
disappears in the semi-classical limit.

As explained before we also regard it as more likely than not, that
the 2D conformal invariance of the classical theory (\ref{i1}) will be
broken in the quantum theory. A-fortiori then also the diffeomorphism
invariance in the two non-Killing coordinates of the Ernst system will
be lost.  Together both remnants of the original 4D diffeomorphism
invariance (i.e.~that it the Killing and in the non-Killing
coordinates) appear to be broken in the quantum theory for dynamical
reasons. Being a field theoretical phenomenon that does not have a
counterpart in systems with finitely many degrees of freedom, the
result may well have significance beyond the symmetry-reduced theory.

The article is organized as follows. In the next section, the
dynamical algebra $\cD$ is introduced and the functional equations
(I), (II) for the matrix elements (\ref{i2}) are derived. In section 3
we discuss the semi-classical limit of the construction and explain
the relation to the phase space of the classical Ernst system. Also
various aspects of the symmetry breaking are detailed. A compilation
of useful action principles for the matter sigma-model is deferred to
appendix A.  In section 4 a solution technique for the functional
equations (I), (II) is described.  We adapt techniques from the
algebraic Bethe ansatz which are summarized in appendix B. In
particular, the apparently new concept of ``sequential'' Bethe roots
and Bethe vectors is introduced.  Finally a list of explicit solutions
for the functions (\ref{i2}) with $\n \leq 4$ is collected in appendix
C.

\newsection{Dynamical algebra and functional equations}

Here we describe the dynamical algebra and derive the functional
equations (I), (II) for the objects (\ref{i2}).  The discussion is
naturally organized into two steps. First an algebra $\DI$ is
introduced giving rise to the functional equations (I). Then $\DI$ is
shown to still contain two-sided ideals; the factor algebra obtained
by dividing out these ideals is the full dynamical algebra $\cD$ and
gives rise to the additional functional equations (II). We begin by
describing $\DI$ and initially keep the data $(R,\beta,*)$ generic.

\newsubsection{\boldmath{$W$}-extended Yangian doubles}

The algebras $\DI$ are centrally extended Yangian doubles
$DY_{\beta}(R)$ with generators $T^\pm(\th)_a^b$ supplemented by
generators $W_a(\th)$. We write $WY(R,\beta,*)$, where $R$ is a is a
solution of the Yang-Baxter equation satisfying unitarity and
crossing. Further $\beta$ is a real parameter and $*$ refers to a
choice of $*$-operation. Lower indices refer to a basis in a finite
dimensional vector space $V$; upper indices refer to the dual basis,
where indices are raised and lowered by means of the constant charge
conjugation matrix $C_{ab}$ and its inverse $C^{ba}$, associated with
the given $R$-matrix. For the moment we only need the data $R$ and
$\beta$; the possible $*$-operations will be discussed below. To any
$R$-matrix and parameter $\beta$ one can assign an associative algebra
$WY$ with unity by postulating the following exchange relations among
its generators:
\bigskip
\eqll{T1}
\bas \jot5mm
&& R^{cd}_{mn}(\th_{12})\,T^\pm  (\th_1)_a^nT^\pm  (\th_2)_b^m
= T^\pm  (\th_2)_n^c T^\pm  (\th_1)_m^d\,R_{ab}^{mn}(\th_{12})\;,\nonum
&& R^{cd}_{mn}(\th_{12})\,T^+(\th_1)_a^nT^-(\th_2)_b^m =
T^-(\th_2)_n^c T^+(\th_1)_m^d\,
R_{ab}^{mn}(\th_{12}+2i\hbar-i\hbar\beta/\pi)\;,
\eas
\eqll{T2}
\bas \jot5mm
&& C_{mn}T^\pm  (\th)_a^m \,T^\pm  (\th-i\hbar)_b^n=C_{ab}\;,\nonum
&& C^{mn}T^\pm  (\th)_m^a \,T^\pm  (\th+i\hbar)_n^b=C^{ab}\;.
\eas
\eqll{TW}
\bas\jot5mm
&& T^-(\th_1)_a^e\,W_b(\th_2)=R^{dc}_{ab}(\th_{12})\,
W_c(\th_2)\,T^-(\th_1)_d^e\;,\nonum
&& T^+(\th_1)_a^e\,W_b(\th_2)=
R^{dc}_{ab}(\th_{12}+i 2\hbar-i\hbar\beta/\pi)\,
W_c(\th_2)\,T^+(\th_1)_d^e\;. 
\eas
\eqll{WW}
\bas \jot5mm
W_a(\th_1)\,W_b(\th_2) \is R^{dc}_{ab}(\th_{12})\;
W_c(\th_2)\,W_d(\th_1) \;,\sspace {\rm Re}\,\th_{12}\neq 0\;,
\eas
with $\th_{12}:= \th_1-\th_2$. The parameter $\hbar$ is included for
later convenience; it can be given any non-zero value by a rescaling
of the $\th$ variables. For convenience we also assume that real
boosts in the $\th$ variables are unitarily implemented, i.e.
$e^{i\lb K} X(\th) e^{-i \lb K} = X(\th + \lb),$ with $\lb \in \R$,
for any generator $X(\th)$ of the algebra. For completeness we also
note the precise form of the Yang Baxter equation
\be
R_{ab}^{nm}(\th_{12})R_{nc}^{kp}(\th_{13})R_{mp}^{ji}(\th_{23})
=R_{bc}^{nm}(\th_{23})R_{am}^{pi}(\th_{13})R_{pn}^{kj}(\th_{12})\;,
\label{r2}
\ee
and the conditions of unitarity and real analyticity 
\be
R_{ab}^{mn}(\th)\,R_{nm}^{cd}(-\th)=\delta_a^d\delta_b^c\;,\sspace
[R_{ab}^{cd}(\th)]^* = R_{ab}^{cd}(-\th^*)\,.
\label{r3}
\ee

We add a few remarks. The algebra $DY_{\beta}(R)$ is a well-known
structure. For $\beta\!=\!2\pi$ it can be viewed as a presentation of
the quantum double of some underlying infinite dimensional Hopf
algebra \cite{Drin86}. The (TW) and (WW) relations are then
characteristic for the intertwining operators between quantum double
modules \cite{LecSmi92,BerLec93}. Particular cases are the Yangian
double or the quantum double of $U_q(\widehat{g})$, in which case the
parameter $\beta$ can be related to the central extension via $c = 2
i(1- \beta/2\pi)$.  Here we do not make use of the co-algebra
structure and always treat $\beta$ as a real numerical parameter.  The
case of the critical level with enlarged center
\cite{ResSem90,FreRes96} in our conventions corresponds to
$\beta\!=\!0$; for the $sl_2$ Yangian case it will be studied in
detail below.

In preparation, we introduce the following quadratic element which
turns to play a decisive role in the $WY$ algebras
\be
D_{ab}(\th) := C_{cd} T^-(\th + i\hbar)_a^c \,T^+(\th)_b^d\,,
\label{Ddef}
\ee
It enjoys the following exchange relations
\begin{subeqnarray}
\nspace 
R_{ab}^{mn}(\th_{21}+i\hbar\beta/\pi -2i\hbar)\,D_{cn}(\th_1)\,W_m(\th_2)
\is R_{ac}^{mn}(\th_{12}+i\hbar)\,W_m(\th_2)\,D_{nb}(\th_1)\;,
\\\nspace \nspace R_{ab}^{mn}(\th_{21})\,D_{cn}(\th_1)\,T^+(\th_2)_m^d
\is R_{ac}^{mn}(\th_{12}-i\hbar + i\hbar\beta/\pi)\,T^+(\th_2)_m^d\,
D_{nb}(\th_1)\,,
\\ \nspace \nspace 
R_{ab}^{mn}(\th_{21}+i\hbar\beta/\pi -2i\hbar)\,
D_{cn}(\th_1)\,T^-(\th_2)_m^d
\is R_{ac}^{mn}(\th_{12}+i\hbar)\,T^-(\th_2)_m^d\,D_{nb}(\th_1)\;,
\\ \nspace \nspace 
R_{ab}^{mn}(\th_{21})R_{mc}^{lk}(\th_{21}+i\hbar\beta/\pi -i\hbar) 
& &\!\!\!\! D_{nk}(\th_1)\,D_{ld}(\th_2) \nonumber \\
= R_{cd}^{mn}(\th_{12}) & &  \!\!\!\!
R_{bn}^{kl}(\th_{12} +i\hbar\beta/\pi -i\hbar)\;
D_{al}(\th_2)\,D_{km}(\th_1)\,.
\label{Drel}
\end{subeqnarray}
In particular, it follows from these relations that the (generalized)
quantum current $D(\th) = C^{ab} D_{ab}(\th)$ lies in the center
of the algebra $WY(R,0,*)$, for any $R$-matrix obeying a standard
crossing relation with a symmetric charge conjugation matrix
$C_{ab}$. Much of the construction described in the rest of the paper
could therefore be transferred to a fairly general class of infinite
dimensional quantum algebras.

With the application to the quantum Ernst system in mind however 
we specialize already at this point to the $sl_2$ Yangian $R$-matrix
\cite{Skly79,FaSkTa79,Drin85}. Its charge conjugation matrix 
is anti-symmetric which enforces a slight modification of the above
scheme. The relevant $R$-matrix is then given by
\be
R_{ab}^{cd}(\th) = r(\th)\left[ - \frac{\th}{i\hbar -\th}
\delta_a^c\delta_b^d + \frac{i\hbar}{i\hbar -\th} 
\delta_a^d\delta_b^c \right]\,,\quad a,b,... =1,2\,,
\label{r1}
\ee
where 
$$
r(\th) = \frac{\ch\frac{\pi\th}{2\hbar} + i \sh\frac{\pi\th}{2\hbar}}%
{\ch\frac{\pi\th}{2\hbar} - i \sh\frac{\pi\th}{2\hbar}}\;
\frac{\Gamma\left(\frac{1}{2} + \frac{\th}{2i\hbar}\right)
\Gamma\left(- \frac{\th}{2i\hbar}\right)}%
{\Gamma\left(\frac{1}{2} - \frac{\th}{2i\hbar}\right)
\Gamma\left(\frac{\th}{2i\hbar}\right)}\,,
$$
satisfies
$$
r(\th)\,r(\th\!-\!i\hbar)=1-\frac{i\hbar}{\th}\;,\sspace
r(\th) = 1 - \frac{i\hbar}{2 \th} + \frac{1}{8}
\left(\frac{i\hbar}{\th}\right)^2 +O\left((i\hbar/\th)^3\right)\,.
$$
Further $r(0)=-1$ and $r(i\hbar + \delta)= -\delta/i\hbar$, $r(-i\hbar +
\delta) =  i\hbar/\delta$, for $\delta \ra 0$. For later reference we
list the main  properties of the $R$-matrix (\ref{r1}). In addition to
the Yang-Baxter equation (\ref{r2}) and (\ref{r3}) one has a   
sign modified crossing invariance
\be
R_{ab}^{dc}(\th) = -C_{aa'}C^{dd'}\,R_{d'b}^{a'c}(i\hbar -\th)\;\;\;
\mbox{with }\;\;\; C_{ab}=i\eps_{ab}\,. \nonumber
\label{r4}
\ee
The $R$-matrix (\ref{r1}) has no poles in the strip 
$0 \leq {\rm Im}\,\th\leq i\hbar$ but a simple zero at $\th = i\hbar/2$. 
At $\th = 0,i\hbar$ it becomes a projector
\ba
R_{ab}^{cd}(0) &=& - \delta_a^d\delta_b^c \nonum
R_{ab}^{cd}(i\hbar) &=& -\delta_a^c\delta_b^d + \delta_a^d\delta_b^c =
C_{ab}C^{cd} \,.
\label{r5}
\ea
The semi-classical expansion is 
\ba
R_{ab}^{cd}(\th)&=&  \delta_a^c\delta_b^d 
- \frac{i\hbar}{\th}\;
   \Omega_{ab}^{cd}
- \left(\frac{i\hbar}{\th}\right)^2
   \left(\delta_a^d\delta_b^c- 
        {\textstyle \frac{5}{8}}\delta_a^c\delta_b^d\right)
+O\left((i\hbar/\th)^3\right)\,,
\label{r6}\\[4pt]
\mbox{with}\quad
\Omega_{ab}^{cd}&=&\delta_a^d\delta_b^c-
             {\textstyle\frac12}\delta_a^c\delta_b^d\;. \nonumber
\ea

Next we determine the possible $*$-operations of the WY-algebras with 
the $R$-matrix (\ref{r1}). Starting with a general linear ansatz one 
finds
\ba 
&& \sigma W_a(\th) = F_a^{a'}W_{a'}(\th^* + i\hbar) \nonum
&& \sigma T^+(\th)_a^b = F_a^{a'} E_{b'}^b\, 
T^-(\th^* + i\hbar\beta/\pi - i\hbar)_{a'}^{b'}\nonum
&& \sigma T^-(\th)_a^b = (E\inv)_{b'}^b\,(F\inv)_a^{a'} 
T^+(\th^* + i\hbar\beta/\pi - i\hbar)_{a'}^{b'}\;,
\label{a1}
\ea
where $E,F$ are $GL(2,\C)$ matrices satisfying
\be
E E^* = \eps \1\,,\;\;\eps \in \{\pm 1\}\;,
\quad F F^* = \1\,,\quad \det E \cdot \det F =1\;.
\label{a1FE}
\ee
Any solution of (\ref{a1FE}) yields a consistent $*$-operation
(\ref{a1}), for any value of $\beta$. We omitted a trivial overall
shift by a purely imaginary number in the arguments on the right hand
side of (\ref{a1}). We also omitted scalar prefactors on the right
hand side which can be removed by a rescaling of the generators. The
operator $D_{ab}(\th)$ is basically hermitian with respect to any of
the $*$-operations (\ref{a1}), (\ref{a1FE})
\ba
&& \sigma D_{ab}(\th) = 
\eps D_{ba} (\th^* + i\hbar\beta/\pi -2i\hbar)\;,\;\;\;
\sigma D(\th) = \eps D(\th^* + i\hbar\beta/\pi -2i\hbar)\;, \nonum
&& \mbox{where}\;\;\; D(\th) := C^{ab}D_{ab}(\th)\;.
\label{a4}
\ea
Clearly, most of these $*$-operations will be equivalent in being
related by an automorphism of the WY-algebra. Such automorphisms are
provided by $SL(2,\C)$ basis transformations $W_a(\th) \ra
f_a^{a'}W_{a'}(\th)$, $T^\pm (\th)_a^b \ra f_a^{a'} g_{b'}^b\,T^\pm
(\th)_{a'}^{b'}$, $f,g\in SL(2,\C)$, under which the $*$-structure
(\ref{a1}) transforms covariantly as
\be
E\mapsto g^{-1}Eg^*\;,\;\;\;\eps\mapsto \eps\;,\sspace 
F \mapsto f^* F f\inv\,.
\label{EFcov}
\ee

It is not hard to classify the inequivalent $*$-structures in the
general case. With regard to the Ernst system however we restrict
attention to real linear transformations and thus require the matrices
$E,F$ to be real.  This leaves four cases for the possible
$*$-operations, corresponding to the sign choices ${\rm sign}(\det F)
= {\rm sign}(\det E)$ and $\eps \in \{\pm 1\}$. Consider first $F$: If
$\det F =-1$ then $F = A \sigma^1$, where $A\in SO(2)$ and $\sigma^j$,
$j =1,2,3$, are the Pauli matrices. If $\det F =1$ then $F =\pm \1$.
In the former case one can achieve $F = \sigma^3$ by a similarity
transformation; in the latter case one can take $F = \1$, because the
sign can be absorbed either into $E$ or into a rescaling of the
generators. It turns out that $\det F=1$ is the case relevant for the
Ernst system, so for brevity we consider the possible $E$'s only for
$\det E =1$.  The general solution of $E^2 = \eps \1$, $E \in
SL(2,\R)$, then is readily worked out. For $\eps =1$ it leaves only $E
= \pm \1$, for $\eps =-1$ one finds a two-parameter family of $E$'s;
by a similarity transformation each of its members can be mapped onto
$E = i\sigma^2$. In summary, we always take $F = \1$ in (\ref{a1}),
which leaves only two inequivalent $*$-structures implemented by real
$E$ matrices, namely
\be
\eps=1\,:\quad E= \1\;\quad \mbox{and}\;\quad 
\eps=-1\,:\quad E = i\sigma^2\;. 
\label{*op1}
\ee
Note that with the second choice the $SL(2,\R)$ basis transformations 
acting on the upper index are restricted to the $SO(2) \subset SL(2,\R)$ 
subgroup leaving $E$ fixed.  

{}From now on $\DI$ will denote the algebra $WY(R,0,*)$ with the
$R$-matrix (\ref{r1}), the parameter $\beta =0$, and the $*$-operation
(\ref{a1}) with $F=\1$ and $E$ given by one of the matrices in
(\ref{*op1}). The case with $\eps =-1$ will turn out to be the one
relevant for the Ernst system. Often we shall treat the $\eps = 1$
case as well in order to emphasize the crucial differences entailed by
the seemingly minor flip. For convenient reference let us note
explicitly
\ba 
&& \sigma W_a(\th) = W_a(\th^* + i\hbar) \nonum
&& \sigma T^+(\th)_a^b = E_{b'}^b\,T^-(\th^* - i\hbar)_a^{b'}\nonum
&& \sigma T^-(\th)_a^b = \eps E_{b'}^b\,T^+(\th^* - i\hbar)_a^{b'}\;,
\label{*op}
\ea
with $E$ as above, as the $*$-operation of $\DI$. $SL(2,\R)$
transformations acting on the lower index are $*$-automorphisms of
$\DI$, and similarly $SO(2)$ rotations acting on the upper
index. Generic $SL(2,\R)$ transformations acting on the upper index in
contrast are automorphisms but do not preserve the
$*$-structure. Rather the matrix $E$ transforms covariantly as $E \ra
g\inv E g$, $g \in SL(2,\R)$.

In addition $\DI$ admits some simple $*$-automorphisms given by 
$\th$-dependent rescalings of the generators. Explicitly
\be
W_a(\th) \rra \omega(\th) \,W_a(\th)\,,\quad
T^{\pm}(\th)_a^b \rra \kappa^{\pm}(\th) \,T^{\pm}(\th)_a^b\,,
\label{*resc1}
\ee
are $*$-automorphisms of $\DI$ provided the scalar functions
$\omega(\th), \kappa^{\pm}(\th)$ obey
\ba
&& \omega(\th)^* = \omega(\th^* + i\hbar)\;,\nonum
&& \kappa^{\pm}(\th)^* = \kappa^{\mp}(\th^* + i\hbar)\;,\quad
\kappa^{\pm}(\th) \kappa^{\pm}(\th \pm i\hbar) =1\,.
\label{*resc2}
\ea 
The last equation in particular entails that $\kappa^{\pm}(\th)$ are
$2i\hbar$ periodic functions.


\newsubsection{Diagonalizing the center at the critical level \boldmath{$\beta =0$}} 

For $\beta =0$ the quantum current is central. Explicitly
\be
[D(\th_1)\,,\,W_a(\th_2)] = 0 =  [D(\th_1)\,,\,T^\pm  (\th_2)_a^b]\;,
\sspace {\rm Re}\,\th_{21} \neq 0\;.
\label{a5}
\ee
The second equation is well known \cite{ResSem90}. The first one follows 
similarly from (\ref{Drel}), which also explains the origin of the 
CDD-like sinh-prefactor in (\ref{r1}).%
\footnote{However, the relations (\ref{a5})
do not imply that the  antisymmetric part of $D_{ab}(\th)$ decouples
algebraically. Defining 
$$
{\cal M}_{ab}(\th) := \frac{i}{2}( D_{ab}(\th) +
D_{ba}(\th))\;,
$$ 
the relations (\ref{Drel}) (at $\beta = 0$) do not hold with
$D_{ab}(\th)$ replaced by ${\cal M}_{ab}(\th)$. We shall see in
section 3 how to separate the symmetric part of $D_{ab}(\th)$ in the
classical limit.}
Since $D(\th)$ is central it is natural to search for representations
of $\DI$ on which $D(\th)$ acts like a multiple of the unit
operator. The Fock space representations of the Yangian double at the
critical level inherited from a free field realization
\cite{Konn97,FreRes96} do not have this property. Experience from other
contexts suggests to search for appropriate representations in terms
of functionals over the algebra $\DI$.

Specifically we consider vector functionals (called ``T-invariant'') 
\be 
\DI \ni X \rra \bra X\ket  = \Obra X \Oket\;,
\label{t1}
\ee
built from a pair of vectors $\Oket$ and $\Obra$ satisfying  
\be 
T^+(\th)_a^b \Oket = \gamma^+(\th)_a^b \Oket \;,\sspace
\Obra T^-(\th)_a^b  = \gamma^-(\th)_a^b\Obra \;,
\label{t2}
\ee 
where $\gamma^\pm (\th)$ are numerical matrices which according to
(T1), (T2) carry one-dimensional representations of the Yangian
algebra $Y({sl}_2)$ respectively. This implies $\gamma^\pm
(\th)=\kappa^\pm (\th)\gamma^\pm $ with $2i\hbar$-periodic scalar
functions $\kappa^\pm (\th)$ and constant matrices $\gamma^\pm $,
satisfying the relations $\kappa^\pm (\th)\kappa^\pm (\th \pm
i\hbar)=1$ and $C_{cd} \gamma^\pm {}_a^c \gamma^\pm {}_b^d =
C_{ab}$. It is natural to supplement the conditions on
$\kappa^{\pm}(\th)$ by the first condition in (\ref{*resc2}). A
rescaling (\ref{*resc1}) of the $T^{\pm}$ generators then allows one
to dispense of the $\th$-dependence of the $\gamma^{\pm}$ matrices in
(\ref{t2}). Henceforth we shall use constant $\gamma^{\pm}$ matrices.
Hermiticity $\bra \sigma (X) \ket = \bra X\ket^*$ then imposes the
conditions
\be
\left[\gamma^+{}_a^b\right]^* = \gamma^-{}_a^{b'} E_{b'}^b\;,\sspace
\left[\gamma^-{}_a^b\right]^* = \eps 
\gamma^+{}_a^{b'} E_{b'}^b\;.
\label{*gam}
\ee
We shall mainly need the combination 
\ba
&& \Gamma_a^b := -C_{mn}\gamma^-{}_a^m\gamma^+{}_k^n C^{kb}\;,\quad\sspace
\mbox{satisfying} \nonum
&& (\Gamma^{-1})_a^b = -C^{bk} C_{mn} \gamma^-{}_k^m\gamma^+{}_a^n\;,
\quad \;\;\; \left[\Gamma_a^b\right]^* = \eps\,(\Gamma\inv)_a^b\;,\nonum
&& \Gamma_a^b + \eps \,[\Gamma_a^b]^* = \Gamma_a^a \1\,,
\makebox[2.4cm]{} \det \Gamma = 1\;.
\label{Gaprop}
\ea
Observe that the value (\ref{t1}) of the central element $D(\th)$ is
given by the trace $\Gamma_a^a$ and reinforces the distinct
features of the $\eps = 1$ and $\eps = -1$ involutions in
(\ref{a1}):
\ba
&& \Obra D(\th) \Oket = \,\Gamma_a^a \Obra \Omega \ket\;,
\nonum
&&  \Gamma_a^a = \left\{ \begin{array}{ll}
-2\;,&  \mbox{if $\eps =\phantom{-}1$ and $\gamma^\pm  $ real} \\
\phantom{-} 0\;,&  \mbox{if $\eps = -1$ and  $\gamma^\pm  $ real\,.} 
\end{array} \right. \label{traceX}
\ea
Natural choices are: $\gamma^\pm  = \1$ for $\eps =1$, and $\gamma^- =
\1,\; \gamma^+ = E$ for $\eps =-1$, in which case $ \Gamma_a^b =
-\delta_a^b$ and $ \Gamma_a^b = -E_a^b $, respectively.

Clearly any $T$-invariant functional (\ref{t1}) 
is uniquely determined by its values on strings of W-generators, 
for which we introduce some extra notation
\be
f_A(\th) := f_{a_\n\ldots a_1}(\th_\n,\ldots,\th_1) := 
\bra W_{a_\n}(\th_\n) \ldots W_{a_1}(\th_1) \ket \;,
\label{t3}
\ee
where ${\rm Re}\,\th_{ij}\neq 0,\;i\neq j$, $\th = (\th_\n,\ldots,\th_1)$,
$A = (a_\n,\ldots,a_1)$. Sometimes also the shorthand $f^{(\n)}$ for the 
value of $\bra\;\cdot  \;\ket$ on a string of $\n$ 
$W$-generators will be used. With these definitions one computes
\ba
&& \bra W_{a_\n}(\th_\n) \ldots W_{a_{k+1}}(\th_{k+1})\, D(\th_0)\,
W_{a_k}(\th_k) \ldots W_{a_1}(\th_1) \ket \label{t4}\\[4pt]
&& \hspace{15em}=~ 
\cT(\th_0|\th)_A^B \,\bra W_{b_\n}(\th_\n) \ldots W_{b_1}(\th_1)\ket\;,
\nonumber
\ea
where $\cT(\th_0|\th)$, $\th = (\th_\n,\ldots,\th_1)$, is basically 
the familiar transfer matrix 
\ba
&\nspace & \cT(\th_0|\th)_A^B ~=~ \Gamma_b^a\,
T_a^b(\th_0 + i\hbar|\th)_A^B\;,\label{t5}\\
&\nspace & 
T_{a}^{b}
(\th_0|\th_{\n},\ldots,\th_1)_{a_\n\ldots a_1}^{b_\n\ldots b_1} ~:=~ 
R_{c_\n a_\n}^{b \;b_\n}(\th_{\n,0})\;
R_{c_{\n-1}a_{\n-1}}^{c_\n\; b_{\n-1}}(\th_{\n-1,0})\;\ldots\;
R_{a\; a_1}^{c_2b_1}(\th_{1,0})\,.\nonumber
\ea
Implicit in (\ref{t5}) are two important features reflecting the fact
that $D(\th_0)$ is central: The right hand side of (\ref{t5}) is $k$
independent and for fixed $\th\in\C^\n$ and varying $\th_0 \in \C$ the
$\cT(\th_0|\th)$ form a one-parameter family of commuting
matrices. Hence they can be simultaneously diagonalized and on the
eigenvectors $D(\th_0)$ will act like a multiple of the unit
operator. We are thus lead to restrict attention to those functionals
(\ref{t1}) (or later a subset thereof) for which the functions
(\ref{t3}) obey
\be 
\cT(\th_0|\th)_A^B \,f_B(\th) ~=~ \tau(\th_0|\th) f_A(\th)\;.
\label{t6}
\ee
We note the following hermiticity properties of the $\cT$-matrices
and their eigenvalues: 
\ba
&& \left[\cT(\th_0|\th)_A^B\right]^* ~=~ \eps\,\cT(\th_0^* - 2i\hbar|
{\th^*}^T + i\hbar)_{A^T}^{B^T}\;,\nonum
&& \tau(\th_0|\th)^* ~=~ \eps \,\tau(\th_0^* 
- 2i\hbar|\th^*{}^T + i\hbar)\;,
\label{t20}
\ea
which follow from (\ref{a4}), (\ref{t4}) and the general hermiticity
condition $\bra \sigma(X)\ket = \bra X\ket^*$. The notation is 
$\th^T = (\th_1,\ldots, \th_\n)$, $A^T = (a_1,\ldots,a_\n)$, etc.
A further important property of the eigenvalues $\tau(\th_0|\th)$ of
(\ref{t6}) is  
\be
\tau(\th_k - i\hbar|\th) \,\tau(\th_k - 2i\hbar|\th) =1\,,\qquad
\;\; k=1,\dots,\n \;.
\label{t21}
\ee
To derive this, consider the matrix $\cT(\th_0|\th)\cT(\th_0 - i\hbar|\th)$,
which describes the action of $D(\th_0) D(\th_0 -i\hbar)$ within the
matrix elements (\ref{t3}). Then 
\be
[\cT(\th_0|\th)\cT(\th_0 - i\hbar|\th)]_A^B\bigg|_{\th_0 = \th_k - i\hbar} = 
\delta_A^B \;,\qquad \;\; k =1, \ldots,\n\;,
\label{t22}
\ee
as may be verified by direct computation from (\ref{t5}).

The matrix $\Gamma_a^b$ in (\ref{t5}) can be thought of as describing the
deviation from the $SL(2,\R)$ symmetry. In particular $\cT(\th_0|\th)$ is
invariant only under the subgroup of $SL(2,\R)$ matrices obeying
\be
\Gamma_a^c\,\Lambda_c^b = \Gamma_c^b\,\Lambda_a^c\;.
\label{norm5}
\ee
For $\eps = 1$ one may take $\Gamma = -\1$ and the condition is empty.
For $\eps = -1$ the solutions of (\ref{norm5}) can be seen to generate
a maximal compact subgroup $SO(2) \subset SL(2,\R)$. This holds
irrespective of any further constraints on the matrix $\Gamma$, but
for the reasons outlined in section 2.3 we take $\Gamma$ to be
real. Of course $\Gamma$ still has to obey the constraints in
(\ref{Gaprop}) and the most general real solution to them may be
parameterized as
\be
\Gamma(\varphi,\nu) := 
{ \;\;\sh \varphi \;\;\;\;\; \nu\, \ch \varphi \choose
-\frac{1}{\nu} \ch \varphi\;\;-\sh \varphi }\;,\;\;\;
0\neq \nu \in \R, \;\varphi \in \R\;.
\label{a2}
\ee
An $SL(2,\R)$ basis transformation in (\ref{t4}) maps $\Gamma$ in
(\ref{t5}) onto $g \Gamma g\inv$, $g \in SL(2,\R)$. The transformed
$\Gamma$ matrix still obeys the constraints in (\ref{Gaprop}) and
hence can be parameterized as in (\ref{a2}), however with different
values for $\varphi$ and $\nu$. The $\Gamma$ matrices therefore define
a conjugacy class in $SL(2,\R)$. The transfer matrix $\cT$ and its
eigenvalues depend on the representative $g \Gamma g\inv$, while the
eigenvalues are class-functions, i.e. depend only on the conjugacy
class. In particular the eigenvalues of $\Gamma$ itself are $\pm i$,
independent of $\varphi$ and $\nu$. For any fixed matrix $\Gamma$ the
solutions of (\ref{norm5}) then generate the missing compact conjugacy
class of $SL(2,\R)$. Explicitly the solutions of (\ref{norm5}) are
given by
\be
\Lambda_a^b(\phi)  = \delta_a^b \cos\phi + \Gamma(\varphi,\nu)_a^b 
\sin\phi\;\;\;\mbox{for}\;\;\;\eps = -1\,,\;\;0 \leq \phi <2\pi\,.
\label{norm8}
\ee 
As anticipated, they generate an $SO(2)$ subgroup of $SL(2,\R)$. 
Furthermore, they satisfy $C^{bm} \Lambda_m^n(\phi) C_{na} =
\Lambda_a^b(-\phi)$.  The invariance of $\cT(\th_0|\th)$ is expressed
by 
\be
\cT(\th_0|\th)_A^C\; \Lambda_{c_1}^{b_1}(\phi)\dots
\Lambda_{c_\n}^{b_\n}(\phi)
= \Lambda_{a_1}^{c_1}(\phi)\dots\Lambda_{a_\n}^{c_\n}(\phi)
\;\cT(\th_0|\th)_C^B\;. 
\label{norm7}
\ee
In particular, (\ref{norm7}) allows one to break up the eigenvalue
problem (\ref{t6}) into subsectors of fixed $SO(2)$ charge;
cf.~appendix B.

The diagonalization (\ref{t6}) of $\cT$ of course is the object of
the Bethe Ansatz. We shall be interested in solutions which are
in addition equivariant with respect to the usual representation of
the permutation group $S_\n$ on the space of tensor-valued functions.
Whence we require
\be
f_A(\th) ~=~ L_s(\th)_A^B \,f_B(s\inv\th) \;,\qquad 
\forall\, s\in S_\n\,.
\label{t7}
\ee
It suffices to specify the action of the generators $s_1,\ldots,
s_{\n-1}$ of $S_\n$:
\ba
&& s_j(\th_\n,\ldots, \th_1) ~=~ 
(\th_\n,\ldots,\th_j,\th_{j+1},\ldots,\th_1)\;,
\;\;\; 1\leq j\leq \n\!-\!1\;,\nonum
&& L_{s_j}(\th)_A^B ~=~ \delta_{a_\n}^{b_\n} \ldots 
\delta_{a_{j+2}}^{b_{j+2}}\,
R_{a_{j+1}a_j}^{b_j\;\,b_{j+1}}(\th_{j+1,j})\,\delta_{a_{j-1}}^{b_{j-1}}
\ldots \delta_{a_1}^{b_1}\;,
\label{t8}
\ea
and the product
\be
L_{s s'}(\th) = L_s(\th) L_{s'}(s\inv\th)\;,\;\;\;
\forall\, s,s'\in S_\n\,.
\label{t9}
\ee
Not surprisingly, the system (\ref{t6}) is compatible with (\ref{t7})
provided the eigenvalues $\tau(\th_0|\th)$ are completely symmetric in
$\th = (\th_\n,\ldots,\th_1)$, which we henceforth assume to be the
case. It suffices to verify the asserted compatibility for the
generators of $S_\n$; this in turn follows from the identities
\be
L_{s_j}(s_j\th)_A^{A'} \,\cT(\th_0|\th)_{A'}^{B'}\,
L_{s_j}(\th)_{B'}^B = \cT(\th_0|s_j\th)_A^B\;,\;\;\;
1\leq j\leq \n\!-\!1\;.
\label{ts}
\ee

\newsubsection{Further properties of the eigenvectors}

The joint solutions (\ref{t6}) and (\ref{t7}) enjoy a number of other
remarkable properties. First they are also solutions to an asymptotic
form of the deformed Knizhnik-Zamolodchikov equations (KZE)
\cite{FreRes92,Smir92}, where in the present conventions $2i(1
-\beta/2\pi)$ parameterizes the level.  The critical level
$\beta\!=\!0$ corresponds to the limiting case where these equations
degenerate into an eigenvalue problem for mutually commuting matrices
$Q_k$, namely (see e.g.~\cite{TarVar96})
\be
Q_k(\th)_A^B\; f_B(\th) = q_k(\th) \,f_A(\th)\;,
\label{t16}
\ee
with
\be
Q_k(\th)_A^B ~:= -\Gamma_c^d\,T^c_{a_k}(\th_k|\th_\n,\ldots,\th_{k+1})%
_{a_\n\ldots a_{k+1}}^{b_\n\ldots b_{k+1}}
T^{b_k}_d(\th_k|\th_{k-1},\ldots,\th_1)%
_{a_{k-1}\ldots a_1}^{b_{k-1}\ldots b_1} \;.
\ee
To see the relation to (\ref{t6}) note 
\be\label{TQ}
\cT(\th_0|\th)_A^B\bigg|_{\th_0 = \th_k - i\hbar} ~=~ Q_k(\th)_A^B \;.
\ee
Hence (\ref{t16}) is a consequence of (\ref{t6}) with $q_k(\th) =
\tau(\th_k - i\hbar|\th)$. The converse is also true as we show in 
appendix B2. The $\cT$ eigenvalue problem (\ref{t6}) and the seemingly 
weaker $Q_k$ eigenvalue problem (\ref{t16}) therefore are equivalent
for $\beta =0$.

For $\beta \neq 0$ the deformed KZE has been found to have
an algebraic counterpart \cite{Nied98}: Supplementing the
relations of $WY_{\beta}(R)$ by 
\be
W_a(\th +i\hbar\beta/\pi) = C_{mn} T^-(\th + i\hbar\beta/\pi)_a^m \,
W_k(\th)\, T^+(\th + i\hbar\beta/\pi - i\hbar)_l^n C^{kl}\;,
\label{M}
\ee
the matrix elements (\ref{t3}) automatically solve the deformed 
KZE with level $i(2 - \beta/\pi)$, or 
equivalently a cyclic equation with shift parameter $\beta$. 
Moreover the relation (\ref{M}) endows the algebra with a
``modular structure'' characteristic for a quantum system at 
finite temperature $1/\beta$. Here we show that much of 
this structure survives in the limit $\beta \ra 0$. 

We begin by showing that the joint solutions of (\ref{t6}) and
(\ref{t7}) also enjoy the following cyclic property
\be
f_A(\th) = \Omega(\th)_A^B f_B(\Omega\inv \th)\;.
\label{C}
\ee
Here $\Omega$ is the Coxeter element $\Omega = s_{\n-1}\ldots s_1 \in
S_\n$, acting by cyclic permutation $\Omega(\th_\n,\ldots, \th_1)
=(\th_1,\th_\n,\ldots,\th_2)$ on elements of $\C^\n$ and
\be
\Omega(\th)_A^B ~:=~ -\tau(\th_\n -2i\hbar|\th) \;
\Gamma_{a_\n}^{b_1}\delta_{a_{\n-1}}^{b_\n} \ldots \delta_{a_1}^{b_2}\;.
\label{om}
\ee
Of course one  could also have used (\ref{t7}) to obtain the 
relation $f_A(\th)= L_{\Omega}(\th)_A^B f_A(\Omega\inv\th)$, 
where $L_{\Omega}(\th)$ is the representation matrix of $\Omega \in S_\n$ 
defined by (\ref{t8}), (\ref{t9}). Consistency is ensured by 
the fact that
\be
L_{\Omega}(\th)_A^B \;f_B(\Omega^{-1}\th) = 
\Omega(\th)_A^B\;f_B(\Omega^{-1}\th)\;,
\label{t13}
\ee
on the joint solutions of (\ref{t6}) and (\ref{t7}). 
In other words, although $\Omega(\th)$ and $L_{\Omega}(\th)$ are
distinct as matrices, they act in the same way on the solutions
of (\ref{t6}) and (\ref{t7}). This follows from
\ba
f_A(\th) 
&\stackrel{(\ref{t21})}{=}& 
\tau(\th_\n \!-\!2i\hbar|\th)\,
\cT(\th_\n \!-\!i\hbar|\th)_A^B \;f_B(\th) \nonum
&\stackrel{(\ref{t5})}{=}& 
-\tau(\th_\n \!-\!2i\hbar|\th)\,\Gamma_{a_\n}^d \,
T_d^{b_\n}(\th_\n|\th_{\n-1},\ldots,\th_1)_%
{a_{\n-1}\ldots a_1}^{b_{\n-1}\ldots b_1} \;f_B(\th) \nonum
&\stackrel{(\ref{t8}),(\ref{t9})}{=}&  
-\tau(\th_\n \!-\!2i\hbar|\th)\,\,\Gamma_{a_\n}^d\,
L_{\Omega\inv}(\Omega\inv \th)_{a_{\n-1}\ldots a_1 d}^B \;f_B(\th) 
\nonum
&\stackrel{(\ref{om})}{=}& 
\Omega(\th)_A^B \; L_{\Omega\inv}(\Omega\inv \th)_B^C \;f_C(\th)\;,
\nonumber
\ea
and together with (\ref{t7}) we obtain (\ref{C}) and (\ref{t13}). 
One may check that consistently the matrix $\cT$ has the cyclic property
\ba
\cT(\th_0|\th)_A^B &=& \cT(\th_0|\Omega^{-1}\th)_%
{a_{\n-1}\ldots a_1 c}^{b_{\n-1}\ldots b_1 d}\,\Gamma^c_{a_\n}\,
(\Gamma\inv)_d^{b_\n}\nonum
&=& 
\Omega(\th)_A^{A'}\,\cT(\th_0|\Omega^{-1}\th)_{A'}^{B'}\,
\Omega^{-1}(\th)_{B'}^{B} \;.
\label{t11}
\ea
>From here one can show that the cyclic equation (\ref{C}) is in fact 
equivalent to the $Q_k$ eigenvalue problem: Specializing (\ref{t11}) 
to $\th_0 = \th_k - i\hbar$ yields cyclicity relations for the 
$Q_k$ matrices and their eigenvalues
\ba
Q_{k-1}(\Omega \th)_A^B &=& Q_k(\th)_{a_{\n-1} \ldots a_1c}%
^{b_{\n-1}\ldots b_1d}\,\Gamma_{a_\n}^c (\Gamma\inv)_d^{b_\n}\,,\nonum
q_{k-1}(\Omega \th) &=& \tau(\th_k - i\hbar|\th) = q_k(\th)\;.
\label{Qcycl}
\ea
In particular modulo the exchange relations (\ref{t7}) the eigenvalue
equation (\ref{t16}) for $k=\n$, say, entails all others. On the other
hand from the computation before (\ref{t11}) one also sees that the
cyclic equation (\ref{C}) and the $Q_\n$ eigenvalue equation are
equivalent. Combining both facts shows the asserted equivalence of
(\ref{t16}) and (\ref{C}).

A further property of the $Q_k$ matrices and their eigenvalues is
unravelled by iterating the cyclic equation (\ref{C})
\ba
\prod_{k=1}^\n q_k(\th) \,f_A(\th) & = &(-)^\n\, 
\Gamma_{a_\n}^{b_\n}\ldots \Gamma_{a_1}^{b_1} f_B(\th)\,,\nonum
[Q_1(\th) \ldots Q_\n(\th)]_A^B & = & (-)^\n \prod_{j=1}^\n 
\Gamma_{a_j}^{b_j}\;,
\label{t25}
\ea
where the second equation follows from the first one together with the
fact that for generic $\th$'s the matrices $Q_k$ have maximal
rank. From (\ref{t21}) one finds $q_k(\th) q_k(\th - i\hbar) =
\eps$. Finally the $q_k(\th)$ have two properties that are readily
seen only in the Bethe ansatz construction relegated to appendix B:
For real $\th$ they are pure phases and the phase is a gradient; see
also \cite{TarVar96,TarVar97}. Explicitly, $q_k(\th)^*=q_k(\th^*)^{-1}$
and
\be
q_k(\th) = e^{i\partial_k \Delta(\th)}\;,\quad \mbox{with} \quad 
\partial_k \Delta(\th) = 
\delta(\Omega^{\n-k}\th )\;,\;\;\;k =1,\ldots,\n\;,
\label{t24}
\ee
where $\delta(\th) = \delta(\th^*)^* := -i\ln\tau(\th_\n -i\hbar|\th)$. 
The fact that the $q_k(\th)$ are pure phases is linked to having $\Gamma$ 
in (\ref{Gaprop}) chosen to be real. 

Real $\Gamma$-matrices are also natural because they allow one to 
introduce a quadratic form on the space of solutions to   
(\ref{t6}), (\ref{t7}). Consider the following quadratic form 
on the space of $V^{\otimes \n}$-valued functions
\be 
\bra f,g \ket = \int d^\n \th \,f_A(\th)^* C^{AB} g_B(\th) \,.
\label{norm1}
\ee
It is manifestly hermitian $\bra f,g\ket^* = \bra g,f\ket$ and
$SL(2,\R)$ invariant. Equations (\ref{t7}), (\ref{t16}), and (\ref{C})
give rise to symmetry operations which are unitary with respect to
$\bra\;,\;\ket$ in the sense that
\begin{subeqnarray}
&& \bra f,g\ket  = \bra f',g'\ket \sspace \mbox{for} \nonum
&& f'_A(\th) = L_s(\th)_A^B\, f_B(s\inv \th)\;,\;\;\;s \in S_\n\,, \\
&& f'_A(\th) = q_k(\th)\inv \,
Q_k(\th)_A^B\, f_B(\th)\;,\;\;\;               \\
&& f'_A(\th) = \Omega(\th)_A^B\, f_B(\Omega\inv\th)\;.
\label{norm2}
\end{subeqnarray}
Hence, for real $\Gamma_a^b$, $\bra \;,\;\ket$ induces a quadratic
form $\bra \;,\;\ket_{sol}$ on the space of functions $f'
\stackrel{!}{=} f$, which are the joint solutions of
(\ref{t7}), (\ref{t6}) $\Longleftrightarrow$ (\ref{t7}),
(\ref{t16}) $\Longleftrightarrow$ (\ref{t7}), (\ref{C}).

Now let us return to the algebraic description. Since for $\beta =0$ 
(and only then) the KZE (\ref{t16}) and the cyclic equation
(\ref{C}) are consequences of the $\cT$-eigenvalue problem 
(\ref{t6}) and (\ref{t7}) one expects that in the algebra 
$\DI$ extra algebraic relations hold which imply
(\ref{t16}), (\ref{C}) for the matrix elements (\ref{t3}).
This is indeed the case and the relevant relations are
\begin{subeqnarray}
&& C_{mn}T^-(\th)_a^m \,  W_k(\th) \,  T^+(\th - i\hbar)_l^n C^{kl}
= - W_a(\th) \,D(\th -i\hbar)\;,\\
&& D(\th - 2 i\hbar) D(\th- i\hbar)W_a(\th)  = W_a(\th)\;.
\label{MD}
\end{subeqnarray}
The first relation can be verified simply by pushing $T^-$ to 
the right using (TW). Equation (\ref{MD}b) then is required for 
compatibility with the $*$-operations (\ref{a1}). Indeed, applying
$\sigma$ to (\ref{MD}a) yields 
\be 
C_{mn}T^-(\th - i\hbar)_k^m \,  W_l(\th) 
\,  T^+(\th - 2 i\hbar)_a^n C^{kl} = - D(\th - 2i\hbar) \,W_a(\th)\;.
\label{m1}
\ee  
Employing associativity and (T2) one recovers (\ref{MD}a) iff (\ref{MD}b) 
holds. Inserting now (\ref{MD}a) into the $k$-th position of a matrix 
element (\ref{t3}) precisely produces the $k$-th KZE-equation 
(\ref{t16}). Using the (WW) relations any of them can be seen to
be equivalent to the cyclic equation (\ref{C}). Finally the 
algebraic consistency relation (\ref{MD}b) amounts to 
(\ref{t22}) when used within the matrix elements (\ref{t3}).

Let us summarize the construction until here. For $\beta\!=\!0$, the
quantum current lies in the center of the algebra $\DI$.  Searching
for T-invariant functionals (\ref{t1}), (\ref{t2}) on which it acts
like a multiple of the unit operator, leads to the eigenvalue problem
(\ref{t6}). We were interested in those solutions which also enjoy the
$R$-matrix exchange relations (\ref{t7}) and found that they have a
number of remarkable bonus properties. Most notably, they satisfy the
asymptotic form (\ref{t16}) of the deformed KZE, which we showed to be
equivalent to the cyclic equation (\ref{C}). Moreover on the space of
permutation equivariant functions (\ref{t7}) the three requirements:
$\cT$ eigenvalue problem (\ref{t6}), $Q_k$ eigenvalue problem, and the
cyclic condition (\ref{C}) are all equivalent.  The actual solution
e.g.~of the $\cT$ eigenvalue problem (\ref{t6}), (\ref{t7}) is
relegated to appendix B. All this was for a fixed number $\n$ of
variables $\th_j$. Next we show that the eigenvectors and eigenvalues
for different $\n$ can naturally be arranged into sequences.

\newsubsection{Sequences of eigenvectors}

Apart from the non-trivial center, the algebra $\DI = WY(R,0,*)$
contains further two-sided ideals which should be divided out.  The
following relations (R) can be checked to arise in this way and we
suspect them to be the only ones%
\footnote{It appears to be a general rule that reducibility of a tensor 
product of fundamental representations is always caused by a pole in the 
$R$-matrix \cite{ChaPre94,FreMuk99}. According to (TW) the $W_a$ play the 
role of intertwiners between two such representations.} 
\vspace{2mm}
\eqll{R}
\bas
&& W_a(\th + i\hbar)\cdot  W_b(\th) = 
\lb(\th)\,C_{ab}\,D(\th- i\hbar)\;,\nonum
&& C^{ab}W_a(\th - i\hbar)\cdot  W_b(\th) =
 \lb(\th-i\hbar)\,D(\th - 2i\hbar)\,,
\eas
where the function $\lambda(\th)$ satisfies $\lambda(\th)^* = \eps
\lambda(\th^*)$.  Under the action of the $*$-automorphism
(\ref{*resc1}) $\lb(\th)$ changes according to
\be
\lb (\th) \rra \lb(\th) \frac{\omega(\th) \omega(\th + i \hbar)}%
{\kappa^-(\th) \kappa^+(\th + i \hbar)}\;.
\label{lam1}
\ee
We will later comment on specific choices for $\lb(\th)$. 
The algebra $\DI$ where in addition the relations (R) are imposed, is our
complete dynamical algebra $\cD$.

The operator product $W_a(\th_1) W_b(\th_2)$ turns out to be singular
as $\th_{21} \ra \pm i\hbar$, with a first order pole. The `$\,\cdot
\,$' in (R) indicates a normal product defined roughly by taking the
residue at the pole. A more precise definition will be given below in
terms of its action within the T-invariant functionals. For the moment
we are only interested in the algebraic properties of the relations
(R).

A stronger, uncontracted version of the second relation is 
\be
W_a(\th -i\hbar)\cdot  W_b(\th) = \lb(\th -i\hbar) D_{ab}(\th- 2 i\hbar)\;.
\label{R1}
\ee
Let us momentarily denote these relations by (R1), (R2)
and (R3), respectively. Recall that the exchange relations
(TW) are valid also for Re$\,\th_{12}=0$, while for the 
(WW) relations these points are a-priori excluded. Consider the following
formal extension of (WW) to $\th_{12} = i\hbar$,
\be
W_a(\th+ i\hbar) \cdot W_b(\th) = - R_{ab}^{cd}(i\hbar)\,
W_d(\th) \cdot W_c(\th + i\hbar)\;,
\label{R2}
\ee  
The extra minus sign of (\ref{R2}) with respect to (WW) is in
accordance with the residue interpretation of the `$\,\cdot \,$'-product.
Then: 
\ba
&& ({\rm R2}) \;\;\,\Longrightarrow \;\; ({\rm R3})\;\;\;
\mbox{by means of}\;\;\;(\ref{MD}),({\rm TW})\;,\nonum
&& (\rm{R1}) \;\;\Longleftrightarrow \;\; ({\rm R2}),({\rm R3})\;\;\;
\mbox{by means of}\;\;\;(\ref{R2})\;.
\ea
The first implication can be seen by starting from $W_a(\th - i\hbar)
\cdot  W_b(\th)$, replacing $W_a(\th - i\hbar)$ in favor of $- D(\th -
2i\hbar)^{-1}C_{mn}T^-(\th - i\hbar)_a^m W_k(\th - i\hbar) T^+(\th -
2i\hbar)_l^n C^{kl}$, then using (TW) to move $T^+$ to the right and
finally applying (R2). Thus (R2) and (R3) are equivalent in
$\cD$ and both are readily seen to be formally equivalent to (R1)
once one is allowed to use (\ref{R2}). In order to avoid potential
troubles with the formal identity (\ref{R2}), however, we postulate
both equations (R) independently, keeping in mind that they are
formally related by (\ref{R2}).

We can now restore topological concepts by calling a T-invariant
functional (\ref{t1}) analytic if: 
\begin{itemize}
\item[(i)] The dependence of the values $f^{(\n)}$ on the parameters
$\th_\n,\ldots,\th_1$ is locally analytic, possibly with branch points
but without cuts.
\item[(ii)] For $X,Y\in \DI$ the expectation value $\bra X\,
W_a(\th_1) W_b(\th_2)\, Y\ket$ has a simple pole at $\th_{12} = i
\hbar$ with residue $\lb(\th_2) \bra X\,C_{ab} D(\th_2 - i\hbar)\, Y\ket$;
similarly $\bra X \,C^{ab} W_a(\th_1) W_b(\th_2) \,Y\ket$ has a simple
pole at $\th_{12} = - i\hbar$ with residue $\lb(\th_2 -i\hbar)\bra X 
D(\th_2 - 2 i\hbar)Y\ket$. In particular this defines the 
`$\,\cdot \,$' product in (R). 
\end{itemize}
{}From now on we assume all functionals (\ref{t1}) to be analytic in
this sense.  The relations (R) then imply the $\n \ra \n\!-\!2$
recursive relations (II) given below for the functions
(\ref{t3}). They link the eigenfunctions and eigenvalues of the
eigenvalue problem (\ref{t6}), (\ref{t7}) in $\n$ and in $\n\!-\!2$
variables.

In summary we arrive at the following system of functional equations:
\bigskip
\medskip
\eqll{I}
\begin{subarray} \jot5mm
&& \cT(\th_0|\th)_A^B \,f_B(\th) ~=~ \tau(\th_0|\th) f_A(\th)\;,
\\
&& f_A(\th) = R_{a_{k+1} a_k}^{\;c\;\;\;\; d}(\th_{k+1,k}) 
f_{a_\n \ldots d c 
\ldots a_1}(\th_\n, \ldots, \th_k ,\th_{k+1},\ldots,\th_1)\,,
\end{subarray}
where $\n$ is fixed, $k=1,\ldots,\n\!-\!1$, and $\cT$ is the transfer
matrix (\ref{t5}). 
The solutions of the functional equations (I) for $\n$ and $\n\!-\!2$
are linked by   
\medskip
\bigskip
\eqll{II}
\smallskip
\begin{subarray} \jot5mm
{\rm Res}_{\,\th_{k+1} = \th_k + i\hbar}\,f_A(\th) &=&
\lb(\th_k) \,\tau(\th_k - i\hbar|p_k\th)\,C_{a_{k+1} a_k} 
f_{p_k A}(p_k \th)\;,
\\
{\rm Res}_{\,\th_{k+1} = \th_k - i\hbar}\,
C^{a_{k+1}a_k}f_A(\th)  &=&
\lb(\th_k -i\hbar) \,\tau(\th_k - 2i\hbar|p_k\th)\,f_{p_kA}(p_k \th)\;,
\\
\tau(\th_0|\th)\bigg|_{\th_{k+1} =\th_k \pm i\hbar} & = &
\tau(\th_0|p_k\th)\;. 
\end{subarray}

Here $k=1,\ldots,\n\!-\!1$, and we adopted the notation 
$p_k \th = (\th_\n,\ldots,\th_{k+2},\th_{k-1},\ldots,\th_1)$, 
$p_k A= (a_\n,\ldots,a_{k+2},a_{k-1},\ldots,a_1)$.
Equation (IIc) arises as consistency condition for (IIa,b) 
whenever their right hand sides are non-vanishing, using the properties
\ba
C_{b_{k+1} b_k}\,\cT(\th_0|\th)_A^B\bigg|_{\th_{k+1} = \th_k + i\hbar}
& = & C_{a_{k+1}a_k} \, \cT(\th_0|p_k \th)_{p_k A}^{p_k B}\;,
\nonum
C^{a_{k+1} a_k}\,\cT(\th_0|\th)_A^B\bigg|_{\th_{k+1} = \th_k - i\hbar}
& = & C^{b_{k+1}b_k} \, \cT(\th_0|p_k \th)_{p_k A}^{p_k B}\;,
\label{t19}
\ea
of the transfer matrix. Observe that specializing here $\th_0$ to 
$\th_k - i\hbar$ yields relations reciprocal to (\ref{TQ}). Similarly
on the level of the eigenvalues (IIc) for $\th_0 = \th_k - i\hbar$
gives 
\ba
q_k(\th)\bigg|_{\th_{k+1}= \th_k + i\hbar} &=& 
\tau(\th_k -i\hbar|p_k\th)\;,
\\ 
q_{k+1}(\th)\bigg|_{\th_{k+1}= \th_k - i\hbar} &=& 
\tau(\th_k -2i\hbar|p_k\th)\;.
\label{qres}
\ea
The restrictions of $\tau(\th_0|\th)$ to $\th_0 = \th_k - i\hbar$ and 
to $\th_{k+1} = \th_k \pm i\hbar$ commute. 
For completeness let us also note the recursive equation implied by
(\ref{R1}) 
\be
{\rm Res}_{\,\th_{k+1} = \th_k - i\hbar} f_A(\th) = 
-\frac{1}{2}\lb(\th_k -i\hbar)\,
C_{b_{k+1}b_k} Q_{k+1}(\th)_A^B \bigg|_{\th_{k+1} = \th_k - i\hbar}
f_{p_k B}(p_k \th)\;.
\label{IIe}
\ee

Clearly, viewed as a projection, $\n \ra \n-2$, the recursive operation
defined through (II) must have a large kernel. Simply counting the
dimensions one expects $2^{\n}$ eigenvectors on the left hand side of
(II) to be mapped onto only $2^{\n-2}$ on the right hand
side. Conversely given a solution of (I) in $\n\!-\!2$ variables it is
non-trivial that the equations (II) can be `integrated' to a function
of $\n$ variables {\em within} the class of solutions of (I). The
explanation why this is possible stems from the underlying algebraic
framework.

Recall the notation $\cD$ for the algebra $\DI= WY(R,0,*)$ supplemented 
by the relations (R). As in section 2.2 we can consider 
``T-invariant'' functionals, now over the algebra $\cD$, i.e.
\ba
&& \bra \;\;\; \ket : \cD \rra \C \,,\;\;\; X \rra \bra X\ket\,,
\nonum
&& \bra T^-(\th_0)_a^b\,X\ket = \gamma^-{}_a^b \,\bra X\ket\;,
\;\;\;\bra X \,T^+(\th_0)_a^b\ket = \gamma^+{}_a^b \,\bra X\ket\;,
\ea
keeping all other notations. Again, such a functional will be
completely determined by its values on strings of $W$-generators. By
construction, it gives rise to a solution of the system of functional
equations (I), (II) with the identification $f_A(\th) := \bra W_A(\th)
\ket$, -- and vice versa. Symbolically we arrive at a one-to-one
correspondence:
\medskip

$$
\begin{tabular}{c} 
T-invariant Functional\\
over $\cD$  
\end{tabular}
\;\;\longleftrightarrow\;\; 
\begin{tabular}{c} 
Sequence of \\ $\;\;$ Solutions of (I), (II) 
\end{tabular}
$$

\newsection{Semi-classical limit and phase space of the Ernst system} 
\label{CLES}

Here we study the semi-classical limit of the regular part of the
dynamical algebra and show that essentially the same Poisson algebra
describes the classical phase space of the Ernst system. In particular
this lends a physical interpretation to the variables
$T^{\pm}(\th)_a^b$ and $W_a(\th)$.

\newsubsection{Semi-classical limit of the dynamical algebra}

We begin with the classical limit of the algebra $\DI$ and assume that 
semi-classically the operator products in $\DI$ behave
like a Moyal product, i.e.
$$
X Y = X^{\rm cl} Y^{\rm cl} - 
\frac{i \hbar}{2} \{ X^{\rm cl}, Y^{\rm cl}\}
+ O(\hbar^2)\,,
$$
where $X^{\rm cl}, Y^{\rm cl}$ are the corresponding functions on
phase space.  Usually we shall drop the superscript ``${\rm cl}$''
when no confusion is possible.  Upon expansion in $\hbar$ the exchange
relations (T1), (TW), (WW) then provide the symplectic structure of a
classical Poisson algebra:
\begin{subeqnarray}
\left\{ T^\pm  (\th_1)_a^d\,,T^\pm  (\th_2)_b^c\right\}
&=& \frac1{\th_{12}}\left(
T^\pm  (\th_1)_e^dT^\pm  (\th_2)_f^c\,\Omega^{ef}_{ab} 
-\Omega^{dc}_{ef}\,T^\pm  (\th_1)_a^eT^\pm  (\th_2)_b^f 
\right) \;,
\\
\left\{ T^+(\th_1)_a^d\,,T^-(\th_2)_b^c\right\}
&=& \frac1{\th_{12}}\left(
T^+(\th_1)_e^dT^-(\th_2)_f^c\,\Omega^{ef}_{ab} 
-\Omega^{dc}_{ef}\,T^+(\th_1)_a^eT^-(\th_2)_b^f 
\right) \;,
\\
\left\{ T^\pm (\th_1)_a^f\,,W_b(\th_2)\right\}
&=& \frac1{\th_{12}}\,T^\pm (\th_1)_c^f \,W_d(\th_2)\;
\Omega^{cd}_{ab}\;, 
\\
\left\{  W_a(\th_1)\,, W_b(\th_2)\right\}
&=& \frac1{\th_{12}}\,
 W_c(\th_1)\, W_d(\th_2)\;\Omega^{cd}_{ab} \;,
\label{WY1}
\end{subeqnarray}
with $\Omega^{cd}_{ab}$ from (\ref{r6}). In particular the   
operators $T^\pm  (\th)_a^b$ have turned into classical $2\times  2$
matrices which due to (T2) have unit determinant
\be
C_{cd}\,T^\pm  (\th)_a^c T^\pm  (\th)_b^d = C_{ab}\;.
\label{WY2}
\ee
The $*$-structure (\ref{*op}) translates into the classical hermiticity
relations 
\be
\left[T^+(\th)_a^b\right]^* = E_{b'}^b\, 
T^-(\th^*)_a^{b'}\;,\;\;\;
\left[T^-(\th)_a^b\right]^* = \eps E_{b'}^b\, 
T^+(\th^*)_a^{b'}\;,\;\;\; 
\left[W_a(\th)\right]^* = W_a(\th^*)\;.
\label{herm}
\ee
The classical analogue of the matrix $D_{ab}(\th)$ from (\ref{Ddef})
is given by
\be
D_{ab}(\th) = C_{cd} T^-(\th)_a^c T^+(\th)_b^d\;,\qquad\mbox{with}\quad
[D_{ab}(\th)]^* = \eps D_{ba} (\th^*)\;.
\label{Dherm}
\ee
Separating the symmetric hermitian and the anti-symmetric part for 
$\eps = -1$ yields 
\ba
&& D_{ab}(\th) =: -i{\cal M}_{ab}(\th) - \frac{1}{2}C_{ab}\, D(\th) \;,
\quad D(\th) = C^{ab}D_{ab}(\th)\;,\nonum
&& \det {\cal M}_{ab}(\th) +\frac{1}{4} D(\th)^2 = 1 
= - \det D_{ab}(\th)\,.
\label{DM}
\ea
Since for $\eps =-1$ $D(\th)$ is purely imaginary, one has in
particular $\det \cM(\th) \geq 1$.

We proceed by showing that the antisymmetric part of $D_{ab}(\th)$
(and hence also $\det {\cal M}_{ab}$) is not a dynamical degree of
freedom of the Poisson algebra (\ref{WY1})--(\ref{herm}). One
indication is the fact that it Poisson-commutes with the generators
$T^\pm (\th)_a^b$ and $W_a(\th)$. More specifically, there exists an
automorphism of the Poisson algebra such that $D(\th)$ is mapped onto
a prescribed numerical constant, e.g.~$D(\th) \equiv 0$ for $\eps
=-1$, yielding $\det\cM(\th) =1$. To find the automorphism consider
first the following rotation:
\ba\label{iso}
T^+(\th)_a^b &\rra & y_{++}T^+(\th)_a^b + y_{+-}T^-(\th)_a^b \;, \nonum
T^-(\th)_a^b &\rra & y_{-+}T^+(\th)_a^b + y_{--}T^-(\th)_a^b \;,\nonum
W_a(\th)     &\rra & W_a(\th) \;,
\ea
which obviously is a homomorphism of the Poisson algebra (\ref{WY1}). 
Note, that the parameters $y_{..}$ may depend on
$\th$ here. If we further specify them to be of the form
\ba
y_{\pm \pm } &=& 
\ft{\cos \alpha_\pm }{\sqrt{1-\frac12D(\th)\sin(2\alpha_\pm )}}
\;,\quad
y_{\pm \mp} ~=~ 
\ft{\sin \alpha_\pm }{\sqrt{1-\frac12D(\th)\sin(2\alpha_\pm )}}
\;,\nonumber
\ea
with $\alpha_\pm =\alpha_\pm (\th)$, the map (\ref{iso}) extends to an
automorphism of the full structure (\ref{WY1}), (\ref{WY2}). The
condition
\be\label{phipm}
\alpha_+(\th) = \eps\, \alpha_-(\th^*)^*\;,
\ee
finally ensures compatibility with hermiticity (\ref{herm}).  For
$\eps =1$ thus $\alpha_+ = \alpha_-=: \alpha$ must be real-valued, as
is $D(\th)$. For $\eps = -1$ both $D(\th)$ and $\alpha_+ = \alpha_- =:
i \alpha$, $\alpha \in \R$ are purely imaginary.  Under this
automorphism the antisymmetric part of the matrix $D_{ab}(\th)$
transforms as
\ba
D(\th) &\rra & D(\th)_\alpha ~:=~ \frac{D(\th)-2\sin(2\alpha)}
  {1-\frac1{2}D(\th)\sin(2\alpha)}\sspace\;\; 
\mbox{for} \;\;\;\eps = 1\,\nonum
D(\th) &\rra & D(\th)_\alpha ~:=~ \frac{D(\th)-2i\sinh(2\alpha)}
  {1-\frac{i}2D(\th)\sinh(2\alpha)}\sspace \mbox{for} \;\;\;\eps = -1\;.
\label{Diso}
\ea
For $\eps\!=\!1$ there are two disjoint orbits of $D(\th)\in \R$ under
the Poisson automorphism (\ref{iso}), the interval $[-2,2]$ and its
complement. In particular, the fixpoints of (\ref{Diso}) at $D\!=\!\pm
2$ are fake and correspond to a noninvertible map (\ref{iso}).  For
$\eps\!=\!-1$, on the other hand, the Poisson automorphism (\ref{iso})
acts transitively on $D(\th)\in i\R$. This means starting from any
non-zero value of $D(\th)$, this automorphism can be used to define
new generators of the Poisson algebra (\ref{WY1})--(\ref{herm}) for
which the new $D(\th) \equiv D(\th)_{\alpha}$ vanishes.  This fact
ensures that one can always work with symmetric $SL(2,\R)$ matrices
\be
\cM_{ab}(\th) := 
iD_{ab}(\th)_{\,\alpha\,\equiv\, 
  -\frac12\mbox{\scriptsize{arcsinh}}\left(\frac12iD(\th)\right)} 
  = \cM_{ba}(\th) \;.
\label{Mdef}
\ee
Further one can assume $\cM(\th)$ to have positive trace. This is
because a vanishing trace would contradict the positive determinant,
so that ${\rm Tr}\cM(\th)$ must be either positive or negative. Since
$T^{\pm}(\th)_a^b \ra \pm T^{\pm}(\th)_a^b,\;W_a(\th) \ra W_a(\th)$ is
a $*$-automorphism of the Poisson algebra (\ref{WY1}), (\ref{WY2}),
(\ref{herm}) one can take ${\rm Tr}\cM(\th) >0$.  The Poisson brackets
of $\cM(\th)$ follow from the classical limit of (\ref{Drel}d)
\ba
\left\{ \cM_{ab}(\th_1)\,, \cM_{cd}(\th_2)\right\} &=&
\frac1{\th_{12}}\bigg(\Omega_{ac}^{mn}\cM_{mb}(\th_1)\cM_{nd}(\th_2)
+  \cM_{am}(\th_1)\cM_{cn}(\th_2) \, \Omega_{bd}^{mn} \bigg)
\label{PBMM}\\
&+&
\frac1{\th_{12}}
\bigg(\cM_{am}(\th_1)\,\Omega_{bc}^{mn}\cM_{nd}(\th_2)\,
+ \cM_{mb}(\th_1)\,\Omega_{ad}^{mn}\cM_{cn}(\th_2) \bigg) \;.\nonumber
\ea
The Poisson algebra (\ref{PBMM}) turns out to provide a direct link to
the phase space of the Ernst system which will be detailed in section
3.2.

We focused on the regular part $\DI$ of the dynamical algebra here
because the various operations invoked: ``Taking the semi-classical
limit'', ``Taking the residue at $\th_{12} = \pm i\hbar$'' and
``Applying the automorphism (\ref{iso})'' are mutually
non-commuting. In particular taking the semi-classical limit of the
relation (R) would require further specifications. However since we
introduced topological concepts only on the level of the matrix
elements, not for the algebra, it is convenient to discuss the
semi-classical limit of the recursive structure directly on the level
of the functional equations (I), (II) and their solutions;
cf.~section 4.

\newsubsection{Phase space of the classical Ernst system}

The classical phase space of the Ernst system can be described in
various ways. A non-redundant parameterization is in terms of gauge
invariant symmetric $SL(2,\R)$ matrices $\cM_{ab}(\th)$, which can be
viewed as the ``scattering data'' from each of which a classical
solution can be reconstructed. These matrices can be shown, starting
from the canonical Poisson brackets associated with the action
(\ref{i1}), to carry the Poisson structure (\ref{PBMM})
\cite{KorSam98}. Hence in this non-redundant parameterization there is
a direct correspondence between the phase space of the Ernst system
and the subsector (\ref{Mdef}), (\ref{PBMM}) of the Poisson manifold
emerging in the classical limit of the dynamical algebra $\cD$.

In the quantum algebra we saw in section 2.2 that the antisymmetric 
part of $D_{ab}(\th)$ does not decouple algebraically. This enforced to 
work with the bigger algebra generated by $T^{\pm}(\th)_a^b,\;W_a(\th)$,
and to implement the decoupling in terms of an eigenvalue problem.
We now show that a Poisson algebra essentially equivalent to (\ref{WY1})
also naturally emerges in the Ernst system. Moreover this lends a
physical interpretation to the variables $T^{\pm}(\th)_a^b$ and 
$W_a(\th)$ in (\ref{WY1}).

We begin by recalling that the scalar sector of the Ernst system is
given by an $SL(2,\R)$ valued matrix $\cV_a^m$ which essentially
contains the vierbein components of the compactified dimensions. The
model is invariant under global $SL(2,\R)$ and local $SO(2)$
transformations
\footnote{We use abstract index notation in this section to indicate
the transformation behavior of the objects: Indices $a,b,\dots$ from
the beginning of the alphabet refer to covariance under $SL(2,\R)$,
whereas indices $k,l,\dots$ from the middle of the alphabet refer to
covariance under local $SO(2)$ rotations.}
\be\label{gh}
\cV_a^m(x) \mapsto g_a^b\,\cV_b^n(x) \, h_n^m(x)\;,\qquad
\mbox{with}\quad g_a^b\in SL(2,\R)\;,\;\;
                 h_n^m(x)\in SO(2)\;.
\ee
This invariance has its roots in the four-dimensional theory,
$SL(2,\R)$ descending from linear diffeomorphisms in the
``compactified'' coordinates, $SO(2)$ being a remnant of the
corresponding part of the local Lorentz group. The bilinear
combination
\be
M_{ab}(x)~=~ \cV_a^m(x)\,\cV_b^n(x)\,\delta_{mn} \;,
\ee
is invariant under local $SO(2)$ transformations and corresponds to
the metric components in the `compactified' dimensions. Note that the
symmetric $\delta_{mn}$ symbol appearing here is invariant only under
the $SO(2)$ subgroup of $SL(2)$.

The dynamics of the Ernst system is captured by a Lax pair
\cite{BelZak78,Mais78} whose spectral parameter -- in contrast to the
flat space integrable systems -- depends explicitly on the space time
coordinates, see \cite{Nico91} for a review. For definiteness, we focus
of the case of cyclindrical gravitational waves. In particular the
worldsheet then has Lorentzian signature and can be covered by
coordinates $x=(t,r),\,t\in \R$ and $r>0$. For the description of the
linear system, light-cone coordinates $x^{\pm} = t\pm r$ are most
convenient. It is then given by
\begin{equation}\label{ls}
\partial_\pm \cVh(x;\gamma) = \cVh(x;\gamma) L_\pm(x;\gamma)\;,
\end{equation}
with
$$
L_\pm(x;\gamma) = Q_\pm(x) + \frac{1\mp\gamma}{1\pm\gamma}\,P_\pm(x)\;.
$$
Here, $Q_\pm$ and $P_\pm$ are the compact and non-compact components
of the $sl(2,\R)$-valued current $\cV^{-1}\partial_\pm\cV$ lying in
$so(2)$ and its orthogonal complement, respectively.  The spectral
parameter $\gamma$ is given by the following explicit function
\be\label{sr}
\gamma(x;\th) = \frac1{r}\left(t-\th-\sqrt{(\th-t)^2-r^2}\;\right)\;,
\ee
of the $2D$ coordinates $x=(t,r)$ and a constant $\th$ which may be
understood as the underlying constant spectral parameter of (\ref{ls}).

The associated monodromy matrices $U(r,r',t\,|\,\th)$ are obtained in
the usual way as path ordered exponentials of the Lax connection
\ba
U(r,r',t\,|\,\th)&:= &\cVh^{-1}(t,r;\gamma(t,r;\th))\:
\cVh(t,r';\gamma(t,r';\th))\label{T}\\[4pt]
&=& \cP \exp \int_r^{r'} \!dz\;
L_1(t,z;\gamma(t,z;\th))\;,\nonumber 
\ea 
which are unique functionals of the connection
$L_\pm=\ft12\left(L_0\!\pm\!L_1 \right)\,.$ The integrand in (\ref{T})
lives on the twofold covering of the complex $\th$-plane where we
choose the branch cut of (\ref{sr}) varying on the real $\th$-axis
while $z$ runs from $r$ to $r'$. The monodromy matrix $U(r,r',t\,|\,\th)$
hence is well defined for $\th\!\notin\!\R$. Under (\ref{gh}) it
transforms as
\be
U(r,r',t\,|\,\th)_m^n \mapsto
h^{-1}(t,r)_m^k\,U(r,r',t\,|\,\th)_k^l\,h(t,r')_l^n \;,
\ee
in particular, it is invariant under the global $SL(2,\R)$
transformations.

Assuming regularity of the currents at the spatial boundaries and
time-independence at spatial infinity -- corresponding to the classical
sector of gravitational waves with regular Ernst potential on the
symmetry axis -- we define the following objects for real values of
the parameter $\th$:%
\begin{subeqnarray}\label{U12}
T^+ {}(\th)_{a}^{b} &:=& i \lim_{\delta\ra0}\,
\cV_0(t)_a^n\,U_n^k(0,\infty,t\,|\,\th\!+\!i\delta )\,
(\cV_\infty)^l_c \,C_{kl}\,C^{cb}\;, \\[6pt]
T^- {}(\th)_{a}^{b} &:=& \lim_{\delta\ra0}\,
\cV_0(t)_a^n\,U_n^k(0,\infty,t\,|\,\th\!-\!i\delta )\,
(\cV_\infty)^l_c \,\delta_{kl}\,C^{cb}\;, 
\nonumber\\[6pt]
W{}_a^m(\th) &:=& \cV_0(t)_a^n\,U_n^m(0,|\th\!-\!t|,t\,|\,\th)\;, 
\end{subeqnarray}
with
$$
\cV_0(t) ~:= ~ \cV(0,t)\;,\qquad 
\cV_\infty ~:= ~ \cV(\infty,t) = \cV(\infty)\;.
$$
The $T^{\pm}(\th)$ are conserved and the $W(\th)$ are conserved 
up to a local gauge transformation
\ba
\label{Wdyn}
\partial_t\,T^\pm {}(\th)_{a}^{b} &=& 0 \\
\partial_t\,W{}_a^m(\th) &=& \left\{ 
\begin{array}{rcl}
W{}_a^n(\th)\; Q_-(t,\th\!-\!t)_n^m & \mbox{for} & \th\!>\!t  \\[1ex]
W{}_a^n(\th)\; Q_+(t,t\!-\!\th)_n^m & \mbox{for} & \th\!<\!t  
\end{array} \;. \right.
\nonumber
\ea
The fact that the last equation is not 2D covariant is due
to the explicit appearance of the time $t$ in the integration boundary
of $W{}_a^m(\th)$. Observe however that according to equation
(\ref{Wdyn}) the time derivative of $W{}_a^m(\th)$ is continuous in
$\th$ since $Q_1(t,r\!=\!0)=0$, as follows from the field equations
derived from (\ref{i1}) and regularity of the vielbein ${\cal V}_a^m$
on the symmetry axis. As indicated we shall usually suppress the
time-dependence of $W_a^m(\th)$ in the notation. Restoring it
momentarily $W_a(\th) \leadsto W_a(t,\th)$ one finds for the time
evolution (e.g.\ for $\th>t$)
\be
W(t,\th) = W(0,\th)\,e^{i\phi(t,\th)\sigma^2}\;,\quad
\phi(t,\th) = -\frac{i}{2}\int_0^t ds \,\Tr[Q_+(s,\th\!-\!s)\sigma^2]\;,
\label{Wevol}
\ee
where $\sigma^2$ is the Pauli matrix and $\phi(t,\th)$ is real for
real $\th$. Note, that in contrast to $W(t,\th)$, the function
$\phi(t,\th)$ is defined by integrating over a null line in spacetime,
and may hence not be considered as canonical object on a fixed time
slice. In particular, this makes it difficult to compute its Poisson
brackets.

The matrices $T^\pm(\th)$ satisfy (\ref{WY2})
whereas the $W(\th)$ obey 
\be
C_{mn}\,W_a^m(\th)W_b^n(\th) ~=~ C_{ab}\;,\qquad
C^{ab}\,W_a^m(\th)W_b^n(\th) ~=~ C^{mn}\;.
\label{Wdet}
\ee
Under the symmetry transformations (\ref{gh}), the matrices
(\ref{U12}) behave as  
\ba
\label{WTgh}
T^\pm {}(\th)_{a}^{b} &\mapsto& g_a^d\,T^\pm {}(\th)_{d}^{c}
                                      \,(g^{-1})_c^b\;,\\
W{}_a^m(\th) &\mapsto& g_a^d\,W{}_d^n(\th)\,h_n^m(t,|\th\!-\!t|)
\;.\nonumber 
\ea
Clearly any gauge invariant quantity build from these monodromy
matrices will be time independent. A gauge invariant object of particular
interest is the bilinear combination
\be\label{RH}
\cM_{ab}(\th) ~=~ i\,T^-(\th)_a^c \,T^+(\th)_b^d\,C_{dc} ~=~ 
W_a^{m}(\th) W_b^{n}(\th)\,\delta_{mn} \;.
\ee
The second equality follows from the definition (\ref{U12}) in the
limit $t\rightarrow\infty$ and shows that the matrix $\cM_{ab}$ is
symmetric in the indices $a,b$, and has positive trace.  In
particular, for the $T^{\pm}(\th)_a^b$ defined by (\ref{U12}), the
combination $D(\th) = C^{ab}C_{cd} \,T^-(\th)_a^c T^+(\th)_b^d$
vanishes automatically. The decomposition of $\cM(\th)$ into the
product of $T^\pm$ corresponds to the Riemann-Hilbert decomposition;
the matrices $T^\pm(\th) $ are holomorphic in the upper resp.~lower
half of the complex $\th$-plane. It may further be shown that
\be\label{contM}
\cM_{ab}(\th\in\R) ~=~ M_{ab}(t\!=\!\th, r\!=\!0)\;,
\ee
i.e.~this matrix coincides with the physical scalar fields on the axis
$r=0$ \cite{BreMai87,KorSam98}. From the viewpoint of the inverse
scattering transform, equation (\ref{contM}) is a striking result.
Usually, the scattering data associated with a given solution ``live
at'' timelike infinity and have no direct relation to the original
field variables. In contrast, for the Ernst system, equation
(\ref{contM}) means that the scattering data live on the symmetry
axis $r=0$ and are directly related to the original vielbein
variables.  The second Gauss-like decomposition of $\cM$ into the
bilinear product of $W$'s in (\ref{RH}) therefore corresponds to the
decomposition of the metric into a spectral-transformed vielbein. As
anticipated by the notation, the matrices (\ref{RH}) can be identified
with (\ref{Mdef}) and the decomposition in (\ref{RH}) may be viewed as
the classical analogue of (\ref{R1}).

\begin{figure}[htbp]
  \begin{center}
    \leavevmode 
\begin{picture}(0,0)%
\epsfbox{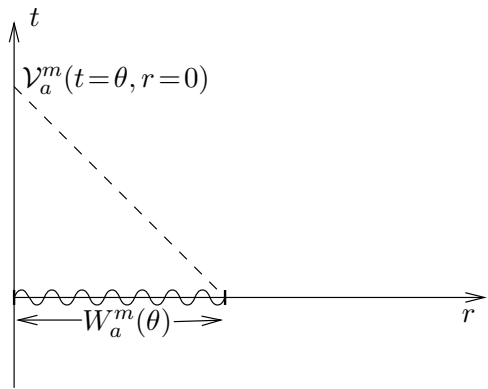}
\end{picture}%
\setlength{\unitlength}{0.00083300in}%
\begingroup\makeatletter\ifx\SetFigFont\undefined%
\gdef\SetFigFont#1#2#3#4#5{%
  \reset@font\fontsize{#1}{#2pt}%
  \fontfamily{#3}\fontseries{#4}\fontshape{#5}%
  \selectfont}%
\fi\endgroup%
\begin{picture}(3015,2391)(1151,-2280)
\put(1599,-1900){\makebox(0,0)[lb]{\smash
{\SetFigFont{11}{13.2}{\rmdefault}{\mddefault}
{\updefault}$W_a^m(\th)$}}}
\put(1221,-384){\makebox(0,0)[lb]{\smash
{\SetFigFont{11}{13.2}{\rmdefault}
{\mddefault}{\updefault}${\cal V}_a^m(t\!=\!\th,r\!=\!0)$}}}
\put(3964,-1851){\makebox(0,0)[lb]{\smash
{\SetFigFont{11}{13.2}{\rmdefault}{\mddefault}{\updefault}$r$}}}
\put(1268, -6){\makebox(0,0)[lb]{\smash
{\SetFigFont{11}{13.2}{\rmdefault}{\mddefault}{\updefault}$t$}}}
\end{picture}
  \end{center}
\caption{The spectral transformed vielbein $W_a^m(\th)$}
\label{VW}
\end{figure}

The fact that the scattering data have an interpretation in terms of
the original field variables imposes an interesting causality
constraint, which also clarifies the structure of the monodromy
matrices $W_a^m(\th)$: Monodromy matrices of the form (\ref{U12}b)
(path ordered integrals over {\em finite} space intervals without any
specification of conditions on the physical fields at the boundary of
the interval) do not arise in the usual flat space integrable
systems. The raison d'\^{e}tre in the Ernst system is the space-time
dependence of the spectral parameter (\ref{sr}). The Lax connection
(\ref{ls}) degenerates at certain points in space-time though the
physical currents remain regular. This happens at the spatial
boundaries $r=0, \infty$, but also, curiously, at $r=|\th\!-\!t|$, --
a point by no means distinguished in the physical space-time. However
this point does have special significance for the causal past of the
point $(\th,0)$ on the symmetry axis. For given $t_0$ and $\th$ (with
$t_0 < \th$, say) consider the intersection of the causal past of
$(\th,0)$ with the $t = t_0$ surface, i.e.~the interval $[0,
\th-t_0]$.  According to causality in the $(t,r)$ Lorentzian space,
the vielbein on the symmetry axis at time $t\!=\!\th$ should be a
functional of the initial data on the interval $[0,\th\!-\!t_0]$
only, which due to the range of integration in (\ref{U12}b) it indeed
is; cf.~Fig.~\ref{VW}.

We proceed by studying the hermitian structure and the Poisson brackets
of the monodromy matrices (\ref{U12}). 
The hermiticity $U(r,r'\,|\,\th)^*=U(r,r'\,|\,\th^*)$ implies
\ba
{}\left[T^\pm(\th)_{a}^{b}\right]^* &=& \mp i\, T^\mp(\th^*)_{a}^c\, 
\cM^\infty_{cd}\, C^{db} 
\;,\label{hhh}\\ 
{}\left[W_a^m(\th)\right]^* &=& W_a^m(\th^*) \;,\nonumber
\ea
where the constant matrix $\cM^\infty_{cd}\equiv (\cV_\infty)_c^{m}
(\cV_\infty)_d^{n}\,\delta_{mn}$ defines a positive definite
bilinear form on $SL(2,\R)$. Due to the explicit appearance of
$\cM^\infty_{ab}$ in (\ref{hhh}), this hermitian structure is not
invariant but transforms covariantly under (\ref{WTgh}). With
$$
E_a^b ~=~ - i\,\cM^\infty_{ac}\,C^{cb}\,,
$$
we recover (\ref{herm}), i.e.~the classical limit of
(\ref{a1}). Fixing the $SL(2,\R)$ symmetry (\ref{WTgh}) by setting
$(\cV_\infty)_a^{m}=\delta_a^{m}$ corresponds to the choice
$E = i\sigma^2$ in (\ref{*op1}). Having done so, one is still left with 
$SL(2,\R)$ basis transformations acting on the lower index
\be
T^\pm {}(\th)_{a}^{b} \mapsto  g_a^d\,T^\pm {}(\th)_{d}^{b}\;,\sspace
W{}_a^m(\th) \mapsto g_a^d\,W{}_d^m(\th)\;.
\label{WTb}
\ee
The linear transformations (\ref{WTb}) provide a $*$-automorphism of
the Poisson algebra even after (\ref{WTgh}) has been ``eaten up'' by
fixing the $*$-structure (e.g.~$\cM_{ac}^{\infty} = \delta_{ac}$) and
even when working with SO(2) gauge fixed $W_a^m(\th)$ matrices. We
shall return to the distinction between (\ref{WTgh}) and (\ref{WTb})
in section 3.3.

The Poisson brackets of the monodromy matrices $T^\pm $ coincide with
(\ref{WY1}a,b), where again the coordinate dependence of the spectral
parameter $\gamma$ plays a crucial role in the computation
\cite{KorSam98}. Evaluating the general formula from \cite{KorSam98},
one similarly obtains for the matrices $W_a^{m}(\th)$
\begin{subeqnarray}\label{WYp}
\left\{ T^\pm (\th_1)_a^e\,,W_b^m(\th_2)\right\}
&=& \frac1{\th_{12}}\,T^\pm (\th_1)_c^e \,W_d^m(\th_2)\;
\Omega^{cd}_{ab}\\
&&{}-\frac{1}{2\th_{12}}\,
 W_b^{l}(\th_2) \,T^\pm (\th_1)_c^e\;
U_a^{k}(\th_2\,|\,\th_1)\,
(U^{-1})_n^c(\th_2\,|\,\th_1)\;E_k^n\,E_l^m\;,   
\nonumber\\[4pt]
\left\{ W_a^m(\th_1)\,,W_b^n(\th_2)\right\}
&=& \frac1{\th_{12}}\, 
W_c^{m}(\th_1)\, W_d^{n}(\th_2)\;\Omega^{cd}_{ab}  \\
&&{}-\frac{\chi(\th_{12})}{2\th_{12}}\,
W_b^{l}(\th_2)\,W_c^{m}(\th_1)\;
U_a^{k}(\th_2\,|\,\th_1)\,
(U^{-1})_j^c(\th_2\,|\,\th_1)\;E_k^j\,E_l^n \nonum
&&{}-\frac{\chi(\th_{21})}{2\th_{12}}\,
W_a^{k}(\th_1)\,W_c^n(\th_2)\;
U_b^{l}(\th_1\,|\,\th_2)\,
(U^{-1})_j^c(\th_1\,|\,\th_2)\;E_k^m\,E_l^j \;,\nonumber
\end{subeqnarray}
with the step function $\chi(\th) = {\rm sign}(\th)$, and the matrix
$E_m^n=\delta_{mk}C^{kn} = (i\sigma^2)_m^n$ here playing the role of an 
$SO(2)$ invariant tensor. Further 
\ba
U_a^{k}(\th_2\,|\,\th_1) &:= &
\cV_0(t)_a^m\,U_m^k(0,|\th_2\!-\!t|,t\,|\,\th_1) \;, \nonum
(U^{-1})_k^a (\th_2\,|\,\th_1) &:= & U_k^m(|\th_2\!-\!t|,0,t\,|\,\th_1)\,
\cV_0(t)_b^n\,C_{mn}\,C^{ba} \;. \nonumber
\ea
Evidently the Poisson structure is not closed but contains the transition
matrices $U(\th_2\,|\,\th_1)$, etc.~on the right hand
side. Nevertheless the Jacobi identities can be verified, and
(\ref{WYp}) defines a consistent Poisson structure on the phase space
of the Ernst system. As remarked earlier, the combination $D(\th) =
C^{ab}C_{cd} \,T^-(\th)_a^c T^+(\th)_b^d$ here vanishes automatically.

Consistency thus requires that $D(\th)$ also Poisson commutes with
$W_a^m(\th)$ (with respect to the brackets (\ref{WYp})), -- which
indeed can be verified to be the case. Moreover it can be shown that
(\ref{WYp}) induces the Poisson structure (\ref{PBMM}) for the gauge
invariant phase space functions $\cM_{ab}(\th)$, i.e. the same as the
somewhat simpler Poisson structure (\ref{WY1}) obtained from the
classical limit of our dynamical algebra.  Thus, up to redundancies,
(\ref{WYp}) can be regarded as equivalent to (\ref{WY1}c,d) in the
sense that both induce the same structure on the space of objects
invariant under the $SO(2)$ gauge transformations in
(\ref{gh}).\footnote{A similar point of view is e.g.~usually adopted
to study the symplectic structures on the moduli space of flat
connections on Riemann surfaces \cite{FocRos92}.}  The extra $U$-terms
in (\ref{WYp}) can be viewed as being a remnant of the gauge dynamics
(\ref{Wevol}).  The problem is that (\ref{WYp}) can not readily be
rewritten as a Poisson bracket structure on the initial data
$W_a^m(0,\th)$. This is because although the time evolution
(\ref{Wevol}) can be viewed as a gauge transformation, the
transformation is a nonlocal functional of the dynamical variable
$Q_+(x)$, so that at some point non-equal-time Poisson brackets would
have to be evaluated.  In principle, however, we view the $W_a(\th)$
in (\ref{WY1}) as being gauge fixed or gauge invariant and
time-independent versions of a linear combination of $W_a^1(\th)$ and
$W_a^2(\th)$. We have not been able so far to properly map the Poisson
brackets (\ref{WYp}) onto (\ref{WY1}) for such a combination.  Our
main argument that it should be possible is, that on gauge invariant
objects like $\cM(\th)$ both induce the same Poisson structure.

One can also check that the counting of degrees of freedom works out:
There are two types of redundancies in (\ref{WYp}). First (\ref{WYp})
is invariant under the symmetry transformations (\ref{WTgh}). In
particular the local gauge transformations $W_a^m(\th) \ra W_a^n(\th)
h_n^m(t,|\th -t|)$ effectively remove one degree of freedom. In
addition (\ref{WYp}) has a one-dimensional Poisson center generated by
the determinant $\det W_a^m(\th)$. Thus (\ref{WYp}) contains only two
physical degrees of freedom for the $W_a^m(\th)$ fields, just as
(\ref{WY1}c,d).

Taking the equivalence of (\ref{WYp}) and (\ref{WY1}c,d) for granted,
the recursive relations (R) can be viewed as a quantum implementation
of the identity (\ref{RH}) and the determinant condition (\ref{Wdet}).
To see this set
\be
W^{\pm}_a(\th) = W_a^m(\th) \,(\Upsilon\inv)^{\pm}_m\,,
\label{Wcorr1}
\ee 
where ${\rm Ad}\Upsilon: SL(2,\R) \ra SU(1,1)$ is the isomorphism
(\ref{act11}) or (\ref{b6}). Then the $W^{\pm}_a(\th)$ are complex
fields transforming as $W^{\pm}_a(\th) \ra W^{\pm}_a(\th) e^{\pm
i\phi(t,|\th - t|)}$ under $SO(2)$ gauge transformations with $h = \cos
\phi \1 + \sin \phi E$. Further, they obey
\be
W^{\pm}_a(\th) W^{\mp}_b(\th) = 
{\textstyle\frac12}\,(\cM_{ab}(\th) \mp C_{ab})\;,
\label{Wcorr2}
\ee
using the definition (\ref{RH}) and the determinant condition
(\ref{Wdet}).  In the quantum theory (\ref{Wcorr2}) will turn into a
singular operator product. Parallel to (R) one can stipulate
\be
W_a^{\pm}(\th - i\hbar)\cdot W_b^{\mp}(\th) ~=~ 
{\textstyle\frac{i}2}\,D_{ab}(\th - 2 i \hbar)\;,
\label{Wcorr3}
\ee
while all `$\,\cdot\,$' products of $W^+$ with itself and of $W^-$
with itself are supposed to vanish. For simplicity we set $\lb =i$
here and only noted the counterpart of (R2), i.e. (\ref{R1}); the
interplay with the other versions is analogous to (R1) -- (R3). The
obvious $*$-operation is $\sigma W^{\pm}_a(\th) = W^{\mp}_a(\th^* +
i\hbar)$. Finally, consider the linear combinations
\be
V_a(\th) = e^{i\phi(\th)} W^+_a(\th) + e^{-i\phi(\th)} W_a^-(\th)\,,
\label{Wcorr4}
\ee  
with parameter $\phi(\th) = \phi(t,\th)$, obeying $\phi(\th)^* =
\phi(\th^* +i\hbar)$. If in addition we take $\phi(\th)$ to be
$i\hbar$-periodic, the `$\,\cdot\,$' products (\ref{Wcorr3}) etc.~for
$W_a^{\pm}(\th)$ imply (R) for $V_a(\th)$ with $\lb =i$. If the
$W^{\pm}$ in (\ref{Wcorr4}) are rescaled by $\omega(\th)$, obeying
$\omega(\th)^* = \omega(\th^*\pm i\hbar)$, the same holds with
$\lb(\th) = i \omega(\th) \omega(\th + i \hbar)$, a form we shall use
later.  In principle we could have developed the entire formalism of
section 2 for an enlarged quantum algebra with a pair of
$W^{\pm}_a(\th)$ generators and exchange relations like those in $\DI$
for both of them. The advantage of working with the linear
combinations (\ref{Wcorr4}) is that the recursive relations are
simpler because the same rule applies to each pair of $V$-generators.
We do not expect substantial difficulties in deriving a system of
functional equations analogous to (I), (II) for matrix elements of
mixed strings of $W^{\pm}$ generators. Observe however that according
to (\ref{Wcorr3}) the recursive structure in any case determines only
the `propagation' of the gauge invariant parts. For example, matrix
elements with only $W^+$'s would only be constrained by the analogue
of the functional equations (I).

As remarked before we prefer to analyze the semi-classical limit
directly on the level of the functional equations (I), (II) and their
solutions.  Observe however that comparing the formal $\hbar \ra 0$
limit of (\ref{Wcorr3}) with (\ref{Wcorr2}) and (\ref{DM}) one can
match the expressions by taking $D(\th) = \pm 2 i$. Clearly the
Poisson algebra (\ref{WYp}) admits an automorphism analogous to
(\ref{iso}). Starting from $D= 0$ one can achieve $D_{\alpha} = \pm 2
i$ by taking $\sh 2\alpha = \pm 1$.  The conjecture that the Poisson
structures (\ref{WY1}) and (\ref{WYp}) are fully equivalent,
translates into one concerning the status of the field $\phi(\th)$. As
long as it is regarded as an independent parameter, the linear
combination (\ref{Wcorr4}) will obey Poisson brackets of the form
(\ref{WY1}), modified by the extra $U$-terms. By allowing $\phi(\th)$
to become dynamical (as in (\ref{Wevol})) one may hope to render the
$V_a(\th)$ time independent and either gauge fixed or gauge
invariant. At the same time the $U$-terms should disappear, making the
correspondence $V_a(\th) \leadsto W_a(\th)$ precise.
 
Summarizing, in the classical limit the dynamical algebra $\DI$ gives
rise to a phase space parametrized by symmetric $SL(2,\R)$ matrices
$\cM$ equipped with the Poisson structure (\ref{PBMM}) and the two
alternative Riemann-Hilbert and Gauss decompositions into matrices
$T^\pm $ and $W$, respectively. The resulting Poisson algebra
(\ref{WY1})--(\ref{herm}) essentially coincides with that deduced from
the fundamental Poisson brackets in the classical Ernst system;
admittedly yet with some loose ends.

\newsubsection{Symmetry breaking}

In retrospect we can now also highlight some aspects of our quantum
formulation. As noted in section 2.1, the antisymmetric part of
$D_{ab}(\th)$ does not decouple algebraically within $\DI$.  Thus in
order to avoid that the quantum theory has an extra dynamical operator
field $D(\th)$, one is forced to go to the critical level $\beta =0$,
where $D(\th)$ becomes central.  However an automorphism of the form
(\ref{iso}), (\ref{Diso}) no longer exists in the quantum theory; the
spectrum of $D(\th)$ on the state space described by the matrix
elements (\ref{t3}) is a characteristic feature of the
system. Remarkably the state space exhibits a (``spontaneous'') {\em
breakdown of the classical $SL(2,\R)$ invariance} that is a is a
remnant of the original four dimensional diffeomorphism invariance in
the Killing coordinates: As we have seen before for $\eps =-1$ the
eigenvalue problem (\ref{t6}) is invariant only under the SO(2)
subgroup, while the algebra itself and the spectrum of $D(\th)$ still
are fully $SL(2,\R)$ invariant. This suggests that, in contrast to the
prevalent assumptions in many approaches to quantum gravity,
diffeomorphism invariance is not sacrosanct; it might be broken for
dynamical reasons.

In view of the possible implications it seems worthwhile to critically
reexamine the line of argument and to see whether, in the context of
the present framework, the conclusion can be avoided.  To address the
issue, it is convenient to introduce shorthands for the different
$SL(2,\R)$ actions involved. Let $SL2_D$ be the $SL(2,\R)$ action
(\ref{WTgh}) inherited via (\ref{gh}) from the linear diffeomorphisms
in the Killing coordinates. Let $SL2_L$ be the $SL(2,\R)$ action
(\ref{WTb}) acting by basis transformations on the lower
index. Finally denote by $SL2_U$ the $SL(2,\R)$ action $W_a^m(\th) \ra
W_a^m(\th),\;T^{\pm}(\th)_a^b \ra g_{b'}^b T^{\pm}(\th)_a^{b'}$. The
latter two have obvious counterparts in $\DI$ and its classical limit,
the $SL2_D$ action then corresponds to the diagonal action ${\rm
diag}(SL2_L \times SL2_U)$. On $D_{ab}(\th)$ both $SL2_D$ and $SL2_L$
act as
\be
D_{ab}(\th) \ra g_a^c g_b^d \,D_{cd}(\th)\;,
\label{SL2BD}
\ee
and similarly for $\cM_{ab}(\th)$. Thus if $SL2_L$, and hence
(\ref{SL2BD}) is broken, also the invariance under $SL2_D$ must be
violated. On the other hand, we saw in section 2 that on the matrix
elements (\ref{t3}) the $SL2_L$ invariance is indeed broken down to
its $SO(2)$ subgroup.

To confirm this result let us reexamine the underlying assumptions.
Technically, the result only hinges on the fact that for $\eps =-1$
the hermiticity conditions (\ref{*gam}) do not permit a solution
yielding $\Gamma \sim \1$. One can also convince oneself that simple
modifications (restoring a $\th$-dependence in $\gamma^{\pm}$, use of
a graded hermiticity requirement, etc.) either do not affect the above
feature or are at odds with the classical limit. In particular,
allowing $\Gamma$ to become complex (and hence to have a nonvanishing
trace by (\ref{Gaprop})) does not help. The solutions $\Lambda$ of
(\ref{norm5}) would then be given by real linear combinations of
$\Gamma + \Gamma^* = 2 \Gamma - \Tr \Gamma \1$ and $\1$, and would
still generate not more than a $SO(2)$ subgroup of
$SL(2,\R)$. Ultimately, the reason why the seemingly kinematical
hermiticity requirement can have such a severe impact on the choice of
the relevant representations is, that the algebra $\cD$ and its
$*$-structure already contain most of the dynamical information about
the system.

The close relation to the hermiticity requirement also explains why a
group averaging procedure does not provide a way out. Since the
eigenvalues in (\ref{t6}) are $SL(2,\R)$ invariant, one might hope
that by averaging over the conjugacy class (\ref{a2}) the invariance
of $\cT$ and its eigenvectors can be restored. Leaving aside the
arbitrariness stemming from the non-uniqueness of a regularized
invariant measure on $SL(2,\R)$, the main problem with this proposal
is that the only $SL(2,\R)$ invariant outcome of the averaging
procedure in $\cT$ can be a $\Gamma^{\rm av} \sim \1$.  This however
violates the hermiticity condition, and one is lead back to the
discussion of the preceding paragraph.

One might also suspect that the lack of $SL(2,\R)$ invariance in
(\ref{t6}) is already present on the level of the (classical or
quantum) algebra through the lack of $SL(2,\R)$ invariance of the
$*$-structure (\ref{herm}) or (\ref{a1}) with respect to the upper
index.  Since the definition of $D(\th)$ involves a contraction over
an upper index pair, one might argue, that the violation of $SL(2,\R)$
invariance already enters at this point. However, this is not the
case. Though not invariant, the $*$-structure (\ref{herm}) is
$SL(2,\R)$ covariant. An $SL2_U$ rotation of the form $W_a(\th) \ra
W_a(\th)$, $T^\pm (\th)_a^b \ra g_{b'}^b\,T^\pm (\th)_a^{b'}$ provides
an automorphism of the algebra $\cD$ under which the $*$-structure
(\ref{a1}) transforms covariantly as in (\ref{EFcov}).  But in the
hermiticity equations (\ref{a4}) the matrix $E$ drops out; both the
operator $D(\th)$ and its hermiticity condition (\ref{a4}) are
invariant under both the $SL2_U$ and the $SL2_L$ actions of the
$SL(2,\R)$ algebra.  However its matrix elements (\ref{t3}) are not!
It is clear that the phenomenon disappears in the classical limit
because then $D(\th)$ can be set to zero. In summary we conclude that
the classical $SL2_L$ symmetry, and by (\ref{SL2BD}) therefore the
$SL2_D$ invariance, being a remnant of the 4D diffeomorphisms in the
Killing coordinates, is ``spontaneously'' broken in the quantum
theory.

Because of the coupling to gravity, no conflict with the
Coleman-Mermin-Wagner theorem \cite{MerWag66,Cole73} arises. In
lattice formulations usually also the compactness of the global
symmetry group is assumed in order to ensure the existence of the
regularized functional integral. If, as in Coleman's version
\cite{Cole73}, the existence of the quantum field theory and the
Noether current is postulated, the result should hold also for
non-compact groups.  However the Poincar\'{e} invariance and the
cluster property are essential for the argument. For the Ernst system
the former is manifestly absent and the latter would at least require
a re-interpretation.  In the statistical mechanics context, one way to
look at the Mermin-Wagner theorem is as a tug of war between entropy
and energy. For a flat space sigma-model in 2D the entropy always
wins, forcing the system to remain in a disordered state even at low
temperatures.  From this perspective the above result indicates that
the coupling to gravity changes this entropy--energy balance such that
a breaking of the $SL(2,\R)$ symmetry becomes possible.

Concerning the worldsheet diffeomorphism invariance, it is clear
that if 2D conformal invariance is broken, so is invariance under
diffeomorphisms.  Classically the constraints induced by fixing the
conformal gauge $h_{\mu\nu} = e^{2 \sigma} \eta_{\mu\nu}$ in the
action (\ref{i1}) are
\be
T_{\pm\pm} = 2 \partial_{\pm}\rho \partial_{\pm} \widehat{\sigma}
-\rho {\rm Tr}[ P_{\pm} P_{\pm}]\;.
\label{conf1}
\ee
Here $\widehat{\sigma}= \sigma - \frac{1}{2}\ln|\partial_+\rho
\partial_-\rho|$ is a conformal scalar and $P_{\mu}$ is the coset part
of the current appearing in (\ref{ls}). Their Poisson brackets form
two commuting copies of the Virasoro-Witt algebra. Further
(\ref{conf1}) can be checked to generate infinitesimal conformal
transformations on gauge invariant objects, otherwise an additional
SO(2) gauge transformation is induced. In particular, for the conserved
charges (\ref{U12}) and Weyl coordinates $\rho =r$ one has
$\{T_{\pm\pm}(t,r)\,,\, T^{\pm}(\th)_a^b \}\! = \!0$, and modulo some 
technical subtleties a similar equation holds for $W_a^m(\th)$. 
Classically the (weakly) vanishing of the constraints is compatible
with the equation of motion for $\sigma$. The latter can be integrated
to express $\sigma$ in terms of $\rho$ and $P_{\mu}$. Off-shell 
the essential dynamical features of $\sigma$ should be captured by the 
quantum counterparts of its Poisson brackets with $T^{\pm}(\th)_a^b$ 
and $W_a^m(\th)$. The latter turn out to be remarkably simple
\be
\{ \sigma(t,0), T^{\pm}(\th)_a^b\} = \partial_{\th} T^{\pm}(\th)\;,
\sspace
 \{ \sigma(t,0), W_a^m(\th)\} = \partial_{\th} W_a^m(\th)\;,
\label{conf2}
\ee
assuming that $\sigma(t,r)$ vanishes for $r \ra \infty$. The obvious
quantum  counterparts are 
\be
e^{i\lb K} T^{\pm}(\th)_a^b e^{-iK\lb} = T^{\pm}(\th + \lb)_a^b\;,\sspace
e^{i\lb K} W_a(\th) e^{-iK\lb} = W_a(\th + \lb)\;. 
\ee
That is, translations in $\th$ are unitarily implemented, and the
generator $K$ is the quantum counterpart of the conformal factor in
the 2D metric.

\newsection{Exact matrix elements}

We return now to the functional equations (I), (II) of section 2.  As
outlined, its solutions are conjectured to describe exact matrix
elements in the quantum theory, without the need for any
renormalization.  We begin by describing a solution algorithm for
(I), (II).

\newsubsection{Solution algorithm}

Before turning to the solution procedure, let us note some simple
structural features of (I) and (II). Clearly a solution of (I) is
determined only up to multiplication by a scalar function completely
symmetric in $\th_\n, \ldots ,\th_1$. The recursive equations (II) cut
down this ambiguity to scalar, symmetric functions solving
\be
P(\th)\bigg|_{\th_{k+1} = \th_k \pm i\hbar} = P(p_k\th)\,.
\label{sol1}
\ee     
In other words, if $(f^{(\n)})_{\n \geq \n_0}$ (with $\n$ even or odd
depending on the starting member) is a sequence solving (I), (II) and
$(P^{(\n)})_{\n \geq \n_0}$ is a sequence solving (\ref{sol1}), then
the sequence obtained by pointwise multiplication,
$(P^{(\n)}f^{(\n)})_{\n \geq \n_0}$, again is a solution of (I),
(II). According to (IIc), the eigenvalues sequences are an example of
a sequence (\ref{sol1}) but there are many others, e.g.~power sums
$\sum_{j =1}^\n e^{s \pi\th_j/\hbar}$, with $s$ odd, or any smooth
scalar function of $\th_{\n} +\ldots + \th_1 - 2(t_1 + \ldots +
t_{\Lambda})$, where $t_{\alpha}$ are the Bethe roots of appendix
B. Usually one will be interested in solutions of (I), (II) which are
`minimal' in the sense that one cannot `naturally' split off a
solution of (\ref{sol1}). Apart from these obvious ambiguities we
expect that associated with each starting member $f^{(\n_0)}$ there is
basically only a single sequence solving (I), (II).

\begin{figure}[h]
\vskip -16mm
\leavevmode
\epsfxsize=135mm
\epsfysize=80mm
\epsfbox{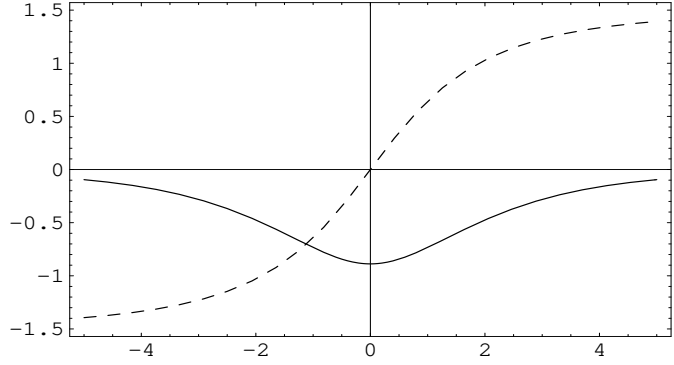}
\vskip -3mm
\caption{Real (solid) and imaginary (dashed) part of 
$\ln[\psi(\frac{\hbar u}{\pi})/\tanh\frac{u}{2}]$ for real $u$.} 
\label{psipic}
\end{figure}

To actually find solutions of (I), (II) we make an ansatz of the form
\be
f_A(\th) = \fbar_A(\th) \,c^{(\n)}(\th) \prod_{k>l} 
\frac{\psi(\th_{kl})}{\th_{kl} - i\hbar}\,,
\label{ansatz}
\ee
where $c^{(\n)}(\th)$ absorbs the $\lb(\th)$ dependence and
$\psi(\th)$ satisfies  
\ba
&& \psi(\th)= r(\th) \psi(-\th)\,,\nonum
&& \psi(-\th)\psi(\th - i\hbar) = -1\;.
\label{ansatz2}
\ea
The rationale for (\ref{ansatz2}) is that the first relation
effectively replaces (Ib) by exchange relations with a rational
$R$-matrix, the second one turns out to achieve the same for the
recursive relations (IIa,b). The solution of (\ref{ansatz2}) analytic
in the strip $-\hbar/2 < {\rm Im}\,\th < \hbar/2$ is given by 
\ba
&& \psi(\th) = \tanh\frac{\pi \th}{2\hbar } \,\exp\left\{
\frac{i}{2}\int_0^{\infty} \frac{dt}{t} \frac{h(t)}{\ch \frac{t}{2}} \,
\sin \frac{t}{2\hbar}(i\hbar + 2\th) \right\}\;,\nonum
&& \mbox{where}\;\;\;\; h(t) = 2\frac{e^{-t/2} + e^{-t}}{1 + e^{-t}}\,.
\label{psi}
\ea
The functional equations (\ref{ansatz2}) are readily verfied by means
of the integral representation 
\be
r(\th) = -\exp\left\{i \int_0^{\infty}\frac{dt}{t} h(t) 
\sin \frac{\th}{\hbar}t \right\}\;.
\label{rint}
\ee 
In addition $\psi(\th)$ has the following properties: It has a simple
pole at $\th = -i\hbar$ with residue $\hbar \psi_0$, where $\psi_0 :=
i \psi(i\hbar) = \lim_{\delta \ra 0} \delta/\psi(\hbar\delta/\pi)
\approx 1.54678$. The only real zero is at $\th =0$. Further it obeys
$\psi(\th)^* = - \psi(\th^*)$ and $\psi(\th)\ra i$ for $\th \ra \pm
\infty$, cf.~Fig.~\ref{psipic}.

It remains to specify the prefactors $c^{(\n)}(\th)$ in (\ref{ansatz}),
supposed to absorb the $\lb(\th)$ dependence and overall constants. 
For definiteness we assume $\lb(\th)$ to be of the form
\be
\lb(\th) = i\,\omega(\th)\omega(\th + i\hbar)\;,
\label{lam2}
\ee
where $\omega(\th)$ satisfies $\omega(\th)^* = \omega(\th^* + i
\hbar)$.  The overall $i$ in (\ref{lam2}) takes care of the $\eps =-1$
hermiticity property.  Comparing with (\ref{lam1}) one sees that
(\ref{lam2}) amounts to fixing the normalization of the $T^{\pm}$
(i.e.~$\kappa^{\pm}(\th) \equiv 1$) while still allowing for
automorphisms (\ref{*resc1}) with a nontrivial $\omega(\th)$. The
restriction to $\kappa^{\pm}(\th) =1$ is not indispensible; however
since their $\hbar \ra 0$ limits can only be $\pm 1$ anyhow, not much
generality is lost. Conversely any prescribed $\lb(\th)$ enjoying
suitable analyticity and fall-off properties can be written in the
factorized form (\ref{lam2}). With $\lb(\th)$ of the form (\ref{lam2})
we then take
\be 
c^{(\n)}(\th) = c^{(\n_0)}\,\omega(\th_\n)\ldots \omega(\th_{\n_0 +1})
\left( \frac{-1}{\psi_0}\right)^{(\n-\n_0)/2}\;,
\quad \n \geq \n_0\;.
\label{sol4} 
\ee
Finally we enter with the ansatz (\ref{ansatz}) into the functional
equations  (I) and (II). The equations (I) translate into
\medskip
\begin{subeqnarray} 
&& {\overline \cT}(\th_0|\th)_A^B \,\fbar_B(\th) ~=~ 
{\overline \tau}(\th_0|\th) \fbar_A(\th)\;,
\\
&& \fbar_A(\th) = {\overline R}_{a_{k+1} a_k}^{\;c\;\;\;\; d}(\th_{k+1,k}) 
\fbar_{a_\n \ldots d c \ldots a_1}(s_k\th)\;,
\label{Ibar}
\end{subeqnarray}
where 
\ba
&&{\overline R}(\th)_{ab}^{cd} := 
\frac{-1}{\th+ i\hbar}[\th \,\delta_a^c \delta_b^d - i\hbar \delta_a^d
\delta_b^c]\;. 
\nonum
&& {\overline \cT}(\th_0|\th) := \prod_{j=1}^\n
(\th_{j0} - 2 i\hbar)\, r(i\hbar - \th_{j0})\;\cT(\th_0|\th)\;. 
\ea
The redefinition of $\cT$ renders the components of
$\overline{\cT}(\th_0|\th)$ polynomial in the
$\th_{j0}\!-\!i\hbar$. The associated eigenvalues
$\overline{\tau}(\th_0|\th)$ turn out to be rational functions of
$\th_j,\,\th_0$ and the Bethe roots, cf.~(\ref{ev}).  The solutions
$\fbar$ in $\n$ and $\n\!-\!2$ variables are now linked by
\begin{subeqnarray} \jot5mm
\fbar_A(\th)\bigg|_{\th_{k+1} = \th_k + i\hbar} &\, =\, &
\overline{\tau}(\th_k - i\hbar|p_k\th)\!
\prod_{j \neq k+1,k} \!(\th_{jk} +i\hbar)\;C_{a_{k+1} a_k}\fbar_{p_k
A}(p_k \th)\;, 
\\
C^{a_{k+1}a_k}
\fbar_A(\th)\bigg|_{\th_{k+1} = \th_k - i\hbar} &\, =\, &
- 2 {\overline \tau}(\th_k - 2 i\hbar|p_k\th)\!
\prod_{j \neq k+1,k} \!(\th_{jk} + 2i\hbar)\;\fbar_{p_kA}(p_k \th)\;,
\\
\overline{\tau}(\th_0|\th)\bigg|_{\th_{k+1} = \th_k + i\hbar} & = &
\th_{k0}(\th_{k0} - 2i\hbar) \,\overline{\tau}(\th_0|p_k\th)\;,
\nonumber\\
\overline{\tau}(\th_0|\th)\bigg|_{\th_{k+1} = \th_k - i\hbar} & = &
(\th_{k0}-i\hbar)(\th_{k0} - 3i\hbar) \,\overline{\tau}(\th_0|p_k\th)\;,
\label{IIbar}
\end{subeqnarray}
where again (\ref{IIbar}c) applies only when the right hand sides of 
(\ref{IIbar}a,b) are nonvanishing. Clearly any one of the eqs
(\ref{IIbar}c)  implies the other, we noted them both for symmetry. 
The system of eqs (\ref{Ibar}), (\ref{IIbar}) for $\fbar$ is now
`quasirational' in the sense that all ingredients are rational
functions in $\th_j,\th_0$ and the Bethe roots, while the Bethe roots
themselves are algebraic functions of the $\th_j$.  

The proper hermiticity requirements are
\ba
[\fbar_A(\th)]^* &=& \fbar_{A^T}({\th^*}^T + i\hbar)\;,\nonum
[\fbar_{e;\sigma_n\ldots \sigma_1}(\th)]^* &=& 
\fbar_{e;\sigma_1\ldots \sigma_n}({\th^*}^T + i\hbar)\;.
\ea 
The first one is just the general hermiticity requirement expressed in
terms of $\fbar$. The second one is the transcription into the
``charged basis'' introduced in appendix B1. In brief one can switch
to a basis in $V^{\otimes \n}$ on which the matrices $\Gamma$ and
$\Lambda$ in (\ref{a2}), (\ref{norm8}) are diagonalized. The
components of $f^{(\n)}$ in the new basis are denoted by
$f_{e;\sigma_\n \ldots \sigma_1}(\th)$, where $\sigma_j \in \{\pm 1
\}$ and $e = \sigma_\n + \ldots + \sigma_1$.  Under a $U(1) \simeq
SO(2)$ rotation (\ref{norm8}) they transform as $f_e(\th) \ra
e^{ie\phi} f_e(\th)$, i.e.~with charge $e$. It is easy to see that the
functional equations (I), (II), or (\ref{Ibar}), (\ref{IIbar}), then
split up into decoupled sectors of fixed charge $e \in \{\n,\n-2,
\ldots,-\n\}$, making the charged basis particularly useful for their
analysis.  In group theoretical terms the basis transformation is
related to the isomorphism $SL(2,\R) \ra SU(1,1)$.

Let us now address the problem of solving the functional equations
(\ref{Ibar}), (\ref{IIbar}), in the charged basis. As indicated
eq.~(\ref{Ibar}a) can be solved by means of the Bethe ansatz; some
details are provided in appendix B. In upshot an eigenvector
$w_{e;\sigma_\n\ldots \sigma_1}(\th)$ is constructed by introducing a
set of auxiliary variables $t_1,\ldots,t_{\Lambda}$, $\Lambda = (\n
-e)/2$, such that a candidate eigenvector $w_{e;\sigma_\n\ldots
\sigma_1}(t|\th) = w_{e;\sigma_\n\ldots
\sigma_1}(t_1,\ldots,t_{\Lambda}|\th_\n,\ldots,\th_1)$ turns into a
proper eigenvector, provided the parameters are turned into
judiciously chosen functions of the $\th_j$'s, i.e.
\be
w_{e;\sigma_\n\ldots \sigma_1}(\th) = 
w_{e;\sigma_\n\ldots \sigma_1}(t(\th)|\th)\;,
\sspace \mbox{ $t(\th)$: solution of BAE (\ref{BAE})}\;.
\label{solw}
\ee   
Here BAE are the Bethe ansatz equations whose solutions are
(complicated) algebraic functions $t_{\alpha}(\th),\;\alpha =
1,\ldots,\Lambda$, completely symmetric in the $\th_j$'s. The
candidate eigenvector can be chosen to be polynomial in the $\th_j$
and the auxiliary parameters $t_{\alpha}$.  A Bethe eigenvector
$w_e(\th)$ solving (\ref{Ibar}a) will not necessarily obey
(\ref{Ibar}b). By (\ref{solsym3}) we know however that
\be
\fbar_{e;\sigma_\n\ldots \sigma_1}(\th) = 
\phi_e(\th)\,\prod_{k >l}\frac{i}{\th_{kl} + i \hbar} 
\,w_{e;\sigma_\n\ldots \sigma_1}(\th)\;,
\label{solsym}
\ee
solves both (\ref{Ibar}a) and (\ref{Ibar}b). Here $\phi_e(\th)$ is a 
completely symmetric function in $\th_j$ not constrained
by (\ref{Ibar}a,b) in any way.

In general a solution of ({\ref{Ibar}a,b) will not satisfy
(\ref{IIbar}).  The generic expression (\ref{solsym}) however contains
two pieces of information left unspecified so far. First the choice of
a specific Bethe eigenvector and second the choice of a specific
symmetric function $\phi_e(\th)$. The obvious way to proceed is to try
adjusting these two ingredients such that also the recursive equations
(\ref{IIbar}) are satisfied. As explained in appendix B the choice of
the proper Bethe eigenvector amounts to a choice of the proper Bethe
root. For the charge $e$ sector one typically expects ${\n \choose
\Lambda}$ independent eigenvectors with distinct eigenvalues (though
there may be some degeneracies). We expect that typically one and only
one of them in addition satisfies the recursive equations
(\ref{IIbar}a,b) in which case the associated eigenvalue satisfies
(\ref{IIbar}c).  Moreover the sequential eigenvector and eigenvalue
can already be identified on the level of the Bethe roots. We call a
solution of the BAE (\ref{BAE}) in $\n$ variables $\th_j$ a
``sequential'' $\Lambda$-tuple of Bethe roots, if it is real for real
$\th_j$ and satisfies
\ba
t_{\alpha}(\th)\Big|_{\th_{k+1}= \th_k + i\hbar} &=& 
t_{\alpha}(p_k\th)\;,
\sspace \alpha =1,\ldots,\Lambda -1\;,\nonum
t_{\Lambda}(\th)\Big|_{\th_{k+1}=\th_k \pm i\hbar} &=& 
\th_k \pm \frac{i\hbar}{2}\;,
\sspace k =1,\ldots, \n\!-\!1\,,
\label{goodroot}
\ea
where on the right hand side of the first equation a sequential
$(\Lambda\!-\!1)$-tuple in $\n\!-\!2$ variables occurs. Observe that
the combination $e = \n\!-\!2\Lambda$ is invariant under $\n \mapsto
\n\!-\!2,\;\Lambda \mapsto \Lambda\!-\!1$; as indicated it can be
identified with the conserved $U(1) \simeq SO(2)$ charge carried by
$f^{(\n)}_e$.  In other words the ``sequential'' solutions to the BAE
can naturally be arranged into sequences of fixed charge $e$ where
consecutive members are linked by (\ref{goodroot}).  In appendix B we
show that such sequential Bethe roots exist and that the eigenvalues
$\tau(\th_0|\th)$ associated with a sequence satisfy the recursive
equations (IIc). Upon restriction to $\th_{k+1} = \th_k \pm i\hbar$
the corresponding eigenvectors should then always become proportional
to the right hand side of (\ref{IIbar}a,b).

Having fixed the Bethe eigenvector in (\ref{solsym}) to be one
associated with a sequential Bethe root, the only freedom left resides
in the symmetric function $\phi_e(\th)$. The aim now is to adjust it
such that eqs (\ref{IIbar}a,b) hold identically. Since in the $\n \ra
\n\!-\!2$ recursion step the Bethe roots $t_{\alpha}, \alpha <
\Lambda$, enter via $\tau(\th_0|p_k\th)$, it is natural to require
that $\phi_e(\th)$ is a rational function in $\th_\n,\ldots,\th_1$ and
$t_1, \ldots ,t_{\Lambda -1}$, invariant under shifts $\th_j \ra \th_j
+ {\rm const}$.  Schematically,
\be
\phi_e(\th) = \phi_e(\th^*)^*\;:\qquad
\begin{array}{l} \mbox{symmetric, shift invariant,} \\
\mbox{rational in $\th_\n,\ldots,\th_1$ and $t_1,\ldots,t_{\Lambda -1}$.}
\end{array}
\label{phiprop}
\ee
We have verified these features for $\n\le4$ by explicit computation and
are confident that they are generic.

In summary we arrive at the following solution procedure for the
recursive functional equations (I), (II) or their `quasirational' form
(\ref{Ibar}), (\ref{IIbar}):
\begin{itemize}
\item[(a)] For given $\n$ and given SO(2) charge $e =
\n\!-\!2,\n\!-\!4,\ldots, -\n\!+\!2$ compute the trial eigenvector
$w_e(t|\th)$ via (\ref{evec}). It contains $\Lambda =
\frac{1}{2}(\n-e)$ Bethe parameters and has ${ \n \choose \Lambda}$
independent components.
\item[(b)] Verify that for any Bethe root satisfying (\ref{goodroot})
the restrictions of $\fbar_e(\th)$ in (\ref{solsym}) to $\th_{k+1} =
\th_k \pm i\hbar$ are proportional to the right hand sides of
(\ref{IIbar}a,b). Multiply by a scalar function $\phi_e(\th)$ of the
type (\ref{phiprop}) such that (\ref{IIbar}a,b) holds.
\item[(c)] When feasible compute the sequential $\Lambda$-tuple of
Bethe roots explicitly. Repeat the procedure for $\n \mapsto
\n\!+\!2$.
\end{itemize}
The explicit form of the sequential Bethe roots will usually be
available only for small $\n$; its existence is ensured by the results
in appendix B.  Together (a) -- (c) produces a solution to the
`quasirational' system (\ref{Ibar}), (\ref{IIbar}). Inserting into
(\ref{ansatz}) then yields a solution of the original system of
functional equations (I), (II).  For clarity's sake let us emphasize
that we have not proven step (b) to work always. However we verified
it on sufficiently non-trivial examples to conjecture with some
confidence that it does. A plausibility argument of course stems from
the very algebraic construction used to derive (I), (II). A
T-invariant functional is an ``$\n$-independent'' object, once it
exists at all it will automatically produce a non-terminating sequence
of functions $f^{(\n)}$ solving (I), (II). The case $e = \pm \n$ is
excluded in (a) because those solutions of (\ref{Ibar}) necessarily
vanish upon restriction to $\th_{k+1} = \th_k \pm i\hbar$. However
they will naturally serve as a starting member $f^{(\n_0)}$ to a
sequence. The above procedure then yields semi-infinite sequences
$(f^{(\n)})_{\n \geq |e|}$, which we expect to be basically uniquely
determined by their starting member $f^{(|e|)}$. To illustrate the
scheme, we have collected the first few members of the charge $e =0,
\pm 1, \pm 2$ sequences in appendix C.


\newsubsection{Semi-classical limit}

Let us now consider the classical limit of these solutions. 
Recalling from (\ref{ansatz}), (\ref{solsym}) 
\be
f_{e;\sigma_\n\ldots \sigma_1}(\th) = c^{(\n)}(\th)\, \phi_e(\th)\, 
w_{e;\sigma_\n\ldots \sigma_1}(\th)
\;\prod_{k > l} \frac{i \psi(\th_{kl})}{\th_{kl}^2 + \hbar^2}\,,
\label{fsol}
\ee
this amounts to studying the $\hbar \ra 0$ limit of the various 
ingredients. For the transcendental function $\psi(\th)$ in (\ref{psi}) 
one simply has (cf.~Fig.~1) 
\be
\psi(\th) = i + O(\hbar)\,.
\label{psicl}
\ee
For the scalar prefactor $\phi_e(\th)$ the explicit results of
appendix C suggest that  
\be
\phi_e(\th) = \phi_e^{\rm cl}(\th) + O(\hbar)\,,
\label{phicl}
\ee
where $\phi_e^{\rm cl}(\th)$ is a ratio of homogeneous polynomials in 
$\th_\n,\ldots,\th_1$. However provided $\phi_e(\th)$ depends on the 
Bethe roots (which will generically be the case) the corresponding 
factor will no longer be symmetric in $\th_\n,\ldots,\th_1$. This feature
is related to the curious behavior of the Bethe roots in the $\hbar \ra 0$ 
limit discussed in appendix B4. Recall for real
$\th_\n,\ldots,\th_1$ the sequential Bethe roots are real-valued and
completely symmetric in all variables. For $\hbar \ra 0$ however one has
\be
t_{\alpha}(\th) = \th_{j(\alpha)} + O(\hbar^2)\;,\quad \alpha = 
1,\ldots,\Lambda\,,
\label{Brootcl}
\ee
where the index $j(\alpha) \in \{1,\ldots,n\}$ of the preferred variable
$\th_{j(\alpha)}$ is determined by the choice of branch in the Bethe root 
and the relative size of the $\th$ variables. Further $j(\alpha) \neq
j(\beta)$ for  $\alpha \neq \beta$. The Bethe vector (\ref{evec})
contains an  explicit power of $\hbar^{\Lambda} = \hbar^{(\n-e)/2}$ 
(since the matrix operator $B(t|\th)$ is $O(\hbar)$). Taking out this 
explicit power we define 
\be
w^{\rm cl}_{e;\sigma_\n\ldots \sigma_1}(\th) = \lim_{\hbar \ra 0}
\hbar^{-\Lambda}\,w_{e;\sigma_\n\ldots \sigma_1}(\th)\,.
\label{w1cl}
\ee
As shown in appendix B4, $w^{\rm cl}_{e;\sigma_\n\ldots
\sigma_1}(\th)$ has only one non-vanishing component
\be
w^{\rm cl}_{e;\sigma_\n\ldots \sigma_1}(\th) = 0 \quad
\mbox{unless} \;\;\; (\sigma_\n,\ldots,\sigma_1) = 
(\eps_\n,\ldots,\eps_1)\,,
\label{w2cl}
\ee
where $(\eps_\n,\ldots,\eps_1)\in \{\pm \}^\n$ is a particular sign
configuration of charge $e$. The sign pattern
$(\eps_\n,\ldots,\eps_1)$ of the non-vanishing component is determined
by the asymptotics (\ref{Brootcl}) of the Bethe roots
$t_1,\ldots,t_{\Lambda}$ as follows: Let $I_{\Lambda} =
\{j(1),\ldots,j(\Lambda)\} \subset \{1,\ldots,\n\}$ be the subset of
indices appearing on the right hand side of (\ref{Brootcl}).  Then
\be
\eps_j = \left\{ \begin{array}{lll} 
\phantom{-}1 & \mbox{if} & j \notin I_{\Lambda}\\
-1 & \mbox{if} & j \in I_{\Lambda}\,.
\end{array}\right.
\label{w3cl}
\ee
Viewed as a scalar function of the $\th$'s the component
$w_{e;\eps_\n,\ldots,\eps_1}(\th)$ is a homogeneous polynomial of degree
$(\n\!-\!1)\Lambda$, however again no longer a symmetric one.    

Combining (\ref{fsol}) -- (\ref{w3cl}) we conclude that for $\n >
\n_0$ the leading term $f^{\rm cl}_{e;\sigma_\n\ldots \sigma_1}(\th)$ in
the $\hbar$-expansion of $f_{e;\sigma_\n\ldots \sigma_1}(\th)$ is
given by
\ba
f^{\rm cl}_{e;\sigma_\n\ldots \sigma_1}(\th) \is
d^{(\n)}_e(\th) \,\phi_e^{\rm cl}(\th)\,w^{\rm cl}_{e;\sigma_\n\ldots
\sigma_1}(\th)\; 
\prod_{k >l} \frac{1}{\th_{kl}^2}\;, \nonum
\mbox{where} \quad
f_{e;\sigma_\n\ldots \sigma_1}(\th) \is \hbar^{\Lambda} \,
f^{\rm cl}_{e;\sigma_\n\ldots \sigma_1}(\th) + O(\hbar^{\Lambda+1})\;, 
\label{fsolcl}
\ea 
where $\Lambda = (N-e)/2$ as before and 
\be
d^{(\n)}_e(\th) =  (-)^{\n(\n-1)/2} c^{(\n)}(\th)^{\rm cl}\,. 
\label{dconst}
\ee
Here we assumed that $\omega(\th)$ in (\ref{lam2}) has a regular
$\hbar \ra 0$ limit $\omega^{\rm cl}(\th)$ in terms of which the limit
$c^{(\n)}(\th)^{\rm cl} = \lim_{\hbar \ra 0} c^{(\n)}(\th)$ is
defined.  As remarked before a fixed relative size of the variables
$\th_\n,\ldots,\th_1$ is implicit in (\ref{fsolcl}), say $\th_\n >
\ldots>\th_1$. The results for other orderings then are compatible
with the classical limit of the exchange relations in (I), i.e.
\be
f^{\rm cl}_{e;\sigma_\n\ldots \sigma_1}(\th_\n,\ldots,\th_1) = 
f^{\rm cl}_{e;\sigma_\n \ldots \sigma_k \sigma_{k+1}\ldots \sigma_1}
(\th_\n,\ldots,\th_k,\th_{k+1},\ldots,\th_1)\,.
\label{Icl}
\ee
In particular $f^{\rm cl}_{e;\sigma_{\n}\ldots \sigma_1}(\th)$ is
separately symmetric in all $\th_j$ with $\sigma_j =1$ and all $\th_j$
with $\sigma_j =-1$. As described before they are also eigenvectors of
the semi-classical transfer matrix (\ref{semiT}) with eigenvalues
(\ref{semit}).

It is then natural to ask whether the $f^{\rm cl}$ also obey a
classical counterpart of the recursive relations (II). This is not
automatic since one cannot a-priori expect the operation of taking the
residue at $\th_{k+1} =\th_k \pm i\hbar$ to commute with the limit
$\hbar \ra 0$. Examples where they don't commute are the function
$\psi(\th_{k+1,k})$, the eigenvalues of the transfer matrix, or the
Bethe roots. Nevertheless for the final solutions it turns out that
both operations do commute, -- up to a numerical factor $Z$:
\be
\lim_{\hbar \ra 0} \left[\hbar^{-\Lambda} 
{\rm Res}_{\,\th_{k+1} = \th_k +i\hbar} 
f_{e;\sigma_{\n}\ldots \sigma_1}(\th)\right] =
Z\,{\rm Res}_{\,\th_{k+1} = \th_k} 
f^{\rm cl}_{e;\sigma_{\n}\ldots \sigma_1}(\th)\;,
\label{clres1}
\ee
and similarly for the appropriate residue at $\th_{k+1} = \th_k -
i\hbar$.  The derivation of (\ref{clres1}) is deferred to Appendix B.
In principle the proportionality constant could depend on the 
solution considered. The explicit checks of (\ref{clres1}) on the 
$\n \leq 4$ solutions however suggest that it is universal and
given by: $Z = (-)^{\n}\psi_0/2$, $\psi_0 \approx 1.54678$. 
Using (\ref{clres1}) in (IIa,b) one finds that both reduce to a single
recurrent relation for $f^{\rm cl}$
\ba
Z\,{\rm Res}_{\,\th_{k+1} = \th_k}\,
f^{\rm cl}_{e;\eps_\n\ldots\eps_1}(\th)  &=& \,\lb^{\rm cl}(\th_k)\,
C_{\eps_{k+1}\eps_k} f^{\rm cl}_{e;\eps_\n\ldots\eps_{k+2}\eps_{k-1}
\ldots \eps_1}
(p_k \th)\bigg
( \!- \!\!\!\!\sum_{j\not=k+1,k}\frac{\eps_j}{\th_{kj}}\bigg)\,.
\label{clres2}
\ea
The last factor on the right hand side equals the leading term,
restricted to $\th_{k+1} =\th_k$ and $\eps_{k+1} +\eps_k =0$, in the
$\hbar$ expansion (\ref{semit}) of the transfer matrix
eigenvalues. This is the consistency condition on (\ref{clres2})
analogous to (IIc). ((IIc) itself cannot naively be expanded in powers
of $\hbar$, due to the non-commutativity of the two operations
mentioned earlier.)

In summary, the solutions of the functional equations (I), (II) admit
a consistent semi-classical expansion. The leading term (\ref{fsolcl})
of this expansion has, for a given ordering of
$\th_{\n},\ldots,\th_1$, only one non-vanishing component; different
orderings being related by (\ref{Icl}). Further these leading terms
are themselves linked by the recurrence relation (\ref{clres2}).  Of
course it is tempting to ask whether the leading terms have a direct
interpretation in the classical theory. This is beyond the scope of
the present paper; however the vertex operator formalism of
\cite{BerJul99,BerReg99} should be the appropriate framework to
address the issue.

\newsection{Conclusions}

Motivated by the fact that none of the conventional field theoretical
techniques presents itself to develop a quantum theory for the Ernst
system, we proposed to bootstrap the quantum theory from
structures linked to its classical integrability. Starting from a few
reasonable assumptions on the nature of the quantum counterparts of
these integrable structures, a very rigid computational framework
emerged. The eventual outcome are meromorphic functions
$f^{(\n)}(\th)$, conjectured to describe exact matrix elements in the
quantum theory, without the need for any renormalization.

On a technical level, our main result is the system of functional
equations (I), (II) for the functions $f^{(\n)}(\th)$. One of the
equations is a standard eigenvalue problem for the transfer matrix,
which is why the functions $f^{(\n)}(\th)$ can be viewed as Bethe
eigenvectors, however very special ones. The special feature is that
they are members of semi-infinite sequences in a similar way the form
factors of an integrable QFT are; consecutive members being linked by
recursive relations. The solution procedure of the functional
equations for ordinary form factors \cite{Smir92} ($\beta = 2\pi$) and
for replica deformed ones \cite{Nied98} ($\beta$ generic) can be
viewed as selecting those special solutions of the deformed KZE
enjoying such extra recursive relations.  Employing the integral
representation for the latter \cite{FreRes92,TarVar96,TarVar97}, the
integrands can be seen to obey recursive relations similar to our
$\beta =0$ ones \cite{Pill99}. This might indicate that the $\beta =0$
system of functional equations found here can serve as a master system,
from whose solutions the solutions of the $\beta \neq 0$ systems can
be obtained by a universal integral transformation.

Physically, the most interesting finding is the ``spontaneous''
breakdown of the $SL(2,\R)$ symmetry that is a remnant of the 4D
diffeomorphism invariance in the `compactified' dimensions. The matrix
elements $f^{(\n)}(\th)$ and hence the state space generated by the
physical operators supposed to underly them are covariant only under
the $SO(2)$ subgroup, despite the fact that the algebraic framework is
fully $SL(2,\R)$ covariant.  Clearly in a next step one should try to
gain a better understanding of the field theoretical meaning of the
matrix elements $f^{(\n)}(\th)$ and how physically interesting
quantities can be computed in terms of them.  To this end it would be
important to develop, at least to some extent, a more conventional
field theoretical formulation.  A combination of perturbative and
semi-classical techniques should be adequate for this purpose.

Non-perturbatively one might aim at developing a dynamical
triangulations approach, using the exact results proposed here, or
quantities computed thereof, as a guideline. In particular, it would
be important to understand the statistical mechanics origin of the
dynamical breaking of the $SL(2,\R)$ symmetry, e.g.~as a tug of war
between energy and entropy as in conventional spin models. The
mechanism may well contain clues for the breaking of diffeomorphism
invariance beyond the symmetry reduced theory.

\vspace{0.9em}
{\bf Acknowledgements:}
We wish to thank H.~DeVega and D.~Korotkin for valuable
discussions.  The work of M.~N. was supported by NSF grant
97-22097; that of H.~S. by EU contract ERBFMRX-CT96-0012.
\newpage


\setcounter{section}{0}

\newappendix{Action principles for coset sigma models}

As outlined in the introduction the 2D matter sector of the Ernst
system is a nonlinear sigma-model whose target space is a (noncompact)
homogeneous space. Such sigma-models are known as coset sigma models,
they are classically integrable and several action principles and
parameterizations turn out to be useful. Although in the bulk of the
paper no direct use of these actions is made, both the
parameterizations employed and the origin of the (gauge) symmetries is
best understood in the terms of these action principles.  For
simplicity we omit the coupling to 2D gravity here, each of the coset
sigma-models can be coupled to gravity as in (\ref{i1}).

Coset sigma models describe the dynamics of generalized harmonic maps
from a 2-dim. spacetime $\Sigma$ to a homogeneous space of the form
$G/H$, where $G$ is a (simple) matrix Lie group and $H$ a maximal
subgroup of $G$.  For the purposes of this appendix, we take $\Sigma$
to be 2-dim. Minkowski space with signature $(+,-)$. There are two
useful action principles for these coset sigma models. A gauge
theoretical one
\be
S[\cV, Q] = \frac{1}{2} \int d^2 x\, {\rm Tr}[D_{\mu} \cV
\,\cV\inv \,D^{\mu} \cV\cV\inv]\;,
\label{act1}
\ee
where $\cV$ is a group-valued field transforming as $\cV \ra \cV h$
under an $H$-valued gauge transformation, and $Q_{\mu}$ is the
associated connection ensuring that $D_{\mu} \cV = \partial_{\mu} \cV
- \cV Q_{\mu}$ transforms covariantly. Clearly the gauge symmetry
removes $\dim H$ degrees of freedom leaving $\dim G/H$ physical ones.
Alternatively, one can choose a non-redundant parameterization of the
coset space by matrices $M \in G$ obeying a suitable quadratic
constraint $M \tau_0(M) = \pm\1$, where $\tau_0$ is an involutive
outer automorphism of $G$.  The subgroup $H$ can then be characterized
as being fixed by a related involutive automorphism $\tau$ of $G$,
given by $\tau(g) = g_0\inv \tau_0(g) g_0$, for some fixed $g_0 \in G$
which likewise satisfies $g_0 \tau_0(g_0) = \pm\1$. Explicitly, the
matrices $M$ can be constructed as $M = \cV g_0\inv \tau_0(\cV^{-1})$;
they are gauge invariant and parameterize the coset space as $\cV$
runs through $G$. Using $\partial_{\mu} M M\inv = 2 D_{\mu} \cV
\cV\inv$, the action (\ref{act1}) becomes
\be
S[M] = \frac{1}{8}\int d^2 x \,{\rm Tr}[ \partial_{\mu}M M\inv 
\,\partial^{\mu} M M\inv]\;,\sspace M \tau_0(M) = \pm\1\;.
\label{act2}
\ee

Specifically we are interested in the case $G/H = SL(2,\R)/SO(2)
\simeq SU(1,1)/U(1)$. The theory with a compact target space
$SU(2)/U(1)$ can be seen to have two useful reformulations. One as a
$U(1)$ gauge theory on the projective space $\P \C^2 =: \C\!\P_1$,
known as the $\C\!\P_1$ model. The second one parameterizes the coset
matrices $M$ in (\ref{act2}) by real 3-dim.~unit vectors and yields
the $S_2$ Heisenberg spin-model. (The latter is also known as the O(3)
nonlinear sigma-model, though $O(3)/O(2)$ would be the proper coset
notation.) The aim in the following is to derive similar
reformulations for the noncompact $SU(1,1)/U(1)$ model.

We begin with the counterpart of the $\C \P_1$ formulation and 
choose an explicit parameterization of the $SU(1,1)$ matrix fields.
Since $\cV \in SU(1,1)$ iff $g \sigma^3 g^{\dagger} = \sigma^3$
one has  
\ba
\cV \is \left( \begin{array}{cc} z_1^* & z_2 \\
z_2^* & z_1 \end{array} \right) \;,\quad |z_1|^2 - |z_2|^2 =1\;,
\nonum
h \is \left( \begin{array}{cc} e^{-i \gamma}& 0 \\
0  & e^{i\gamma} \end{array} \right) \;,\quad \gamma \in \R\;.
\label{act3}
\ea
Here and below $\sigma^j, j=1,2,3$, are the Pauli matrices.  The
automorphism $\tau$ in the case at hand is given by $\tau(g) =
\sigma^3 g \sigma^3$, $g \in SU(1,1)$, i.e. $g_0 = -i \sigma^3$ and
$\tau_0 = id$ in the general framework outlined before.  On the
component fields $\tau$ acts as $\tau(z_1) = z_1, \tau(z_2) = -z_2$.
The condition $h \in H \simeq U(1)$ iff $\tau(h) = h$ then yields the
above parameterization of $h$. The gauge transformations $\cV \ra \cV
h$ simply amount to $z_1 \ra z_1 e^{i\gamma}$, $z_2 \ra z_2
e^{i\gamma}$ and $A_{\mu} \ra A_{\mu} + \partial_{\mu}\gamma$, where
$Q_{\mu} = - i A_{\mu} \sigma^3$. Inserting this parameterization of
$\cV$ and $Q_{\mu}$ into the action (\ref{act1}) one finds
\ba
&& S[z,A] =  \int d^2 x 
(\partial^{\mu} z^{\dagger} + i A^{\mu} z^{\dagger})\cdot
(\partial_{\mu} z - i A_{\mu} z)\;,
\sspace z^{\dagger} \cdot z =1\;. 
\nonum
&& \mbox{where} \;\;\;
z^{\dagger} = (z_1^*,z_2^*)\;,\quad z= {z_1 \choose z_2}\;,
\quad w^{\dagger} \cdot z = w_1^* z_1 - w_2^* z_2\;. 
\label{act4}
\ea
This is a $U(1)$ gauge theory on the `Lorentzian' projective space 
$\P \C^{1,1}$, and is a non-compact analogue of the $\C\!\P_1$ model.
The $SU(1,1)/U(1)$ coset theory can therefore be viewed as a matrix 
version of (\ref{act4}).

The second reformulation of the $SU(1,1)/U(1)$ theory starts from the
coset action (\ref{act2}) and yields a hyperbolic counterpart of the
$S_2$ spin-model.  In upshot it simply amounts to parameterizing the
matrices $M$ by elements of the 2-dim.~hyperboloid $H_2$, replacing
the sphere $S_2$ in the compact case. Following the general
construction we first compute the gauge invariant matrix
\be
M = \cV i \sigma^3 \cV^{-1} = \left(
\begin{array}{cc} i(|z_1|^2 + |z_2|^2) & -2 i z_2 z_1^*\\
2 i z_2^* z_1 & -i(|z_1|^2 + |z_2|^2)  \end{array}
\right)\;,\sspace
M^2 =- \1\;.
\label{act5}
\ee    
Elements $Y$ of the Lie algebra $su(1,1)$ are characterized by 
$-\sigma^3 Y = Y^{\dagger} \sigma^3$. A convenient basis is 
$\tau^0 = i \sigma^3, \tau^1 = \sigma^1, \tau^2 = \sigma^2$. 
For any real triplet $n = (n^0,n^1,n^2)$ the matrix 
$\sum_j n^j \tau^j$ defines an $U(1,1)$ matrix which squares
to the multiple $- (n^0)^2+ (n^1)^2 + (n^2)^2$ of the unit matrix.
The relations 
\be
M = \sum_j n^j \tau^j\quad \Longleftrightarrow \quad 
i n^j = z^{\dagger} \cdot \tau^j z\,,\;\;j=0,1,2\,,
\label{act6}
\ee
(where the `$\;\cdot\;$' is that of (\ref{act4})) then provide an 
isomorphism 
\ba
&& SU(1,1)/U(1) \rra H_2\;,\nonum
&& H_2 = \{ n \in \R^{1,2} \,|\, n \cdot n = 
(n^0)^2 - (n^1)^2 - (n^2)^2 = 1\;,\;\; n^0 > 0 \} \;,
\ea 
where $R^{1,2}$ is the ambient Lorentzian vector space of 
signature $(+,-,-)$. Substituting (\ref{act6}) into the coset action 
(\ref{act2}) one obtains
\be
S[n] = -\frac{1}{4}\int d^2 x \,\partial_{\mu} n \cdot \partial^{\mu} n\;,
\quad n \cdot n =1\;,
\label{act7}
\ee 
where the bilinear form `$\;\cdot \;$' is that of $R^{1,2}$.
An alternative way of arriving at (\ref{act7}) would have been to vary
(\ref{act4}) with respect to $A_{\mu}$ and substitute the (algebraic)
equation of motion back into the action. We preferred the above route in 
order to have the link (\ref{act6}) to the coset matrices $M$ at 
our disposal. The action (\ref{act7}) is also convenient to 
verify that the energy density of the system is positive definite
\be
{\cal H} = -\frac{1}{2}[\partial_0 n \cdot \partial_0 n + 
\partial_1 n \cdot \partial_1 n] > 0\;.
\label{act9}
\ee
To see (\ref{act9}) one differentiates the constraint and uses
Schwarz inequality  
\be
(\partial_{\mu} n^0)^2 = \frac{1}{(n^0)^2}
\left(\sum_{j=1}^2 n^j \partial_{\mu} n^j \right) <
(\partial_{\mu}n^1)^2 + (\partial_{\mu}n^2)^2 \;,\quad \mu =0,1\,.
\label{act10}
\ee 
It is essential here that the hyperboloid is timelike,
for a spacelike hyperboloid the energy density would be indefinite. 

In the above we started with the complex $SU(1,1)/U(1)$ coset space 
because it is the description most convenient in the bulk of the paper.
The dimensional reduction of the 4D Einstein-Hilbert action however
initially yields a $SL(2,\R)/SO(2)$ matter sigma-model \cite{Gero71}. 
The relation is given by the isomorphism
\ba
{\rm Ad} \Upsilon\; :\; SL(2,\R) & \rra & SU(1,1)\;,\sspace\; 
\Upsilon = \frac{1}{\sqrt{2}}
{ \,1 \, -i \choose 1\;\;\phantom{-} i}\;,\nonum  
\left( \begin{array}{cc} a & b \\ c & d \end{array} \right) & \rra & 
\left( \begin{array}{cc} z_1^* & z_2 \\ z_2^* & z_1 \end{array} \right)\;,
\quad
\begin{array}{c} z_1 = \frac{1}{2}[(a+d) - i (b-c)]\\[1mm] 
z_2 = \frac{1}{2}[(a-d) - i (b+c)]\end{array}\;. 
\label{act11}
\ea
For the $H_2$ variables one gets 
\be
n^0 = \frac{1}{2}(a^2 + b^2 + c^2 + d^2) \;,\quad
n^1 = -(ac + bd) \;,\quad
n^2 = \frac{1}{2}(a^2 + b^2 - c^2 - d^2) \;.
\label{act12}
\ee 
The automorphism $\tau$ for $SL(2,\R)$ is inner and is given by $\tau(g) = (g^T)\inv =
(i\sigma^2)^{-1} g (i\sigma^2)$, i.e. by $\tau(a) = d\,,\tau(b) = - c$ on the above 
components. It evidently selects the proper $SO(2)$ subgroup which
is mapped onto the diagonal $SU(1,1)$ matrices in (\ref{act3}). On the 
gauge invariant $H_2$ variables $\tau$ (in either the $SL(2,\R)$ or the 
$SU(1,1)$ version) acts as $\tau(n^0) = n^0,\,\tau(n^1) = - n^1,\,
\tau(n^2) = - n^2$, which maps $H_2$ onto itself and has only the trivial
fixed point $n =(1,0,0)$. In the coset space $SL(2,\R)/SO(2)$ one can pick
representatives given by upper triangular $SL(2,\R)$ matrices. A 
conventional parameterization is 
\be
\cV = \left( \begin{array}{cc} \Delta^{1/2} & B \Delta^{-1/2} \\ 
0 & \Delta^{-1/2} \end{array} \right) \;,\quad \Delta^{1/2} > 0\,,\;\;
B \in \R\,.
\label{act13}
\ee
The combination $\cE = \Delta + i B$ then is the ``Ernst potential''
used in the general relativity literature. In this parameterization
the `hyperbolic spins' read
\be
n^0 = \frac{\Delta^2 + B^2 +1}{2\Delta}\;,\quad
n^1 = -\frac{B}{\Delta}\;,\quad 
n^2 = \frac{\Delta^2 + B^2 -1}{2\Delta}\;. 
\label{act14}
\ee
In particular the Einstein-Rosen waves have $B\equiv 0$, i.e.~$n^1
\equiv 0$, and are thus described by a $O(1,1)$ matter sigma-model.
For the gauge invariant $M$ matrix in the $SL(2,\R)$ description one obtains
\be
M = \cV i\sigma^2 \cV^{-1} = \cV \cV^T i\sigma^2 = 
\left( \begin{array}{cc} n^1 & n^2 + n^0 \\
n^2 - n^0 & -n^1 \end{array} \right)\;, 
\label{act15}
\ee
related to the $SU(1,1)$ version (\ref{act5}),~(\ref{act6}) by 
${\rm Ad}\Upsilon$, as it should. Contact to the parameterization
used e.g.~in (\ref{contM}) is made by lowering one index,
$M_{ab} = i M_a^{b'} C_{b'b}$, etc. 

In the case of the compact coset space $SU(2)/U(1)$ the counterpart
of the action (\ref{act4}) yields the classical $\C\!\P_1$ model, 
the counterpart of (\ref{act7}) yields the $O(3)$ nonlinear sigma-model.
Heuristically one can then think of the $n$-fields as ``mesonic bound states'' 
of the ``quark-doublets'' $z$, via $n^j = z^{\dagger} \sigma^j z$. The
analogy is correct in the sense that in the quantum field theory only
the $n$-field generates scattering states.

\newappendix{Diagonalization of \boldmath{$\cT$}}

Here we describe the solution of the eigenvalue problem (\ref{t6}),
(\ref{t7}) for case $\eps =-1$, where only the ${\rm SO}(2) \subset
{\rm SL}(2,\R)$ invariance remains. The remaining abelian symmetry is
nevertheless useful in that it still allows one to decompose the full
problem into pieces of lower dimensionality.

\newappsection{Decomposition into charge \boldmath{$e$} sectors}

To this end we switch to a basis in $V^{\otimes \n}$ diagonalizing
$\Gamma$. With the choice $\Gamma = \Gamma(\varphi,\lb)$ in (\ref{a2}) the
eigenvalues are $\pm  i$ and we simply label the components with respect
to the new basis by the sign of the corresponding eigenvalue, i.e.
\be
w_{\sigma_\n\ldots \sigma_1}(\th) = 
\Upsilon_{\sigma_\n}^{a_n} 
\ldots \Upsilon_{\sigma_1}^{a_1} \,w_{a_\n\ldots a_1}(\th)\;,
\;\;\;\sigma_j \in \{\pm \}\,.
\label{b5}
\ee
For the inverse matrix we use the index staggering
$(\Upsilon\inv)_a^{\sigma}$, such that in $\Gamma_{\sigma}^{\tau} =
\Upsilon_{\sigma}^a \Gamma_a^b (\Upsilon\inv)_b^{\tau}$ the
nonvanishing components of $\Gamma_{\sigma}^{\tau}$ are $\Gamma_+^+ =
i = - \Gamma_-^-$. Explicitly, one finds for $\nu >0$
\be
\Upsilon \Gamma \Upsilon^{-1} =  
{ i \;\;\phantom{-}0 \choose 0\;\; -i}\,,\;\;\;
\mbox{with}\;\;\; 
\Upsilon = \Upsilon^a_{\sigma} = \frac{1}{\sqrt{2 \nu \ch \varphi}} 
\left( \begin{array}{cc} \ch\varphi & \nu(-i + \sh \varphi) \\
                         \ch\varphi & \nu(i + \sh \varphi) 
\end{array} \right)\,.
\label{b6}
\ee
It obeys $\det \Upsilon =i$ and $(\Upsilon_{\sigma}^a)^* =
\Upsilon_{-\sigma}^a$.  Clearly, the same transformation also
diagonalizes the matrix $\Lambda(\phi)$ in (\ref{norm8}),
i.e. $\Lambda_{\sigma}^{\tau}(\phi) = \delta_{\sigma}^{\tau}\,
e^{\frac{i}{2}(\sigma + \tau) \phi}$. The charge conjugation matrix
becomes $C_{\sigma\tau} = - \eps_{\sigma \tau}$. For simplicity we
shall refer to the basis (\ref{b6}) as the ``charged basis''
(regarding the phase $e^{\pm i \phi}$ as associated with a U(1)
charge) and to the original basis as the ``real'' basis.  We write $e$
for the U(1) charge of $w_{\sigma_\n\ldots\sigma_1}(\th)$, i.e. $e =
\sigma_\n + \ldots + \sigma_1$ is the number of $+$ minus the number
of $-$ in a multi-index.

In group theoretical terms ${\rm Ad}\Upsilon$ yields a two-parameter
family of automorphisms $SL(2,\R)\newline \ra SU(1,1)$ generalizing
(\ref{act11}).  It suffices to verify this on the level of the Lie
algebras. Let $\tau^0 = i\sigma^3,\;\tau^1 = \sigma^1,\;\tau^2 =
\sigma^2$ denote the basis of $su(1,1)$ used before. The matrices
$\Upsilon\inv \tau^j \Upsilon$, $j=0,1,2$, then are readily checked to
be real and trace-free, and hence can serve as a (non-standard) basis
of $sl(2,\R)$.

Returning to the eigenvalue problem (\ref{t6}) it follows from
(\ref{norm7}) that 
\be
\cT(\th_0|\th)_{\sigma_\n\ldots \sigma_1}^{\tau_\n\ldots \tau_1} =0\;,
\;\;\;\mbox{unless}\;\;\;\sigma_\n + \ldots + \sigma_1 = 
\tau_\n + \ldots +\tau_1\;.
\label{b7}
\ee
In the charged basis (\ref{t6}) therefore decomposes into 
decoupled sectors of dimension 
\be
m_e(\n) = { \n \choose \Lambda} = { \n \choose \n\!-\!\Lambda} =
m_{-e}(\n)\;,\sspace \Lambda := \frac{1}{2}(\n\!-\!e)\,.
\label{bdim}
\ee 
Explicitly we write
\be 
\overline{\cT}_e(\th_0|\th) w_e(\th) = 
\overline{\tau}_e(\th_0|\th) w_e(\th)\;,
\;\;\; e = \n, \n\!-\!2, \ldots , -\n\!+\!2, -\n\;, 
\label{b8}
\ee
where $w_e(\th)$ is a column vector of length $m_e(\n)$. 
For convenience we split off here a scalar factor 
\be
\cT(\th_0|\th) = \prod_{j=1}^\n \frac{r(\th_{j0} -
i\hbar)}{\th_{j0}-2i\hbar}\; 
\overline{\cT}(\th_0|\th)\;,
\label{b9}
\ee
which renders the components of $\overline{\cT}(\th_0|\th)$
polynomial in the $\th_{j0} - i\hbar$. 
Similarly the KZE eigenvalue problem (\ref{t16}) splits up
into decoupled sectors with fixed charge. The eigenvalues  
$q_{e;k}(\th)$ in the charge $e$ sector are phases and in addition 
obey $\prod_{k=1}^\n q_\n(\Omega^{k-\n}\th) = (-)^\n i^e$,
which follows from (\ref{t25}). Again it is convenient to 
split off a scalar function from the $Q_k(\th)$ matrices and write
\be 
Q_k(\th) = -i \prod_{j\neq k} r(\th_{jk})\,\overline{Q}_k(\th)\;,\quad
\overline{\cT}(\th_k - i\hbar|\th) = 
\hbar \prod_{j\neq k} (\th_{jk} - i\hbar)
\,\overline{Q}_k(\th)\,.
\label{b10}
\ee
The reduced eigenvalues in the charge $e$ sector defined by 
\be
\overline{Q}_{e;k}(\th)\, w_e(\th) = \overline{q}_{e;k}(\th)\, w_e(\th)\;,
\;\;\;{\rm i.e.}\;\;\; \overline{q}_{e;k}(\th) = \frac{1}{\hbar} 
\prod_{j \neq k} \frac{1}{\th_{jk} - i \hbar}\,
\overline{\tau}_e(\th_k - i\hbar|\th)\;,
\label{b12}
\ee
will then satisfy 
\be\label{prodq}
\prod_{k=1}^\n \overline{q}_{e;k}(\th) ~=~ i^{e-\n}\;.
\ee
The ${\overline \cT}$ eigenvalue
problem (\ref{b8}) is in fact equivalent to the seemingly weaker
${\overline Q}_{e;k}$ eigenvalue problem (\ref{b12}). To see this
note that ${\overline \cT}(\th_0|\th)$ is a polynomial in $\th_0$ 
of degree $N-1$, due to the tracelessness of $\Gamma$. The second
relation in (\ref{b10}) therefore amounts to $\n$ linear equations 
for the $\n$ matrix-valued coefficients of this polynomial. Provided
all variables $\th_j$ are distinct these equations turn out to be 
independent, so that the ${\overline \cT}(\th_0|\th)$ matrix is
uniquely determined by the $\overline{Q}_k(\th)$ matrices. By expanding 
the ${\overline \cT}$ eigenvalue problem into powers of $\th_0$
one sees that also the eigenvalues ${\overline \tau}(\th_0|\th)$ must 
be polynomials of degree $\n-1$ in $\th_0$. By the same token they
will be uniquely determined by their values at $\n$ points, i.e. by the
eigenvalues ${\overline q}_{e;k}(\th)$. Explicitly one finds 
\be
{\overline \tau}_e(\th_0|\th)  =  \hbar \sum_{j=1}^\n  
{\overline q}_{e;j}(\th)  
\prod_{k\not=j}
\frac{(\th_{0k}\!+\!i\hbar)(\th_{kj}\!-\!i\hbar)}{\th_{jk}}\,. 
\label{tauqbar}
\ee

Finally let us note a number of useful involution properties linking the
spectrum in the charge $e$ and the charge $-e$ sector. Since 
\be
\cT(\th_0|\th)_{\sigma_\n \ldots \sigma_1}^{\tau_\n \ldots \tau_1} =
-\cT(\th_0|\th)_{-\sigma_\n \ldots -\sigma_1}^{-\tau_\n \ldots -\tau_1}\;,
\label{eflip1}
\ee
one infers: 
\be
\begin{array}{lclc}
{\rm If} \;\;\; & \tau_e(\th_0|\th) \in {\rm Spec}\,\cT_e(\th_0|\th)
\;\;\; & {\rm then} \;\;\; 
& -\tau_e(\th_0|\th) \in {\rm Spec}\,\cT_{-e}(\th_0|\th)\;,
\\
{\rm If} \;\;\; & q_{e;k}(\th) \in {\rm Spec}\,Q_{e;k}(\th)
\;\;\; & {\rm then} \;\;\; 
& - q_{e;k}(\th) \in {\rm Spec}\,Q_{-e;k}(\th)\,.
\end{array}
\label{eflip2}
\ee
In particular for charge $e =0$ this means all eigenvalues come in pairs 
differing only by a sign. For $e \neq 0$ the sign flip $ e \mapsto - e$
amounts to $ \Lambda \mapsto \n\!-\!\Lambda$ so that by (\ref{bdim}) one
expects, at least for generic rapidities, the eigenvalues to be in 
1-1 correspondence. Under complex conjugation one has 
$[w_{\sigma_\n \ldots \sigma_1}(\th)]^* = w_{-\sigma_1 \ldots
-\sigma_\n} ({\th^*}^T\!+\!i\hbar)$ and a counterpart of the first
equation in (\ref{t20}). Combined with (\ref{eflip1}) this implies 
\be
\begin{array}{lclc}
{\rm If} \;\; & \tau_e(\th_0|\th) \in {\rm Spec}\,\cT_e(\th_0|\th)
\;\;\; & {\rm then} \;\;\; 
& \tau_e(\th_0^* - 2 i \hbar|\th^* + i\hbar)^* \in 
{\rm Spec}\,\cT_e(\th_0|\th)\,.
\end{array}
\label{eflip3}
\ee
For the corresponding eigenvectors one can choose normalizations such that
\be
[w_{\sigma_\n \ldots \sigma_1}(\th)]^* = w_{\sigma_1 \ldots \sigma_\n}
({\th^*}^T + i\hbar)\;,
\label{eflip4}
\ee
which also matches the properties of the Bethe Ansatz vectors
(\ref{evec}) below.

\newappsection{Bethe ansatz equations}

The aim in the following is to compute the eigenvalues and
eigenvectors in (\ref{b8}) explicitly. For small $\n$ this can be done
by brute force but for generic $\n$ it is useful to parameterize the
solutions in terms of the roots of the Bethe equations. The literature
on the Bethe Ansatz is enormous, some guidance can be obtained from
the book \cite{KoBoIz93} and e.g.~the following papers
\cite{KirRes86,Babu93,MaiSan96,Fadd95}. Transferred to the present
context the construction can be outlined as follows: Denote by
\be
\Omega_A := (\Upsilon\inv)_{a_{\n}}^+(\Upsilon\inv)_{a_{\n\!-\!1}}^+\dots
(\Upsilon\inv)_{a_1}^+\;, 
\ee
a cyclic vector on which the operator $\Gamma T$ from (\ref{t5}) in the 
charged basis acts as follows
\be\label{Bv}
\Upsilon_\sigma^a \Gamma_a^b (\Upsilon^{-1})_c^\tau \;
\Tbar_b^c(\th_0\!+\!i\hbar|\th)_A^B 
\;\Omega_B ~~=~ \left(\begin{array}{cc}
i\prod_j(\th_{0j}\!+\!2i\hbar)\,\Omega_A &*\\
0&-i\prod_j(\th_{0j}\!+\!i\hbar)\,\Omega_A
                                 \end{array} \right)\,.
\ee
Here, $\Tbar_a^b(\th_0|\th)_A^B$ denotes the monodromy matrix
(\ref{t5}) with the same prefactor taken out as in (\ref{b9}).
Following the Bethe Ansatz procedure we generate candidate eigenstates
from $\Omega$ by the repeated action of $B(t|\th) := \Upsilon_+^a
\Gamma_a^b (\Upsilon^{-1})_c^-\; \Tbar_b^c(t\!+\!i\hbar|\th)$. The
matrix operators $B(t|\th)$ are commuting for different values of $t$
and each $B(t_\alpha|\th)$ lowers the $SO(2)$ charge of a candidate
eigenstate of $\cT$ by the two units.  The candidate eigenstates can
be made proper eigenstates by turning the parameters $t_{\alpha}$ into
judiciously chosen functions of the $\th_j$. In upshot one obtains
eigenvectors
\be\label{evec}
w_{e}(\th)~=~\prod_{\alpha=1}^{\Lambda} B(t_\alpha|\th) \Omega \;,\qquad
\Lambda:=\ft12(N\!-\!e)\;,
\ee
with eigenvalues
\be
\overline{\tau}_e(\th_0|\th) ~=~
\frac{i\prod_j(\th_{j0}\!-\!2i\hbar)
     \prod_{\alpha}(\th_0\!-\!t_\alpha\!+\!i\hbar/2)}
     {\prod_{\alpha}(\th_0\!-\!t_\alpha +3 i\hbar/2)} ~-~
\frac{i\prod_j(\th_{j0}\!-\!i\hbar)
     \prod_{\alpha}(\th_0\!-\!t_\alpha\!+\!5 i\hbar/2)}
     {\prod_{\alpha}(\th_0\!-\!t_\alpha + 3i\hbar/2)} \;,
\label{ev}
\ee
where the Bethe roots $t_\alpha$ are solutions of the following 
Bethe Ansatz equations (BAE) 
\be\label{BAE}
\prod_{j=1}^\n \frac{\th_j\!-\!t_\alpha\!-\!i\hbar/2}
                   {\th_j\!-\!t_\alpha\!+\!i\hbar/2}  ~=~ -
\prod_{\beta\not=\alpha} \frac{t_\beta\!-\!t_\alpha\!-\!i\hbar}
                              {t_\beta\!-\!t_\alpha\!+\!i\hbar}\;,
\sspace \alpha = 1,\ldots,\Lambda\,.
\ee
The only modification of the BAE as compared to the standard case
$\eps=1$ is the sign on the r.h.s.~which comes from the ratio of
the eigenvalues of $\Gamma$. These equations ensure that
$\overline{\tau}_e(\th_0|\th)$ is indeed a polynomial in $\th_0$
of degree $\n -1$, as anticipated in section B1:
\ba
\overline{\tau}_e(\th_0|\th) &=& 
(-1)^{\n-1}\hbar\, \th_0^{\n-1}\,(\n\!-\!2\Lambda)\\[1mm]
&&{}-(-1)^{\n-1}\hbar\,\th_0^{\n-2} 
\Big[(\n\!-\!2\Lambda\!-\!1)\sum_j (\th_j - 3i\hbar/2)
+ 2\sum_\alpha (t_\alpha - 3i\hbar/2)\Big] \nonumber\\
&&{}+\dots \nonum
&=:& \sum_{p=0}^{\n-1} \th_0^p \,{\overline \tau}_{e;p}(\th)\;, 
\label{taupoly}
\ea
where the coefficients obey ${\overline \tau}_{e;p}(\th)^* = 
{\overline \tau}_{e;p}(\th^* + 3i\hbar)$. 
For $e =\n$ no Bethe roots are required and (\ref{evec}), (\ref{ev})
should be interpreted as $w_{\n;A}(\th) = \Omega_A$ and
\be
{\overline\tau}_\n(\th_0|\th) = i \prod_j (\th_{j0} - 2 i\hbar) - i
\prod_j (\th_{j0} - i \hbar)\,,
\ee
from which one computes $\overline{q}_{\n;k} =1$, $k=1,\ldots,\n$.
Since for $e =\n$ the eigenvalue problem (\ref{b8}) is one-dimensional
it follows from (\ref{eflip2}) that ${\overline \tau}_{-\n}(\th_0|\th)
= - {\overline \tau}_\n(\th_0|\th)$ and $\overline{q}_{-\n;k}=-1$. In
the expressions (\ref{ev}) for the eigenvalues coming out of the Bethe
Ansatz however this is not obvious as now $\n$ Bethe roots are
required, rather than none as in the charge $\n$ sector. More generally
the charge $e$ and charge $-e$ sector enter asymmetrically in the
Bethe Ansatz construction, though by (\ref{eflip2}) they are
practically identical.

The eigenvalues $\overline{q}_{e;k}(\th)$ of the asymptotic qKZE
equations are accordingly given by
\be\label{qBethe}
{\overline q}_{\pm \n;k} = \pm 1\;,\sspace \overline{q}_{e;k}(\th) ~=~ 
\prod_{\alpha}\frac{\th_k\!-\!t_\alpha\!-\!i\hbar/2}
      {\th_k\!-\!t_\alpha\!+\!i\hbar/2}\,,\;\;\; |e| \leq  \n\!-\!2\,.
\ee
The explicit form (\ref{qBethe}) allows to directly check many of the
properties which we have derived in the main text.  Inspection of the
BAE (\ref{BAE}) shows that for real $\th_j$ the Bethe roots appear in
(possibly degenerate) complex conjugate pairs $(t^{\phantom{*}}_\alpha\,,
t_\alpha^*)$. The $\overline{q}_{e;k}(\th)$ therefore indeed
are pure phases, consistent with (\ref{t24}).  Assuming the Bethe roots
to be completely symmetric functions of the $\th_j$ eq.~(\ref{qBethe})
also makes manifest the cyclic property (\ref{Qcycl}) of the
$\overline{q}_{e;k}(\th)$.  For the product of the eigenvalues
$\overline{q}_{e;k}(\th)$ we obtain with (\ref{BAE}) and
(\ref{qBethe}):
\be
\prod_k \overline{q}_{e;k}(\th) ~=~ (-)^\Lambda 
\prod_{\beta\not=\alpha} \frac{t_\beta\!-\!t_\alpha\!-\!i\hbar}
                              {t_\beta\!-\!t_\alpha\!+\!i\hbar} ~=~ 
i^{\n-e}\;,
\ee
confirming (\ref{prodq}) and thus (\ref{t25}).

Finally the logarithmic derivative of the BAE yields
\ba
\frac{i\hbar}{(\th_k\!-\!t_\alpha\!-\!i\hbar/2)
              (\th_k\!-\!t_\alpha\!+\!i\hbar/2)} 
&=&\sum_{\beta\not=\alpha} 
\frac{2i\hbar\,\partial_k(t_\beta-t_\alpha)}
     {(t_\beta\!-\!t_\alpha)^2-(i\hbar)^2} \\
&& {}+\sum_j\frac{i\hbar\,\partial_k t_\alpha}
     {(\th_j\!-\!t_\alpha\!-\!i\hbar/2)(\th_j\!-\!t_\alpha\!+\!i\hbar/2)}\,,   
\nonumber
\ea
which may be used to prove that
\be
\partial_k\, \ln \overline{q}_{e;l}(\th) ~=~ 
\partial_l\, \ln \overline{q}_{e;k}(\th)\;,
\ee
as claimed in (\ref{t24}) and in accordance with
\cite{TarVar96,TarVar97}. 

Let us also briefly comment on the solutions to the BAE. In analogy to
the homogeneous case (all $\th_j$ equal \cite{Fadd95}) one expects
that only the solutions with $t_{\alpha} \neq t_{\beta}$, $\alpha \neq
\beta$, are relevant. Assuming that $t_{\alpha} -t_{\beta} \neq 0,\pm
i\hbar$ the eqs (\ref{BAE}) can be rewritten in polynomial
form. Specifically they constitute a system of $\Lambda$ polynomial
equations of degree $\Lambda\!+\!\n\!-\!1$ for the unknowns
$t_{\alpha} - \frac{1}{\n}(\th_\n +\ldots + \th_1)$, $\alpha =1,
\ldots, \Lambda$, whose coefficients are symmetric polynomials in
$\widehat{\th}_j= \th_j - \frac{1}{\n}(\th_\n + \ldots +\th_1)$,
$j=1,\ldots,\n$.  The point of adding and subtracting the `center of
mass term' is that the $\widehat{\th}_j$'s are boost invariant. In
particular it follows that
\be
t_{\alpha} - \frac{1}{\n}(\th_\n + \ldots +\th_1) \;\;\;\mbox{are boost
invariant}\,, 
\label{boost}
\ee
i.e.~are (completely symmetric) functions of the differences
$\th_{jk}$ only.   

It may be instructive to exemplify the construction for the 
simplest case $\n=2$:
\ba
& e = 2:& \mbox{Bethe roots:}\;\;\; \emptyset \nonum
&& \overline{\tau}_2(\th_0|\th_2,\th_1) = \hbar(\th_{20} 
+ \th_{10} - 3 i \hbar)\;,
\sspace \overline{q}_{2;2} = \overline{q}_{2;1} = 1\;,
\nonum
& e = 0:& \mbox{Bethe roots:}\;\;\; t = \frac{1}{2}\Big[\th_1 + \th_2  
\pm \sqrt{\hbar^2 + \th_{12}^2}\,\Big]\,,\nonum
&& \overline{\tau}_0(\th_0|\th_2,\th_1) = \pm \hbar\sqrt{\hbar^2 +
\th_{21}^2}\,, 
\sspace \overline{q}_{0;2}(\th_2,\th_1) = 
\pm i\sqrt{ \frac{i \hbar + \th_{12}}{i\hbar - \th_{12}} }
= \overline{q}_{0;1}(\th_1,\th_2)\,,
\nonum
& e = -2:& \mbox{Bethe roots:} \;\;\;
t_1 = \frac{1}{2}\Big[\th_1 +\th_2 +\sqrt{-\hbar^2
+\th_{12}^2}\,\Big]\, \nonum
&& \bspace \;\;\;\;\;
t_2 = \frac{1}{2}\Big[\th_1 +\th_2 -\sqrt{-\hbar^2
+\th_{12}^2}\,\Big]\, \nonum
&& \overline{\tau}_{-2}(\th_0|\th_2,\th_1) = -\hbar(\th_{20} + \th_{10}
- 3 i \hbar)\;, 
\sspace \overline{q}_{-2;2} = \overline{q}_{-2;1} = -1\;.
\label{ntwo}
\ea

Let us now return to the eigenvectors (\ref{evec}).  Clearly any
eigenvector is only determined up to multiplication by an arbitrary
scalar function, or the corresponding linear combinations in the case
with degeneracies.  The Bethe eigenvectors (\ref{evec}) will in
general not obey the exchange relations
\ba 
&& \fbar_{e;\sigma_n\ldots\sigma_1}(\th) = 
{\overline R}_{\sigma_{k+1}
\sigma_k}^{\,\tau_k\;\;\tau_{k+1}}(\th_{k+1,k})  
\fbar_{e;\tau_n\ldots \tau_1}(s_k\th)\;,\nonum
&&{\overline R}(\th)_{++}^{++} = 
\frac{-\th + i\hbar}{\phantom{-}\th +i\hbar} \;,\;\;
{\overline R}(\th)_{+-}^{+-} = \frac{-\th}{\th + i\hbar} \;,\;\;
{\overline R}(\th)_{+-}^{-+} = \frac{i\hbar}{\th + i\hbar}\;,
\label{solsym1}
\ea
going along with the redefined monodromy matrix
$\overline{T}_{\tau}^{\sigma}(\th_0|\th)$.  However, it is not
difficult to modify them so that they do. Due to the symmetry
${\overline\tau}_e(\th_0|s\th) = {\overline\tau}_e
(\th_0|\th),\;\forall s \in S_\n$, a joint solution of (\ref{b8}),
(\ref{solsym1}) can be obtained simply by symmetrizing with the
$\overline{R}$-matrix: Let $S_\n \ni s \ra \overline{L}_s(\th)$ be the
representation of the permutation group analogous to (\ref{t8}),
(\ref{t9}), just with $R$ replaced by $\overline{R}$. The identity
(\ref{ts}) remains valid with $L$ replaced by $\Lbar$ and implies that
\be
\fbar_{e;\sigma_\n\ldots \sigma_1}(\th) \sim \frac{1}{\n!} \sum_{s \in S_\n} 
{\overline L}_s(\th)_{\sigma_\n \ldots \sigma_1}^{\tau_\n \ldots \tau_1}\, 
w_{e;\tau_\n\ldots \tau_1}(s\inv \th)\;,
\label{solsym2}
\ee
is an eigenvector obeying also (\ref{solsym1}). Here $\sim$ indicates
that one is still free to multiply by a completely symmetric function
in $\th_j$ without affecting (\ref{b8}), (\ref{solsym1}). Observe that
since $w_e(s\th) = w_e(t(s\th)|s\th) = w_e(t(\th)|s\th)$ the
operations: ``${\overline R}$-symmetrization'' and ``Inserting the
Bethe roots $t_{\alpha} =t_{\alpha}(\th)$'' commute. It is clearly
convenient to first perform the symmetrization and then insert the
Bethe roots. Taking once more advantage of the equivariance properties
analogous to (\ref{ts}) one finds that the symmetrization just results
in a scalar prefactor.  Explicitly, for any given Bethe eigenvector
(\ref{evec}) the product
\be
\fbar_{e;\sigma_\n\ldots\sigma_1}(\th) \sim  
\prod_{k > l}\frac{i}{\th_{kl} + i
\hbar}\,w_{e;\sigma_\n\ldots\sigma_1}(\th)\;, 
\label{solsym3}
\ee
solves both (\ref{b8}) and (\ref{solsym1}). It is manifestly
polynomial in  the Bethe roots and rational in the $\th_j$'s.

\newappsection{Sequential Bethe roots}

Let us examine the behavior of the Bethe ansatz equations and their
solutions under pinching $\th_{k+1} \ra \th_k \pm i\hbar$ of the insertions
$\th_\n,\ldots,\th_1$. The relations (II) imply that the $SO(2)$ charge
of the eigenvectors is conserved under $\th_{k+1} \ra \th_k \pm i\hbar$,
i.e.
\be
\n\ra \n\!-\!2\;,\quad e\ra e \;,\quad \Lambda\ra\Lambda\!-\!1\;.
\ee
This suggests that the Bethe roots describing these (special)
eigenvectors might  likewise be related. Indeed the BAE (\ref{BAE})
are consistent with the following  $\n\ra \n\!-\!2$ reduction of its
solutions  
\ba\label{tred}
t_\Lambda(\th)\bigg|_{\th_{k+1} = \th_k \pm i\hbar} &=& \th_k
\pm \frac{i\hbar}{2}\, \nonum
t_\alpha(\th)\bigg|_{\th_{k+1} = \th_k \pm i\hbar} &=& t_\alpha(p_k\th)\;,
\quad\mbox{for } \alpha<\Lambda\;,
\ea
Since the Bethe roots are symmetric in all $\th_j$ it suffices to
verify (\ref{tred}) for the $\th_{k+1} = \th_k + i\hbar$ case. The
other then formally follows from applying the $\th'_{k+1}=\th'_k +
i\hbar$ reduction to $\th' = (\th_\n,\ldots,\th_{k+1},\th_k-i\hbar,
\th_{k-1},\ldots,\th_1)$.  It is easy to verify that with (\ref{tred})
the BAE (\ref{BAE}) for $\alpha<\Lambda$ reduce to the BAE with
$\n\!-\!2$ insertions for the $t_\alpha(p_k\th)$. The equation for
$\alpha = \Lambda$ is a bit more subtle and requires to specify the
limit in which the pinched configuration is approached. Entering
with the ansatz
\ba\label{tlred}
t_\Lambda(\th) = \th_k\!+\!\frac{i\hbar}{2}+\delta /Z(\th) +
o(\delta) \,, \qquad \mbox{for}\quad \th_{k+1} = \th_k\!+\!i\hbar +
\delta\;, 
\ea
into the $\alpha = \Lambda$ BAE one obtains for $\n > 2$
\ba\label{Z}
\prod_{\beta<\Lambda}
\frac{t_\beta\!-\!\th_k -3i\hbar/2}{t_\beta\!-\!\th_k\!+\!i\hbar/2}
&=& -\left(\frac{-i\hbar}{-\delta/Z}\right)
\left(\frac{\delta\!-\!\delta/Z}{i\hbar}\right)
\prod_{j\not=k,k+1}\frac{\th_j\!-\!\th_k\!-\!i\hbar}{\th_j\!-\!\th_k}
\\[8pt]
&=&(1\!-\!Z)\,\prod_{j\not=k,k+1}
\frac{\th_j\!-\!\th_k\!-\!i\hbar}{\th_j\!-\!\th_k}\;.\nonumber
\ea
This can be taken to define $Z=Z(\th)$ in (\ref{tlred}), showing the
consistency of the reduction rule for $t_{\Lambda}(\th)$ as $\delta
\ra 0$.

Of course, not every solution of the BAE will satisfy (\ref{tred}), in
fact the vast majority will not.  The argument shows however that
under the same `genericity assumptions' under which solutions exist at
all, there also exists at each recursion step $\n -2 \mapsto \n$ at
least one $\Lambda$-tuple of Bethe roots enjoying the property
(\ref{tred}).  In addition (\ref{tred}) is compatible with the
following reality condition
\be
t_{\alpha}(\th)^* = t_{\alpha}(\th^*)\;,
\qquad \alpha =1,\ldots,\Lambda\,.
\label{treal}
\ee 
Here we refer to the observation after (\ref{qBethe}) that the
solutions of the BAE come in pairs
$(t_{\alpha}(\th)^*,t_{\alpha}(\th^*))$, where in general
$t_{\alpha}(\th)^* \neq t_{\alpha}(\th^*)$.  We call a solution of the
BAE a ``sequential'' tuple of Bethe roots, if all roots are distinct, real
in the sense of (\ref{treal}), and satisfy (\ref{tred}).%
\footnote{The concept appears to be new. The only article we are aware of 
where a pinching of inhomogeneities is considered, is \cite{KirRes86}.}

To justify the terminology let us consider the behavior of the
eigenvalues $\tau(\th_0|\th)$ under $\th_{k+1} \ra \th_k + i\hbar$. From
(\ref{tred}) one finds
\ba
&& \nspace \overline{\tau}_e(\th_0|\th)\;\; 
\stackrel{\th_{k+1} \ra \th_k + i\hbar}%
{-\!\!\!-\!\!\!-\!\!\!-\!\!\!\longrightarrow}\;\; 
i\th_{k0}(\th_{k0} - 2 i \hbar) \prod_{\alpha < \Lambda}
(\th_0 - t_{\alpha}+3i\hbar/2)\inv \times 
\nonumber\\
&& \sspace \times \left[
\prod_{j\not=k,k+1}(\th_{j0}\!-\!2i\hbar)
     \prod_{\alpha<\Lambda}(\th_0\!-\!t_\alpha\!+\!i\hbar/2) \,-\!
\prod_{j\not=k,k+1}(\th_{j0}\!-\!i\hbar)
     \prod_{\alpha<\Lambda}(\th_0\!-\!t_\alpha\!+\!5i\hbar/2) \right]
\nonumber\\[10pt]
&& \sspace = \th_{k0}(\th_{k0} - 2 i \hbar) 
\,{\overline \tau}_e(\th_0|p_k\th)\;,
\ea
and similarly for $\th_{k+1} = \th_k - i\hbar$. 
Hence for any $\Lambda$-tuple of Bethe roots satisfying (\ref{tred}) the 
associated eigenvalues satisfy the recursive relation (IIc) from
section 2.4 
\be
\tau_e(\th_0|\th)\Big|_{\th_{k+1}=\th_k \pm i\hbar}= \tau_e(\th_0|p_k\th)\;.
\ee
One can also easily verify (\ref{qres}) i.e.~the fact that the limits
$\th_0\ra\th_k\!-\!i\hbar$ and $\th_{k+1}\ra\th_k\!+\!i\hbar$ of
$\tau(\th_0|\th)$ commute.

\newappsection{Semi-classical limit}\label{Asemicl}

The semiclassical limit of the transfer matrix $\cT$ (\ref{t5}) follows
directly from (\ref{r6}): 
\ba
\cT(\th_0|\th)_A^B &=&
i\hbar\,\sum_k \frac{\Gamma_k}{\th_{0k}} + (i\hbar)^2\,\left(
-\frac12 \sum_k \frac{\Gamma_k}{\th^2_{0k}}+\sum_k \frac{H_k}{\th_{0k}}
\right) + O(\hbar^3)\;, \label{semiT}\\[10pt]
\mbox{with}\quad (\Gamma_k)_A^B &=& 
\delta_{a_\n}^{b_\n}\dots \Gamma_{a_k}^{b_k} \dots\delta_{a_1}^{b_1} \nonum 
H_k &=& \sum_{l\not=k}\frac{\Omega_{kl}\,(\Gamma_k+\Gamma_l)}{\th_{kl}}
\;, \qquad
(\Omega_{kl})_A^B ~=~ \delta_{a_\n}^{b_\n}\dots \Omega_{a_ka_l}^{b_kb_l} 
                      \dots\delta_{a_1}^{b_1}\,,
\nonumber
\ea
and $\Omega_{ab}^{cd}$ from (\ref{r6}). This expansion is valid either
in the sense of a formal power series in $\hbar$ or, with a numerical
$\hbar$, in the region ${\rm Im}\,\th_{0k} \gg \hbar$, ${\rm
Im}\,\th_{lk} \gg \hbar$, $l\neq k$, in order to prevent a mixing of
different powers of $\hbar$.  The absence of a term of order $\hbar^0$
in (\ref{semiT}) is due to the tracelessness (\ref{traceX}) of
$\Gamma$ and distinguishes this case from the usual situation $\eps=1$
(see e.g.~\cite{Babu93,FeFrRe94}). The same fact implies
$\Gamma_k^2=-1$ and furthermore that the operators $\Gamma_k$ and
Hamiltonians $H_k$ form a family of mutually commuting operators. (In
fact the $H_k$ can be viewed as the Hamiltonians of an abelian $SO(2)$
Knizhnik-Zamolodchikov system). Simultaneous diagonalization of the
$\Gamma_k$ yields eigenvectors with only one nonvanishing component
$w_{\eps_\n\dots\eps_1}$, $(\eps_\n,\ldots,\eps_1) \in \{\pm\}^\n$,
in the charged basis, $\sum_j\eps_j=e$. On these eigenvectors the $H_k$
already act diagonally. Thus the first terms in the semiclassical
expansion of the eigenvalues $\tau$ are given by:
\ba\label{semit}
\tau(\th_0|\th) &=& -\hbar\sum_k \frac{\eps_k}{\th_{0k}} 
               +i \hbar^2 \Bigg(\frac{1}{2} \sum_k \frac{\eps_k}{\th^2_{0k}}
                {}-\sum_{k\not=l\atop\eps_k=\eps_l}
                        \frac{\eps_k}{\th_{0k}\th_{kl}}\Bigg) 
                    +O(\hbar)^3\; 
\ea

This phenomenon can also be understood in terms of the Bethe ansatz.
Examination of the explicit solutions of the Bethe roots for
$\n=2,3$ indicates that the symmetry in $\th_\n,\ldots,\th_1$ gets
lost in the limit $\hbar \ra 0$ and that they typically behave
like 
\be
t_{\alpha}(\th) = \th_{j(\alpha)} + (i\hbar)^2 \, s_{\alpha}(\th) 
+ O(\hbar^3)\;,
\label{rootcl}
\ee 
for some $j(\alpha) \in \{1,\ldots,\n\}$ with $j(\alpha) \neq
j(\beta)$ for $\alpha \neq \beta$. This curious behavior is 
directly linked to the seemingly innocent sign flip in the Bethe ansatz 
equations (\ref{BAE}). Indeed, entering with (\ref{rootcl}) into the BAE
and matching coefficients in powers of $\hbar$ one finds at $O(\hbar)$
\be
s_{\alpha}(\th) = \frac{1}{4} 
\sum_{j \neq j(\alpha)}\frac{1}{\th_j - \th_{j(\alpha)}}
- \frac{1}{2} 
\sum_{\beta \neq \alpha} \frac{1}{\th_{j(\beta)} -\th_{j(\alpha)}}\;.
\label{root2cl}
\ee
Generally one can show that the Bethe roots admit a power series expansion
in $\hbar$ (in the region ${\rm Im}\,\th_{kl} \gg \hbar,\,k \neq l$) whose
coefficients are uniquely determined by the assignment $\alpha \ra 
j(\alpha)$ in (\ref{rootcl}).

Expanding the BA expression for the eigenvalues (\ref{ev}) one obtains
\ba
\tau_e(\th_0|\th) \is \hbar \left(
\sum_j \frac{1}{\th_{j0}} -2 \sum_{\alpha} \frac{1}{t_{\alpha} - \th_0}
\right) + O(\hbar^2) \;,\nonum
\is \hbar \left(
\sum_{j \notin I_{\Lambda}} \frac{1}{\th_{j0}} - 
\sum_{j \in I_{\Lambda}} \frac{1}{\th_{j0}} \right)+ O(\hbar^2)\;.
\label{root3cl}
\ea
In the second line we inserted (\ref{rootcl}) and denoted by
$I_{\Lambda} = \{ j(1), \ldots,j(\Lambda)\}\subset \{1,\ldots,\n\}$ the
subset of $j$'s appearing on the right hand side of
(\ref{rootcl}). Comparing now with the result (\ref{semit}) we
conclude that $\eps_j = 1$ if $j\notin I_{\Lambda}$ and $\eps_j =-1$ if
$j \in I_{\Lambda}$. A similar computation then yields
\be
\qbar_{e;k}(\th) = \eps_k + O(\hbar)\;,
\ee
which one can also check to be consistent with (\ref{tauqbar}).

\newappendix{Explicit solutions for \boldmath{$\n \leq 4$}}

Here we illustrate the solution procedure for the functional equations
(I), (II) outlined in section 4 and list the first few members of the
charge $e =0, \pm 1, \pm 2$ sequences.  The eigenvectors will be given
in the charged basis (\ref{b5}); we denote by $\fbar_{e;\sigma_n
\ldots \sigma_1}(\th),\;\sigma_j \in \{\pm \}$ the $\n$-th member of
the charge $e$ sequence in this basis. Taking advantage of the duality
described in appendix B one can restrict attention to positive
charges.

Let us begin with the charge $e=1$ sector.     
For $\n=1$ one will naturally take $f_{1;\sigma} = \fbar_{1;\sigma}$
be prescribed non-zero constants which serve as the starting member of 
the sequence. For later convenience we take $\fbar_{1;\sigma} = 
\delta_{+,\sigma}$ and $c^{(1)} =1$. To determine the $\n=3$ member we 
follow the procedure (a) -- (c) described in section 4. The components of 
the Bethe trial vector (\ref{evec}) are
\ba
&& w_{1;++-}(\th) = \hbar u_3 u_2\;,\nonum
&& w_{1;+-+}(\th) =\hbar u_3(u_1 - i\hbar)\;,\nonum
&& w_{1;-++}(\th) = \hbar(u_2 - i\hbar)(u_1 -i\hbar)\;,
\label{e=11}
\ea
where $u_j = \th_j -t + i\hbar/2$ and $t$ is the Bethe parameter.
The ansatz (\ref{solsym}) reads 
\be
\fbar_{1;\sigma_3\sigma_2\sigma_1}(\th) = \phi_1(\th)
\frac{i^3}{(\th_{21} + i\hbar)(\th_{31} + i\hbar)(\th_{32}+ i\hbar)}
\;w_{1;\sigma_3\sigma_2\sigma_1}(\th)\;,
\label{e=13}
\ee
For step (b) one first verifies that the consistency conditions 
\ba
\fbar_{1;\sigma_3\sigma_2\sigma_1}(\th)\bigg|_{\th_3= \th_2 +i\hbar;
t= \th_2 +i\hbar/2}  
& \sim & C_{\sigma_3\sigma_2} \fbar_{1;\sigma_1}\;,\nonum
C^{\sigma_3\sigma_2}\fbar_{1;\sigma_3\sigma_2\sigma_1}(\th)
\bigg|_{\th_3= \th_2 -i\hbar;t = \th_2 - i\hbar/2} 
& \sim & \fbar_{1;\sigma_1}\;,
\label{e=14}
\ea
(and a similar pair for $k=1$) are obeyed. Following (\ref{phiprop}) the 
symmetric function $\phi_1(\th)$ searched for to turn (\ref{IIbar}) into
identities should be a rational, boost invariant function of the $\th_j$. 
Naturally one will select the one with the smallest possible numerator 
and denominator degrees. Since 
$\overline{\tau}(\th_0|\th_1) = \hbar$ this fixes
\be
\fbar_{1;\sigma_3\sigma_2\sigma_1}(\th) =  
2i \frac{\th_3^2 +\th_2^2 +\th_1^2 -\th_3\th_2 -\th_3\th_1 -\th_2\th_1
+3 \hbar^2} 
{(\th_{21} + i\hbar)(\th_{31} + i\hbar)(\th_{32}+ i\hbar)}
\;w_{1;\sigma_3\sigma_2\sigma_1}(\th)\;.
\label{e=15}
\ee
So far only the existence of the sequential Bethe root, i.e.~its
defining properties (\ref{goodroot}) have been used. For $\n=3$ one can
still find it explicitly, a presentation valid for $\th \in \R^3$ is
\ba
\label{good3}
&& \nspace t(\th) = \frac{1}{3}(\th_1+\th_2+\th_3) - \frac{1}{6} i \,s^{1/3}
+ \frac{i}{6} s^{-1/3} [9\hbar^2 + 4(\th_{12}\th_{13} + 
\th_{23}\th_{21} + \th_{31} \th_{32})]\;,
\nonum
&& \nspace  s := 4i(\th_{13} + \th_{23})(\th_{12} + \th_{32})(\th_{21}
+ \th_{31}) \\ 
&&  + \Big[(9 \hbar^2 + 4(\th_{12}\th_{13} + \th_{23}\th_{21} +
\th_{31}\th_{32}))^3 
-16(\th_{13} +\th_{23})^2(\th_{12} + \th_{32})^2(\th_{21} +
\th_{31})^2 \Big]^{1/2}\,,
\nonumber
\ea   
where the expression under the square root is positive for all $\th
\in \R^3$. Further $s^{1/3}$ is defined to be the cube root of $s$
that is real for $\th_3 = \frac{1}{2}(\th_1 + \th_2)$ (and cyclic) and
equals the positive square root of $[9 \hbar^2 + 4(\th_{12}\th_{13}
+\th_{23}\th_{21} + \th_{31}\th_{32})]$. With this choice one has
\be
[s^{1/3}]^* = [9 \hbar^2 + 4(\th_{12}\th_{13} +\th_{23}\th_{21} +
\th_{31}\th_{32})]\, 
s^{-1/3}\,, \;\;\;\mbox{for} \;\;\th \in \R^3\,,
\ee
and $t(\th)$ is indeed real for $\th \in \R^3$.  It is also
instructive to study the branch points of this Bethe root. They are
located at the zeros of the square root in (\ref{good3}) and have no
intersection with the strip $|{\rm Im}\,\th_{ij}|<\hbar$. E.g.,~as a
function of $\th_3$ the Bethe root has 4 branch points of order 2 such
that moving $\th_3$ around two of them interchanges the two
non-sequential Bethe roots whereas the other two separate the
sequential Bethe root from the non-sequential ones. Under pinching
$\th_2\ra\th_1\!+\!i\hbar$ the latter two vanish at complex infinity
which is just in agreement with the desired behavior (\ref{goodroot}).

Having illustrated the procedure for the charge $e =1$ case we now just 
present the results for the $e=2$ and $e =0$ series. For $e=2$ we
take $\fbar_{++}(\th) = -i(\th_{21} - i\hbar)$ with $c^{(2)} =1$ 
as the starting member. The $n=4$ member is conveniently expressed 
in terms of the Bethe trial vectors, which for $\n=4$, $e=2$ read
\ba
&& w_{2;+++-}(\th) = \hbar u_4u_3u_2\;,\nonum
&& w_{2;++-+}(\th) = \hbar u_4 u_3(u_1-i\hbar)\;,\nonum
&& w_{2;+-++}(\th) = \hbar u_4(u_2-i\hbar)(u_1-i\hbar)\;,\nonum
&& w_{2;-+++}(\th) = \hbar (u_3-i\hbar)(u_2-i\hbar)(u_1-i\hbar)\;,
\ea
with $u_j = \th_j - t + i\hbar/2$. The symmetric rational function
$\phi_2(\th)$  is conveniently described in terms of a basis of boost
invariant symmetric polynomials 
\bas
&& \tau_2 = \thh_4\thh_3 + \thh_4\thh_2 + \thh_4\thh_1 + \thh_3\thh_2 + 
\thh_3 \thh_1 + \thh_2 \thh_1\;,\nonum
&& \tau_3 = \thh_4 \thh_3\thh_2 + \thh_4\thh_3\thh_1 + \thh_4\thh_2\thh_1 +
\thh_3 \thh_2\thh_1\;,\nonum
&& \tau_4 = \thh_4\thh_3\thh_2\thh_1\;,
\label{tau4}
\eas
where $\thh_j = \th_j - \frac{1}{4} \sum_k \th_k$. Explicitly it is
given by 
\be
\phi_2(\th)  = -16 \tau_3[12 \tau_4 + \tau_2^2 - 8\hbar^2 \tau_2 + 7
\hbar^4]\;, 
\ee
and satisfies
\be
\phi_2(\th)\bigg|_{\th_4 = \th_3 + i\hbar} = 2(\th_{31} + \th_{32} +
i\hbar) 
(\th_{21}^2 + \hbar^2)(\th_{32} + 2 i\hbar)(\th_{31} + 2 i\hbar)
(\th_{32} - i\hbar)(\th_{31} - i\hbar)\;.
\label{v}
\ee
The final result is 
\be
\fbar_{2;\sigma_4\sigma_3\sigma_2\sigma_1}(\th) =
\phi_2(\th)\,\prod_{k > l} \frac{i}{\th_{kl} + i\hbar}
\;w_{2;\sigma_4\sigma_3\sigma_2\sigma_1}(\th)\;.
\ee

For the $e=0$ sequence one has two options, it can start at $\n_0 =0$ or
at $\n_0 =2$.  Of course already the $\n =2$ members will be different and
accordingly two distinct sequences will emerge. For the
$\n_0=0$ series one naturally takes $f = \fbar = \hbar$ with $c^{(0)}
=1$ as the starting member. The next member of the series is then
given by
\ba
&& \fbar_{0;+-}(\th) = \frac{2 i}{\th_{21} + i\hbar}\, \hbar u_2\;,\nonum
&& \fbar_{0;-+}(\th) = \frac{2 i}{\th_{21} + i\hbar}\, \hbar(u_1 - i\hbar)\;.
\ea
Alternatively one can consider an $e=0$ series starting at $\n_0 =2$.
An appropriate starting member then is
\ba
&& \fbar_{0;+-}(\th) = i(\th_{21} - i\hbar)\, \hbar u_2\;,\nonum
&& \fbar_{0;-+}(\th) = i(\th_{21} - i\hbar)\, \hbar(u_1 - i\hbar)\;,
\ea
and we take $c^{(2)} =1$. Equivalently this amounts to having
$\phi_0(\th_2,\th_1)= 2$ for the $\n_0 =0$ series and $\phi_0(\th_2,\th_1) =
\th_{21}^2 + \hbar^2$ for the $\n_0 = 2$ series.

To describe the $\n=4$ members of both sequences we again first note
the the Bethe trial vectors. For $\n=4$, $e=0$ there are two Bethe
parameters $t_1,t_2$. We set $u_j := \th_j - t_1 + i\hbar/2$, $v_j :=
\th_j - t_2 + i\hbar/2$, in terms of which the Bethe trial vectors
come out to be
\smallskip
\ba
&& w_{0;++--}(\th) = \hbar^2 v_4v_3u_4u_3\left(u_2v_1+u_1v_2
-i\hbar(u_1\!+\!v_2)-\hbar^2\right) \nonum 
&& w_{0;+-+-}(\th) = \hbar^2v_4u_4\Big(
u_3u_2(v_2\!-\!i\hbar)v_1 + 
(u_2\!-\!i\hbar)(u_1\!-\!i\hbar)(v_3\!-\!i\hbar)v_2 \Big)\nonum
&& \phantom{w_{0;+-+-}(\th) = }{}
   -\hbar^4v_4u_4u_3(u_1\!-\!i\hbar) \nonum
&& w_{0;+--+}(\th) = \hbar^2v_4(v_1\!-\!i\hbar)u_4(u_1\!-\!i\hbar)
\left(u_3v_2+u_2v_3-i\hbar(u_2\!+\!v_3)-\hbar^2\right)\nonum
&& w_{0;-++-}(\th) = \hbar^2(v_3\!-\!i\hbar)(v_2\!-\!i\hbar)v_1u_4u_3u_2
+\hbar^2(v_4\!-\!i\hbar)v_3v_2(u_3\!-\!i\hbar)(u_2\!-\!i\hbar)(u_1\!-\!i\hbar)
\nonum
&& \phantom{w_{0;+-+-}(\th) = }{}
-\hbar^4(v_3\!-\!i\hbar)u_4u_3(u_1\!-\!i\hbar)
-\hbar^4v_2u_4(u_2\!-\!i\hbar)(u_1\!-\!i\hbar) \nonumber\\[2mm]
&& w_{0;-+-+}(\th) = \hbar^2(v_1\!-\!i\hbar)(u_1\!-\!i\hbar)
\Big((v_3\!-\!i\hbar)v_2u_4u_3 + 
(v_4\!-\!i\hbar)v_3(u_3\!-\!i\hbar)(u_2\!-\!i\hbar) \Big) \nonum
&& \phantom{w_{0;+-+-}(\th) = }{}
-\hbar^4(v_1\!-\!i\hbar)(u_1\!-\!i\hbar)u_4(u_2\!-\!i\hbar)\nonum
&& w_{0;--++}(\th) = \hbar^2(v_2\!-\!i\hbar)(v_1\!-\!i\hbar)
                          (u_2\!-\!i\hbar)(u_1\!-\!i\hbar)
\left(u_4v_3+u_3v_4-i\hbar(u_3\!+\!v_4)-\hbar^2\right) \nonumber
\ea
As required they are invariant under $u_j \leftrightarrow v_j$ and enjoy
the property (\ref{eflip4}).

The symmetric multiplier functions $\phi_0(\th)$ are now given by the
product of a symmetric polynomial $v_0(\th)$ and a factor $u_0(\th)$
depending on the  first Bethe root. Explicitly
\ba
&\mbox{$\n_0 =0$ series:}&\quad \phi_0(\th) = u_0(\th)[12 \tau_4 + \tau_2^2 - 
8\hbar^2 \tau_2 + 7 \hbar^4]\;,\label{phi0}\\[3mm]
&\mbox{$\n_0 =2$ series:}&\quad \phi_0(\th) = u_0(\th)[16 \tau_4\tau_2
-18 \tau_3^2  
- 4 \tau_2^3 + 18 \hbar^2 \tau_2^2 - 40 \hbar^2 \tau_4 
-24 \hbar^4 \tau_2 + 10 \hbar^6]\;,\nonumber
\ea
where 
\ba
&& u_0(\th) = \frac{2}{\hbar} \frac{U_+ + U_-}{U_+U_-}
\left[i( U_+ - U_-) + \frac{1}{2\hbar}(U_+ + U_-)(4 t_1 - (\th_4
+\th_3 + \th_2 +\th_1)) \right]\;,\nonum
&&\sspace \quad \mbox{with}\quad U_\pm  = \prod_{j=1}^4 (\th_j - t_1
\pm i\hbar/2)\;.  
\ea 
$u_0(\th)$ is completely symmetric, boost invariant and real for real
$\th$'s. Using 
$$
2 i\hbar \frac{U_+ - U_-}{U_+ + U_-}\Bigg|_{\th_4 = \th_3 +i\hbar} = 
3 \th_3 + i\hbar -\th_1 -\th_2 \mp \sqrt{\th_{21}^2 + \hbar^2}\;,
$$
one verifies 
$$
u_0(\th)\bigg|_{\th_4 = \th_3 + i\hbar} = \frac{\pm 16 \sqrt{\th_{21}^2 +
\hbar^2}}%
{\Big(3i\hbar + \th_{32} +\th_{31} \mp \sqrt{\th_{21}^2 + \hbar^2}\Big)
\Big(-i\hbar + \th_{32} +\th_{31} \mp \sqrt{\th_{21}^2 + \hbar^2}\Big)}\;,
$$
and further (\ref{IIbar}). Finally the $\n=4$ member of the two $e=0$
series is given by 
\be
\fbar_{0;\sigma_4\sigma_3\sigma_2\sigma_1}(\th) =
\phi_0(\th) \prod_{k > l} \frac{i}{\th_{kl} + i\hbar}
\;w_{0;\sigma_4\sigma_3\sigma_2\sigma_1}(\th)\;,
\ee
with $\phi_0(\th)$ given in (\ref{phi0}).

The semi-classical limit of these $\n \leq 4$ solutions is readily
taken, and one can verify the general pattern described in section
4.2. Specifically let us verify the semi-classical residue equation
(\ref{clres2}), and along the way determine the constant $Z$ in 
(\ref{clres1}).  It is convenient to work with the reduced
functions $\fbar$. So, in a first step we note the counterparts of
equations (\ref{fsolcl}) -- (\ref{clres2}) in terms of $\fbar$. One
finds
\ba
\fbar^{\rm cl}_{e;\sigma_\n\ldots \sigma_1}(\th) \is
\phi_e^{\rm cl}(\th)\,w^{\rm cl}_{e;\sigma_\n\ldots \sigma_1}(\th)\;
\prod_{k >l} \frac{i}{\th_{kl}}\;,\quad \mbox{where} \nonum
\fbar_{e;\sigma_\n\ldots \sigma_1}(\th) \is \hbar^{\Lambda} \,
\fbar^{\rm cl}_{e;\sigma_\n\ldots \sigma_1}(\th) + O(\hbar^{\Lambda
+1}) \;. 
\label{fbarcl}
\ea 
If we assume analogously to (\ref{clres1})
\be
\lim_{\hbar \ra 0} \left[ \hbar^{-\Lambda} 
\fbar_{e;\sigma_{\n}\ldots \sigma_1}(\th)
\Big|_{\th_{k+1} = \th_k +i\hbar} \right] =
\overline{Z}\;\fbar^{\rm cl}_{e;\sigma_{\n}\ldots \sigma_1}(\th)\,
\Big|_{\th_{k+1} = \th_k +i\hbar}\;.
\label{clresbar1}
\ee
and similarly for $\th_{k+1} = \th_k-i\hbar$, the recursive equations 
(\ref{IIbar}) turn into 
\ba
\overline{Z} \,
\fbar^{\rm cl}_{e;\eps_\n\ldots\eps_1}(\th)\bigg|_{\th_{k+1}=\th_k} &=&
C_{\eps_{k+1}\eps_k}\,
\fbar^{\rm cl}_{e;\eps_\n\ldots\eps_{k+2}\eps_{k-1}\ldots \eps_1}(p_k \th)
\,\prod_{l \neq k+1,k} \th_{kl}^2 
\bigg( \!\!- \!\!\!\sum_{j\not=k+1,k}\frac{\eps_j}{\th_{kj}}\bigg)\,.
\label{clresbar2}
\ea
On the other hand $f$ and $\fbar$ are related by (\ref{ansatz}), while
$f^{\rm cl}$ and $\fbar^{\rm cl}$ are related by 
\be
f^{\rm cl}_{e;\sigma_\n\ldots\sigma_1}(\th) =
\fbar^{\rm cl}_{e;\sigma_\n\ldots\sigma_1}(\th)\, 
d_e^{(\n)}(\th) \prod_{k>l} \frac{-i}{\th_{kl}}\;.
\ee
Matching (\ref{clres1}) against (\ref{clresbar1}) one finds
\be
Z = (-)^{\n -1} \psi_0 \,\overline{Z}\;,\quad \psi_0 \approx 1.54678\,.
\ee 
It remains to verify (\ref{clresbar2}) and to determine
$\overline{Z}$. To this end we first note
\be
\lim_{\hbar \ra 0} \left[ \phi_e(\th)\Big|_{\th_{k+1} = \th_k \pm
i\hbar} \right] 
= \Big[\lim_{\hbar \ra 0}\phi_e(\th)\Big]_{\th_{k+1} = \th_k}\;,
\label{phires}
\ee
which for the $\phi_e(\th)$ involving Bethe roots is note quite
automatic. It follows however from the observation that the classical
limit of the $t_\alpha$ (\ref{rootcl}) and the pinching operation
(\ref{tred}) commute in the relevant situations: (\ref{phires}) is
only relevant when the right hand side of (\ref{clresbar2}) is
non-vanishing; that is when $\eps_{k+1} \neq \eps_k$, and when a
branch of the Bethe roots is selected by having all but $\th_{k+1}$
real, say. With these specifications one can choose a labeling of the
Bethe roots such that $j(\alpha) \neq k,k+1$ for all $\alpha <
\Lambda$. Indeed, since either $k\in I_\Lambda$ or $k\!+\!1\in
I_\Lambda$, only one of the corresponding $\th$'s appears on the right
hand side of (\ref{rootcl}), which one can label to be
$\th_{j(\Lambda)}$. This ensures the asserted commutativity with the
pinching operation (\ref{tred}).

A similar argument can then be applied to the remainder
$\fbar_e(\th)/\phi_e(\th)$.  The components of the Bethe vectors are
symmetric polynomials in the Bethe roots, and after canceling common
terms against the $\prod_{k>l} 1/\th_{kl}$ numerator, the operation to
be performed on the left hand side of (\ref{clresbar1}) is known to
have a regular limit. The result must thus be proportional to the
right hand side.  A proportionality constant different from $1$ can
arise as a remnant of the before-mentioned cancellations. In
principle, the constant $\overline{Z}$ could depend on the solution
considered. However the explicit evaluation for the $\n \leq 4$
solutions suggests the universal value $\overline{Z} = - 1/2$.
        
To see this, note e.g. 
\ba
\phi_1^{\rm cl}(\th)\Big|_{\th_3 = \th_2} \is -2(\th_{21})^2\;,\nonum
\phi_2^{\rm cl}(\th)\Big|_{\th_4 = \th_3} \is -2(\th_{13}+\th_{23})
(\th_{32}\th_{31}\th_{21})^2\;,\nonum
\phi_0^{\rm cl}(\th)\Big|_{\th_4 = \th_3} \is -4(\th_{21})^3 (\th_{32})^2
\;,\quad \mbox{$\n_0 = 2\;$ series}\,,\nonum
\phi_0^{\rm cl}(\th)\Big|_{\th_4 = \th_3} \is -4\th_{21}(\th_{32})^2
\;,\quad \quad \,\mbox{$\n_0 =0\;$ series}\,.
\ea
{}From here one can readily verify (\ref{clresbar2}) with 
$\overline{Z} = - 1/2$.

\bigskip
\bigskip
\bigskip


\end{document}